\documentclass[11pt,a4paper,english]{article}
\usepackage[T1]{fontenc}
\usepackage[latin9]{inputenc}
\usepackage{color}
\usepackage{babel}
\usepackage{amsmath}
\usepackage{amssymb}
\usepackage{setspace}
\usepackage[unicode=true,pdfusetitle,
 bookmarks=false,
 breaklinks=false,pdfborder={0 0 0},backref=false,colorlinks=true]
 {hyperref}
\usepackage{breakurl}

\makeatletter

\providecommand{\tabularnewline}{\\}

\numberwithin{equation}{section}

\usepackage[a4paper,vmargin={20mm,20mm},hmargin={30mm,30mm}]{geometry}
\usepackage{txfonts}
\DeclareMathAlphabet{\mathcal}{OMS}{cmsy}{m}{n}
  \usepackage{hyperref}
  \hypersetup{colorlinks=true,citecolor=cyan,linkcolor=magenta}
\usepackage[dvips]{pstricks}% This command sets the scale of all the diagrams
\psset{xunit=.5pt,yunit=.5pt,runit=.5pt}
\newcommand{\sfrac}[2]{{\textstyle\frac{#1}{#2}}}
\newcommand{\half}{\sfrac{1}{2}}
\date{}

\makeatother

\begin{document}
\begin{center}
\begin{flushright}\footnotesize \texttt{DESY-2012-127}\\ \texttt{ZMP-HH/12-14}\\ \texttt{HBM-432}\\ \texttt{HU-Math-09/12}\\ \texttt{HU-EP-12/21}\\	\texttt{AEI-2012-071}\\ \vspace{0.5cm} \end{flushright}
\par\end{center}

\begin{center}
{\huge From Baxter Q-Operators to Local Charges}
\par\end{center}{\huge \par}

\begin{center}
\vspace*{1.2cm}

\par\end{center}

\begin{center}
{\large Rouven Frassek\textsuperscript{a,b} \& Carlo Meneghelli\textsuperscript{c,d}}
\par\end{center}{\large \par}

\begin{singlespace}
\begin{center}
\textit{\textsuperscript{a} Institut fr Mathematik und Institut
fr Physik, Humboldt-Universitt zu Berlin }\\
\textit{Johann von Neumann-Haus, Rudower Chaussee 25, 12489 Berlin,
Germany }
\par\end{center}

\begin{center}
\textit{\textsuperscript{b} Max-Planck-Institut fr Gravitationsphysik,
Albert-Einstein-Institut}\\
\textit{Am Mhlenberg 1, 14476 Potsdam, Germany }
\par\end{center}

\begin{center}
\textit{\textsuperscript{c} Fachbereich Mathematik, Universitt Hamburg
Bundesstrae 55, 20146 Hamburg, Germany }
\par\end{center}

\begin{center}
\textit{\textsuperscript{d} Theory Group, DESY, Notkestrae 85, D-22603,
Hamburg, Germany }
\par\end{center}
\end{singlespace}

\begin{center}
rfrassek@physik.hu-berlin.de\\
carlo.meneghelli@gmail.com\vfill{}

\par\end{center}
\begin{abstract}
We discuss how the shift operator and the Hamiltonian enter the hierarchy
of Baxter Q-operators in the example of $\mathfrak{gl}(n)$ homogeneous
spin-chains. Building on the construction that was recently carried out by
the authors and their collaborators, we find that a reduced set of
\mbox{Q-operators} can be used to obtain local charges. The mechanism relies
on projection properties of the corresponding $\mathcal{R}$-operators
on a highest/lowest weight state of the quantum space. It is intimately
related to the ordering of the oscillators in the auxiliary space.
Furthermore, we introduce a diagrammatic language that makes these
properties manifest and the results transparent. Our approach circumvents
the paradigm of constructing the transfer matrix with equal representations
in quantum and auxiliary space and underlines the strength of the
Q-operator construction.\vfill{}

\newpage{}

\tableofcontents{}
\end{abstract}

\section{Introduction\label{sec:Introduction}}

In 1971 R.J. Baxter introduced the Q-operator along with the celebrated
Baxter equation/TQ relation \cite{Baxter1972a} in order to calculate
exactly the partition function of the eight-vertex model. The method
of functional relations and commuting transfer matrices originated
in this work. It plays a fundamental role in the theory of integrable
quantum systems. 

One of the most important developments in the field since then is
the framework of the quantum inverse scattering method (QISM), see
e.g. \cite{Faddeev2007}. This approach builds on the existence of
Lax operators which are solutions to the Yang-Baxter equation and
can be seen as the generating objects of the model. It employs the
idea of an auxiliary space to construct transfer matrices which define
one parameter families of commuting operators, labeled by the spectral
parameter $z$. Here the transfer matrix built out of Lax matrices
with the same representation in quantum and auxiliary space is especially
important. The Hamiltonian, along with all other local charges of
the spin-chain, arises as a logarithmic derivative around the shift
point of this particular transfer matrix \cite{Tarasov1983}. As the
local charges belong to the family of commuting operators their spectrum
can be obtained from the algebraic Bethe ansatz \cite{Kulish1983a}. 

The link between Q-operators and the Hamiltonian as well as higher
local charges is rather indirect. From the point of view of the algebraic
Bethe ansatz the link is established by applying the Hamiltonian to
Bethe vectors. Following this procedure one obtains the eigenvalues
of the Hamiltonian in terms of Bethe roots \cite{Tarasov1983}. By
identifying the Baxter Q-functions with the polynomial which vanishes
at the Bethe roots one may write the eigenvalues of the local integrals
of motion for the Heisenberg chain as \cite{Faddeev1995} 
\begin{equation}
I_{k}=\frac{\partial^{k}}{\partial z^{k}}\ln\left.\frac{Q(z+\delta)}{Q(z-\delta)}\right|_{z=0}\,.\label{eq:intofm}
\end{equation}
The first two charges $I_{0}$ and $I_{1}$ correspond to the momentum
and the nearest-neighbor Hamiltonian, respectively. The label $\delta$
is the $\mathfrak{sl}(2)$ weight of the local vacuum in the algebraic
Bethe ansatz. As mentioned in \cite{Faddeev1995}, where integrable
spin-chains emerged in the context of high energy QCD \cite{Lipatov1994571-574a},
the relation between the integrals of motion and the Q-functions can
be extended to the operatorial level. From the viewpoint of the analytic Bethe ansatz, expressing the relevant T-function in terms of Q-functions, see e.g.~\cite{Frassek2011}, in principle allows to derive higher rank analogues of the expression (\ref{eq:intofm}) using functional relations and analyticity properties\footnote{To our knowledge this procedure has never been carried out beyond the class of representations known as Kirillov-Reshetikhin modules \cite{Kirillov1990}. Even for this class of representations, and importantly their infinite dimensional analogues considered in this paper, it is hard to find the general case in the literature.}. Here, we neither make an ansatz for the wave function nor use functional relations to obtain local charges. The goal of the present article
is to provide a more \textit{direct} connection between Q-operators
and local charges.

In a series of papers \cite{Frassek2011,Bazhanov2010a,Bazhanov2010,Frassek2010}
Q-operators were constructed for $\mathfrak{gl}(n)$ homogeneous spin-chains
from fundamental principles. The construction follows the quantum
inverse scattering method employing degenerate solutions of the Yang-Baxter
equation as generating objects. While the quantum space of these Lax
operators is determined by the model, the auxiliary space is infinite-dimensional.
It is realized by a set of oscillator algebras as pioneered in \cite{Bazhanov1999}.
All functional relations involving Q-operators and transfer matrices
of the type mentioned above follow from remarkable fusion/factorization
properties of the degenerate Lax operators. The hierarchy of Q-operators
is most easy to construct. It is best illustrated using Hasse diagrams
\cite{Tsuboi2009}. In \cite{Frassek2011} the nearest-neighbor Hamiltonian
was obtained solving the Yang-Baxter equation for the R-matrix with
equal representation in quantum and auxiliary space, see also \cite{MacKay1991},
while the dispersion relation was derived for some finite-dimensional
representations from the functional relations of the Q-operators and generalized transfer matrices.
Since $\mathcal{R}$-operators for Q-operators are conceptually simple solutions to
the Yang-Baxter equation, it is desirable to obtain the nearest-neighbor
Hamiltonian and also higher local charges from Q-operators avoiding
this detour. 

In the present article we show how local charges can be extracted
directly from the Q-operators built in \cite{Frassek2011}. No reference
to the transfer matrix mentioned above is required. This avoids the
notoriously complicated construction of the transfer matrix with equal
representation in quantum and auxiliary space. In section~\ref{sec:Review}
we give a brief review of the recent construction of Q-operators for
$\mathfrak{gl}(n)$ homogeneous spin-chains and establish some notation.
In section~\ref{sec:Alternative-presentation-of} we introduce an
opposite product on the auxiliary space and obtain alternative presentations
of the degenerate solutions used for the construction of Q-operators,
hereafter referred to as $\mathcal{R}$-operators. Section~\ref{sec:Projection-property-of}
is dedicated to the extremely important projection properties of the
degenerate Lax operators. Their alternative presentation obtained
in section~\ref{sec:Alternative-presentation-of} is essential in
order to fully exploit these properties. After developing these techniques
we introduce a convenient diagrammatic language for $\mathcal{R}$-operators
which extends to Q-operators in section~\ref{sec:Diagrammatics-and-local}.
It further concerns the derivation of the shift operator and the Hamiltonian
from Q-operators. Equation (\ref{eq:haction}) is one of the main
results of this paper. It defines the Hamiltonian density in terms
of the novel $\mathcal{R}$-operators for Q-operators introduced in
\cite{Frassek2011}. Section~\ref{sub:conclusion} offers
a summary of our results and suggestions for further studies. Furthermore,
we provide several appendices where specific examples are studied.
They also contain definitions and properties of the operatorial shifted
weights frequently used in this paper.

\section{Review\label{sec:Review}}

In a series of papers \cite{Bazhanov2010a,Bazhanov2010,Frassek2011}
new solutions to the Yang-Baxter equation were derived. They allow
to construct Baxter Q-operators for $\mathfrak{gl}(n)$ invariant
spin-chains. These so called $\mathcal{R}$-operators are of remarkably
compact form and can be written as%
\footnote{For reasons that will become clear in the next section we explicitly
denoted the product by {}``$\cdot$''.%
}
\begin{equation}
\mathcal{R}_{I}(z)\,=e^{\bar{\mathbf{a}}_{c}^{\dot{c}}\, J_{\dot{c}}^{c}}\cdot\mathcal{R}_{0,I}(z)\cdot e^{-\mathbf{a}_{\dot{c}}^{c}\, J_{c}^{\dot{c}}}\,,\label{eq:RI}
\end{equation}
with 
\begin{equation}
\mathcal{R}_{0,I}(z)=\,\rho_{I}(z)\,\prod_{k=1}^{\vert\bar{I}\vert}\,\Gamma(z-\sfrac{\vert\bar{I}\vert}{2}-\hat{\ell}_{k}^{\bar{I}}+1)\,.\label{eq:R0}
\end{equation}
These equations require some explanations. The letter $I$ denotes
a subset of the set $\{1,\ldots,n\}$ of cardinality $\vert I\vert$.
The undotted indices take values from the set $I$ and the dotted
ones from its conjugate $\bar{I}$
\begin{equation}
a,b,c\in I,\quad\quad\dot{a},\dot{b},\dot{c}\in\bar{I},\quad\quad A,B,C\in I\cup\bar{I}\,.\label{eq:sets}
\end{equation}

The $\mathcal{R}$-operators are composed
out of $\vert I\vert\cdot\vert\bar{I}\vert$ families of oscillators
\begin{equation}
[\mathbf{a}_{\dot{b}}^{a},\bar{\mathbf{a}}_{d}^{\dot{c}}]=\delta_{d}^{a}\delta_{\dot{b}}^{\dot{c}}\label{eq:osc_alg}
\end{equation}
and \textbf{$\mathfrak{gl}(n)$ }generators $J_{B}^{A}$ with 
\begin{equation}
[J_{B}^{A},J_{D}^{C}]=\delta_{B}^{C}J_{D}^{A}-\delta_{D}^{A}J_{B}^{C}\,.\label{eq:gln_alg}
\end{equation}
The choice of the set $I$ naturally identifies a subalgebra $\mathfrak{gl}(I)$
of $\mathfrak{gl}(n)$, i.e.~the subalgebra spanned by $J^a_b$, see (\ref{eq:sets}). The precise definition of the operators $\hat{\ell}_{k}^{\bar{I}}$
entering (\ref{eq:R0}) can be found in the appendix~\ref{sec:Shifted-weights},
they are operatorial shifted weights of the subalgebra $\mathfrak{gl}(\bar{I})$.
The operators $\mathcal{R}_{I}$ are elements of a suitable extension
of the product space $\mathcal{U}(\mathfrak{gl}(n))\otimes\mathcal{H}^{(I,\bar{I})}$
in the following denoted by $\mathfrak{A}_{I}$. The normalization
$\rho_{I}$ is not determined by the Yang-Baxter relation and is discussed
in the next sections.

The $\mathcal{R}$-operators above satisfy the Yang Baxter equation

\begin{equation}
\mathcal{L}(z_{1})\,\mathbf{L}_{I}(z_{2})\cdot\mathcal{R}_{I}(z_{2}-z_{1})=\mathcal{R}_{I}(z_{2}-z_{1})\cdot\mathbf{L}_{I}(z_{2})\,\mathcal{L}(z_{1})\,.\label{eq:YBE}
\end{equation}
Here $\mathbf{L}_{I}$ denotes the operator $\mathcal{R}_{I}$ with
fundamental representation in the $\mathfrak{gl}(n)$ part
\begin{equation}
\mathbf{L}_{I}(z)=\begin{pmatrix}z\delta_{b}^{a}+H_{b}^{a} & \bar{\mathbf{a}}_{b}^{\dot{a}}\\
-\mathbf{a}_{\dot{b}}^{a} & \delta_{\dot{b}}^{\dot{a}}
\end{pmatrix}\quad\text{for}\quad I=\{1,\ldots,\vert I\vert\}\,,\label{eq:R-fundamental}
\end{equation}
 with%
\footnote{Here we only consider the \textit{minimal} case discussed in \cite{Frassek2011}.%
} $H_{b}^{a}=-\bar{\mathbf{a}}_{b}^{\dot{c}}\mathbf{a}_{\dot{c}}^{a}-\half\delta_{\dot{c}}^{\dot{c}}\delta_{b}^{a}$
where the summation over the dotted indices is understood. The operator
$\mathcal{L}$ denotes the well-known Lax matrix 
\begin{equation}
\mathcal{L}(z)=\begin{pmatrix}z\delta_{b}^{a}+J_{b}^{a} & J_{b}^{\dot{a}}\\
J_{\dot{b}}^{a} & z\delta_{\dot{b}}^{\dot{a}}+J_{\dot{b}}^{\dot{a}}
\end{pmatrix}\,.\label{eq:Lax}
\end{equation}
Baxter Q-operators are constructed as regularized traces over the
oscillator space of monodromies built from the operators $\mathcal{R}_{I}$.
Following \cite{Frassek2011}, they are given by 

\begin{equation}
\mathbf{Q}_{I}(z)=e^{iz\,\phi_{I}}\,\widehat{\text{Tr}}_{{\mathcal{H}}^{(I,\bar{I})}}\big\{\mathcal{D}_{I}\,\mathcal{R}_{I}(z)\otimes\ldots\otimes\mathcal{R}_{I}(z)\big\}\quad\text{with}\quad\phi_{I}=\sum_{a\in I}\phi_{a}\,.\label{eq:Qop}
\end{equation}
Here the quantum space consists out of $L$ sites and will be denoted
by $\mathcal{V}=\mathcal{V}_{L}\otimes\ldots\otimes\mathcal{V}_{1}$.
In the following each $\mathcal{V}_{i}$ corresponds to the same representation
$\Lambda$ of $\mathfrak{gl}(n)$. The regulator in (\ref{eq:Qop})
is defined as 
\begin{equation}
\mathcal{D}_{I}=\exp\Big\{-i\,\sum_{a,\dot{b}}\phi_{a\dot{b}}\,\mathbf{h}_{a\dot{b}}\Big\}\,,\label{eq:regulator}
\end{equation}
where we introduced the twist parameters $\phi_{a\dot{b}}=\phi_{a}-\phi_{\dot{b}}$,
the number operator $\mathbf{h}_{a\dot{b}}=\bar{\mathbf{a}}_{a}^{\dot{b}}\mathbf{a}_{\dot{b}}^{a}+\sfrac{1}{2}$
and the normalized trace 
\begin{equation}
\widehat{\text{Tr}}_{\mathcal{H}}\big\{ e^{-i\phi\textbf{h}}\star\big\}\,=\frac{\text{Tr}_{\mathcal{H}}\big\{ e^{-i\phi\textbf{h}}\star\big\}}{\text{Tr}_{\mathcal{H}}\big\{ e^{-i\phi\textbf{h}}\big\}}\,.\label{eq:norm_trace}
\end{equation}
The operators $\mathbf{Q}_{I}$ generate a large family of commuting
operators. These operators are functionally dependent, they satisfy
certain quadratic equations known as QQ-relations. Those functional
relations can be regarded as {}``off-shell'' Bethe equations. The
hierarchy of the $2^{n}$ Q-operators can be graphically exposed in
the Hasse diagram \cite{Tsuboi2009}.

\section{Alternative presentation of $\mathcal{R}$-operators: Reordering
oscillators\label{sec:Alternative-presentation-of}}

The solution (\ref{eq:RI}) to the Yang-Baxter equation (\ref{eq:YBE})
is presented as a normal ordered expression in the oscillators $\{\bar{\mathbf{a}}_{c}^{\dot{c}},\mathbf{a}_{\dot{c}}^{c}\}$.
For reasons that will become clear in the next section we are also
interested in its expression which is anti-normal ordered in the oscillators
of the auxiliary space. The anti-normal ordered form of the $\mathcal{R}$-operators
can be obtained either from the Yang-Baxter equation or directly by
reordering the oscillators in (\ref{eq:RI}). As we will see the approach
from the Yang-Baxter equation will be very powerful to obtain the
desired expressions. However, it is not possible to fix the relative
normalization by this method.

\subsection{Yang-Baxter approach\label{sub:Yang-Baxter-approach}}

To derive the expression for the anti-normal ordered $\mathcal{R}$-operators
directly from the Yang-Baxter equation it turns out to be convenient
to introduce an opposite product on $\mathfrak{A}_{I}$. Let $\mathcal{O}\in\mathfrak{A}_{I}$
be written as 
\begin{equation}
\mathcal{O}=\sum_{k}a(k)\otimes b(k)\,,\label{eq:opdecomp}
\end{equation}
with $a(k)\in\mathcal{U}(\mathfrak{gl}(n))$ and $b(k)\in\mathcal{H}^{(I,\bar{I})}$.
Given two elements of $\mathfrak{A}_{I}$ the product used in (\ref{eq:YBE})
is defined as 
\begin{equation}
\mathcal{O}_{1}\cdot\mathcal{O}_{2}=\sum_{k,l}a_{1}(k)\, a_{2}(l)\otimes b_{1}(k)\, b_{2}(l)\,.\label{eq:prod}
\end{equation}
We now define the opposite product $\circ$ as 
\begin{equation}
\mathcal{O}_{1}\circ\mathcal{O}_{2}=\sum_{k,l}a_{1}(k)\, a_{2}(l)\otimes b_{2}(l)\, b_{1}(k)\,.\label{eq:prod_op}
\end{equation}
One can easily check that this product is associative. The Yang-Baxter
equation (\ref{eq:YBE}) can then be written as
\begin{equation}
\mathcal{L}(z_{1})\,\mathcal{R}_{I}(z_{2}-z_{1})\circ\mathbf{L}_{I}(z_{2})=\mathbf{L}_{I}(z_{2})\circ\mathcal{R}_{I}(z_{2}-z_{1})\,\mathcal{L}(z_{1})\,.\label{eq:YBE_opposite}
\end{equation}
We like to stress that this is exactly the same equation as (\ref{eq:YBE})
only rewritten in terms of the opposite product. Now, in analogy to
\cite{Frassek2011} we substitute the ansatz 
\begin{equation}
\mathcal{R}_{I}(z)\,=\, e^{\bar{\mathbf{a}}_{c}^{\dot{c}}\, J_{\dot{c}}^{c}}\,\circ\,\tilde{\mathcal{R}}_{0,I}(z)\,\circ\, e^{-\mathbf{a}_{\dot{c}}^{c}\, J_{c}^{\dot{c}}}\label{eq:RI_anti}
\end{equation}
into (\ref{eq:YBE_opposite}) and as before obtain four sets of defining
relations for $\tilde{\mathcal{R}}_{0,I}$. As there is some redundancy
in these equations we only present one of them 
\begin{equation}
\tilde{\mathcal{R}}_{0,I}(z)\left((z+\frac{\bar{I}}{2})J_{b}^{\dot{a}}-J_{b}^{c}J_{c}^{\dot{a}}\right)=J_{b}^{\dot{a}}\,\tilde{\mathcal{R}}_{0,I}(z)\,.\label{eq:antinormalybe}
\end{equation}
Comparing this equation to \cite{Frassek2011} it is easy to recognize
that $\tilde{\mathcal{R}}_{0,I}(z)$ satisfies the same defining relation
as $\mathcal{R}_{0,\bar{I}}^{-1}\,(z+\sfrac{n}{2})$. We conclude
that 

\begin{equation}
\tilde{\mathcal{R}}_{0,I}(z)=\,\tilde{\rho_{I}}(z)\,\prod_{k=1}^{\text{\ensuremath{\vert}}I\vert}\,\frac{1}{\Gamma(z+\sfrac{\vert\bar{I}\vert}{2}-\hat{\ell}_{k}^{I}+1)}\,.\label{Rt0-2}
\end{equation}
The ratio of $\rho_{I}$ and $\tilde{\rho}_{I}$ entering (\ref{eq:R0})
and (\ref{Rt0-2}) can be determined by requiring that (\ref{eq:RI})
and (\ref{eq:RI_anti}) are the \textit{same} operators. It is investigated
in the next subsection.

\subsection{Direct approach\label{sub:Direct-approach}}

The $\mathcal{R}$-operators (\ref{eq:RI}) and (\ref{eq:RI_anti})
can be written as 
\begin{eqnarray}
\mathcal{R}_{I}(z) & = & \sum_{n,m=0}^{\infty}\frac{(-1)^{m}}{n!\, m!}\,\bar{\mathbf{a}}_{a_{1}}^{\dot{a}_{1}}\cdots\bar{\mathbf{a}}_{a_{n}}^{\dot{a}_{n}}\,\mathbf{a}_{\dot{b}_{1}}^{b_{1}}\cdots\mathbf{a}_{\dot{b}_{m}}^{b_{m}}\, J_{\dot{a}_{1}}^{a_{1}}\cdots J_{\dot{a}_{n}}^{a_{n}}\,\mathcal{R}_{0,I}(z)\, J_{b_{1}}^{\dot{b}_{1}}\cdots J_{b_{m}}^{\dot{b}_{m}}\,,\label{Rliexpand}\\
\mathcal{R}_{I}(z) & = & \sum_{n,m=0}^{\infty}\frac{(-1)^{m}}{n!\, m!}\,\mathbf{a}_{\dot{b}_{1}}^{b_{1}}\cdots\mathbf{a}_{\dot{b}_{m}}^{b_{m}}\,\bar{\mathbf{a}}_{a_{1}}^{\dot{a}_{1}}\cdots\bar{\mathbf{a}}_{a_{n}}^{\dot{a}_{n}}\, J_{\dot{a}_{1}}^{a_{1}}\cdots J_{\dot{a}_{n}}^{a_{n}}\,\tilde{\mathcal{R}}_{0,I}(z)\, J_{b_{1}}^{\dot{b}_{1}}\cdots J_{b_{m}}^{\dot{b}_{m}}\,.\label{Rtliexpand}
\end{eqnarray}
Here we expanded the exponentials using the definition of the products
in (\ref{eq:prod}) and (\ref{eq:prod_op}), respectively. To obtain
the relation between $\mathcal{R}_{0,I}$ and $\tilde{\mathcal{R}}_{0,I}$
we have to reorder the oscillators in one of the two expressions.
We find that for each pair $\bar{\mathbf{a}}_{c}^{\dot{c}}$, $\mathbf{a}_{\dot{c}}^{c}$
that is reordered in (\ref{Rliexpand}) $\mathcal{R}_{0,I}$ is {}``conjugated''
by the corresponding $\mathfrak{gl}(\{c,\dot{c}\})$ generators as
\begin{equation}
\mathcal{R}_{0,I}\longrightarrow\sum_{k=0}^{\infty}\frac{1}{k!}\,(J_{\dot{c}}^{c})^{k}\,\mathcal{R}_{0,I}(z)\,(J_{c}^{\dot{c}})^{k}.\label{eq:R_conjugation}
\end{equation}
This relation is obtained using (\ref{eq:reorder2}) and does not
rely on the precise form of $\mathcal{R}_{0,I}$. After subsequent
conjugation of $\mathcal{R}_{0,I}$ with all $\vert I\vert\cdot\vert\bar{I}\vert$
$\mathfrak{gl}(\{c,\dot{c}\})$ subalgebra generators of $\mathfrak{gl}(n)$
one obtains an expression for $\tilde{\mathcal{R}}_{0,I}$:
\begin{equation}
\tilde{\mathcal{R}}_{0,I}(z)=\sum_{\{k_{c\dot{c}}\}=0}^{\infty}\prod_{c\in I,\dot{c}\in\bar{I}}\frac{1}{\sqrt{k_{c\dot{c}}!}}\,(J_{\dot{c}}^{c})^{k_{c\dot{c}}}\,\mathcal{R}_{0,I}(z)\,\prod_{c\in I,\dot{c}\in\bar{I}}\frac{1}{\sqrt{k_{c\dot{c}}!}}(J_{c}^{\dot{c}})^{k_{c\dot{c}}}.\label{eq:R_conjugation-1}
\end{equation}
This fixes the ratio of the prefactors $\rho$ and $\tilde{\rho}$
appearing in (\ref{eq:R0}) and (\ref{Rt0-2}). Naively, (\ref{eq:R_conjugation-1})
appears rather different from (\ref{Rt0-2}). However, they must coincide
since they satisfy the same defining relations. This is explicitly
demonstrated for the case of $\mathfrak{gl}(2)$ in appendix~\ref{sec:Example:gl2}.
See also section~\ref{sec:Projection-property-of} for a clarifying
discussion on the normalization.

\section{Projection properties of $\mathcal{R}$-operators\label{sec:Projection-property-of}}

The construction of local charges in the conventional QISM relies
on the \textit{fundamental} R-matrix $\mathbf{R}$ for which the auxiliary
space is the same as the quantum space at each site. It requires the
existence of a special point $z_{*}$ where the R-matrix reduces to
the permutation operator 
\begin{equation}
\mathbf{R}(z_{*})=\mathbf{P}\,,\label{eq:regularity}
\end{equation}
see e.g.~\cite{Faddeev2007}. This property is often referred to as regularity condition. The construction
of local charges presented here bypasses the use of the fundamental
R-matrix and is based on remarkable properties of the $\mathcal{R}$-operators
in (\ref{eq:RI}) for special values of the spectral parameter. For
the example of the fundamental representation in the quantum space
we find that the operator \textbf{$\mathbf{L}_{I}$ }given in (\ref{eq:R-fundamental})
degenerates at \textit{two} special points. Using \textit{both} products
introduced in the previous section we find that 
\begin{equation}
\mathbf{L}_{I}\left(+\sfrac{\vert\bar{I}\vert}{2}\right)=\begin{pmatrix}\bar{\mathbf{a}}_{b}^{\dot{c}}\\
\delta_{\dot{b}}^{\dot{c}}
\end{pmatrix}\cdot\begin{pmatrix}-\mathbf{a}_{\dot{c}}^{a} & \delta_{\dot{c}}^{\dot{a}}\end{pmatrix},\quad\quad\mathbf{L}_{I}\left(-\sfrac{\vert\bar{I}\vert}{2}\right)=\begin{pmatrix}\bar{\mathbf{a}}_{b}^{\dot{c}}\\
\delta_{\dot{b}}^{\dot{c}}
\end{pmatrix}\circ\begin{pmatrix}-\mathbf{a}_{\dot{c}}^{a} & \delta_{\dot{c}}^{\dot{a}}\end{pmatrix}\,.\label{eq:degenerate_fund}
\end{equation}
Interestingly, properties similar to (\ref{eq:degenerate_fund}) appear quite naturally in the derivation of the Baxter equation, see e.g.~\cite{Baxtereq}. This in turn is strictly connected to Baxter's original idea \cite{Baxter1972a}. The degeneration can be understood from the spectral parameter
dependent part of the $\mathcal{R}$-operators, compare to (\ref{eq:RI}),
(\ref{eq:RI_anti}). For the fundamental representation, it originates
from a reduction of the rank of $\mathcal{R}_{0,I}$ and $\tilde{\mathcal{R}}_{0,I}$
at the special points $\hat{z}=+\sfrac{\vert\bar{I}\vert}{2}$ and
$\check{z}=-\sfrac{\vert\bar{I}\vert}{2}$, respectively. Of distinguished
importance are $\mathbf{L}_{I}$-operators with $\vert\bar{I}\vert=1$.
In this case the rank of the oscillator independent part reduces to
$1$. 

Relations of the type (\ref{eq:degenerate_fund}) hold for any highest/lowest
weight representation of $\mathfrak{gl}(n)$. Their precise form can
be obtained via a careful analysis of the spectrum of the shifted
weight operators $\hat{\ell}_{i}^{K}$ entering $\mathcal{R}_{I}$.
However, the analysis is technically involved. Details can be found
in the appendix~\ref{sec:Shifted-weights}. In the following we restrict
to representations corresponding to rectangular Young diagrams and
their infinite dimensional generalization. 

A rectangular Young diagram is labeled by two parameters $(s,a)$
with $s,a\in\mathbb{N}$ according to 

\begin{equation}
%LaTeX with PSTricks extensions
%%Creator: inkscape 0.48.1
%%Please note this file requires PSTricks extensions
\psset{xunit=.5pt,yunit=.5pt,runit=.5pt}
\begin{pspicture}[shift=-47](94.35027313,76.98535919)
{
\newrgbcolor{curcolor}{0 0 0}
\pscustom[linewidth=1,linecolor=curcolor]
{
\newpath
\moveto(33.85028,76.48535657)
\lineto(53.85028,76.48535657)
\lineto(53.85028,56.48535657)
\lineto(33.85028,56.48535657)
\closepath
}
}
{
\newrgbcolor{curcolor}{0 0 0}
\pscustom[linewidth=1,linecolor=curcolor]
{
\newpath
\moveto(53.85028,76.48535657)
\lineto(73.85028,76.48535657)
\lineto(73.85028,56.48535657)
\lineto(53.85028,56.48535657)
\closepath
}
}
{
\newrgbcolor{curcolor}{0 0 0}
\pscustom[linewidth=1,linecolor=curcolor]
{
\newpath
\moveto(73.85028,76.48535657)
\lineto(93.85028,76.48535657)
\lineto(93.85028,56.48535657)
\lineto(73.85028,56.48535657)
\closepath
}
}
{
\newrgbcolor{curcolor}{0 0 0}
\pscustom[linewidth=1,linecolor=curcolor]
{
\newpath
\moveto(33.85028,56.48535657)
\lineto(53.85028,56.48535657)
\lineto(53.85028,36.48535657)
\lineto(33.85028,36.48535657)
\closepath
}
}
{
\newrgbcolor{curcolor}{0 0 0}
\pscustom[linewidth=1,linecolor=curcolor]
{
\newpath
\moveto(53.85028,56.48535657)
\lineto(73.85028,56.48535657)
\lineto(73.85028,36.48535657)
\lineto(53.85028,36.48535657)
\closepath
}
}
{
\newrgbcolor{curcolor}{0 0 0}
\pscustom[linewidth=1,linecolor=curcolor]
{
\newpath
\moveto(73.85028,56.48535657)
\lineto(93.85028,56.48535657)
\lineto(93.85028,36.48535657)
\lineto(73.85028,36.48535657)
\closepath
}
}
{
\newrgbcolor{curcolor}{0 0 0}
\pscustom[linestyle=none,fillstyle=solid,fillcolor=curcolor]
{
\newpath
\moveto(21.80596548,61.53359502)
\curveto(21.80596548,59.22899059)(20.65366326,57.97193362)(17.24913399,56.61012191)
\curveto(17.09200187,56.55774454)(16.98724712,56.40061242)(16.98724712,56.19110293)
\curveto(16.98724712,55.92921606)(17.0396245,55.82446131)(17.30151136,55.71970657)
\curveto(21.80596548,53.99125324)(21.80596548,52.21042255)(21.80596548,50.3248371)
\lineto(21.80596548,41.10641938)
\curveto(21.85834285,40.21600403)(22.32973921,38.95894707)(23.79630567,37.96377697)
\curveto(25.47238162,36.86385213)(28.14362767,36.18294627)(28.45789191,36.18294627)
\curveto(28.77215615,36.18294627)(28.87691089,36.28770102)(28.87691089,36.60196526)
\curveto(28.87691089,36.9162295)(28.82453352,36.9162295)(28.30075979,37.07336162)
\curveto(27.35796706,37.44000324)(24.73909839,38.48755071)(24.21532466,40.53026827)
\curveto(24.11056991,40.94928726)(24.11056991,41.00166463)(24.11056991,42.25872159)
\lineto(24.11056991,50.42959185)
\curveto(24.11056991,52.15804517)(24.11056991,54.61978172)(18.66332307,56.19110293)
\curveto(24.11056991,57.86717888)(24.11056991,60.11940594)(24.11056991,62.00499138)
\lineto(24.11056991,70.17586164)
\curveto(24.11056991,72.00906971)(24.11056991,73.73752303)(28.6674014,75.41359898)
\curveto(28.87691089,75.46597635)(28.87691089,75.72786322)(28.87691089,75.78024059)
\curveto(28.87691089,76.09450483)(28.77215615,76.25163695)(28.45789191,76.25163695)
\curveto(28.19600504,76.25163695)(23.5344188,75.15171211)(22.32973921,72.95186243)
\curveto(21.80596548,71.95669233)(21.80596548,71.64242809)(21.80596548,70.22823901)
\closepath
\moveto(21.80596548,61.53359502)
}
}
{
\newrgbcolor{curcolor}{0 0 0}
\pscustom[linestyle=none,fillstyle=solid,fillcolor=curcolor]
{
\newpath
\moveto(56.05446011,20.85471568)
\curveto(57.69511365,20.85471568)(60.74204167,20.85471568)(63.32021153,14.44835421)
\curveto(63.55459061,13.82334334)(63.55459061,13.74521698)(64.02334876,13.74521698)
\curveto(64.17960148,13.74521698)(64.49210692,13.74521698)(64.57023328,14.05772241)
\curveto(67.2265295,20.85471568)(69.96095208,20.85471568)(72.77350101,20.85471568)
\lineto(86.52374026,20.85471568)
\curveto(89.25816284,20.93284204)(90.74256367,23.1203801)(91.28944819,23.97977005)
\curveto(92.77384901,26.08918176)(93.9457444,30.30800516)(93.9457444,30.85488968)
\curveto(93.9457444,31.32364784)(93.63323897,31.47990055)(93.24260717,31.47990055)
\curveto(92.77384901,31.47990055)(92.69572265,31.24552148)(92.6175963,30.85488968)
\curveto(90.1175528,24.29227549)(87.6175093,24.29227549)(84.80496036,24.29227549)
\lineto(72.6172483,24.29227549)
\curveto(69.88282572,24.29227549)(66.52339226,24.29227549)(64.02334876,16.24526048)
\curveto(61.36705255,24.29227549)(58.47637725,24.29227549)(55.35132287,24.29227549)
\lineto(43.16361081,24.29227549)
\curveto(40.58544095,24.29227549)(37.92914473,24.29227549)(35.50722759,30.69863696)
\curveto(35.27287351,31.24552148)(35.27287351,31.47990055)(34.80409036,31.47990055)
\curveto(34.41343356,31.47990055)(34.17907948,31.32364784)(34.17907948,30.85488968)
\curveto(34.17907948,30.77676332)(35.42910123,24.21414913)(38.71040833,21.87035835)
\curveto(40.1166828,20.85471568)(41.21045183,20.85471568)(43.08548445,20.85471568)
\closepath
\moveto(56.05446011,20.85471568)
}
}
{
\newrgbcolor{curcolor}{0 0 0}
\pscustom[linestyle=none,fillstyle=solid,fillcolor=curcolor]
{
\newpath
\moveto(6.62500367,59.01786041)
\curveto(6.25000367,59.76786041)(5.68750367,60.29911041)(4.78125367,60.29911041)
\curveto(2.46875367,60.29911041)(0.00000367,57.36161041)(0.00000367,54.45536041)
\curveto(0.00000367,52.58036041)(1.09375367,51.26786041)(2.62500367,51.26786041)
\curveto(3.03125367,51.26786041)(4.03125367,51.36161041)(5.21875367,52.76786041)
\curveto(5.37500367,51.92411041)(6.09375367,51.26786041)(7.03125367,51.26786041)
\curveto(7.75000367,51.26786041)(8.18750367,51.73661041)(8.53125367,52.36161041)
\curveto(8.84375367,53.08036041)(9.12500367,54.29911041)(9.12500367,54.33036041)
\curveto(9.12500367,54.54911041)(8.93750367,54.54911041)(8.87500367,54.54911041)
\curveto(8.68750367,54.54911041)(8.65625367,54.45536041)(8.59375367,54.17411041)
\curveto(8.25000367,52.89286041)(7.90625367,51.70536041)(7.09375367,51.70536041)
\curveto(6.53125367,51.70536041)(6.50000367,52.23661041)(6.50000367,52.61161041)
\curveto(6.50000367,53.04911041)(6.53125367,53.23661041)(6.75000367,54.11161041)
\curveto(6.96875367,54.92411041)(7.00000367,55.14286041)(7.18750367,55.89286041)
\lineto(7.90625367,58.67411041)
\curveto(8.03125367,59.23661041)(8.03125367,59.26786041)(8.03125367,59.36161041)
\curveto(8.03125367,59.70536041)(7.81250367,59.89286041)(7.46875367,59.89286041)
\curveto(6.96875367,59.89286041)(6.68750367,59.45536041)(6.62500367,59.01786041)
\closepath
\moveto(5.34375367,53.86161041)
\curveto(5.21875367,53.48661041)(5.21875367,53.45536041)(4.93750367,53.11161041)
\curveto(4.06250367,52.01786041)(3.25000367,51.70536041)(2.68750367,51.70536041)
\curveto(1.68750367,51.70536041)(1.40625367,52.79911041)(1.40625367,53.58036041)
\curveto(1.40625367,54.58036041)(2.03125367,57.01786041)(2.50000367,57.95536041)
\curveto(3.12500367,59.11161041)(4.00000367,59.86161041)(4.81250367,59.86161041)
\curveto(6.09375367,59.86161041)(6.37500367,58.23661041)(6.37500367,58.11161041)
\curveto(6.37500367,57.98661041)(6.34375367,57.86161041)(6.31250367,57.76786041)
\closepath
\moveto(5.34375367,53.86161041)
}
}
{
\newrgbcolor{curcolor}{0 0 0}
\pscustom[linestyle=none,fillstyle=solid,fillcolor=curcolor]
{
\newpath
\moveto(66.90625767,7.68750041)
\curveto(66.37500767,7.65625041)(65.96875767,7.21875041)(65.96875767,6.78125041)
\curveto(65.96875767,6.50000041)(66.15625767,6.18750041)(66.59375767,6.18750041)
\curveto(67.03125767,6.18750041)(67.50000767,6.53125041)(67.50000767,7.31250041)
\curveto(67.50000767,8.21875041)(66.65625767,9.03125041)(65.12500767,9.03125041)
\curveto(62.50000767,9.03125041)(61.75000767,7.00000041)(61.75000767,6.12500041)
\curveto(61.75000767,4.56250041)(63.21875767,4.28125041)(63.81250767,4.15625041)
\curveto(64.84375767,3.93750041)(65.87500767,3.71875041)(65.87500767,2.62500041)
\curveto(65.87500767,2.12500041)(65.43750767,0.43750041)(63.03125767,0.43750041)
\curveto(62.75000767,0.43750041)(61.21875767,0.43750041)(60.75000767,1.50000041)
\curveto(61.53125767,1.40625041)(62.03125767,2.00000041)(62.03125767,2.56250041)
\curveto(62.03125767,3.00000041)(61.68750767,3.25000041)(61.28125767,3.25000041)
\curveto(60.75000767,3.25000041)(60.15625767,2.84375041)(60.15625767,1.93750041)
\curveto(60.15625767,0.81250041)(61.31250767,0.00000041)(63.00000767,0.00000041)
\curveto(66.25000767,0.00000041)(67.03125767,2.40625041)(67.03125767,3.31250041)
\curveto(67.03125767,4.03125041)(66.65625767,4.53125041)(66.40625767,4.75000041)
\curveto(65.87500767,5.31250041)(65.28125767,5.43750041)(64.40625767,5.59375041)
\curveto(63.68750767,5.75000041)(62.90625767,5.90625041)(62.90625767,6.81250041)
\curveto(62.90625767,7.37500041)(63.37500767,8.59375041)(65.12500767,8.59375041)
\curveto(65.62500767,8.59375041)(66.62500767,8.43750041)(66.90625767,7.68750041)
\closepath
\moveto(66.90625767,7.68750041)
}
}
\end{pspicture}\,.\label{eq:tableaux}
\end{equation}
For representations of this type, also known as Kirillov-Reshetikhin modules \cite{Kirillov1990}, there exist two values of the spectral
parameter such that $\mathcal{R}_{0,I}$ and $\tilde{\mathcal{R}}_{0,I}$
respectively are projectors on a highest weight state%
\footnote{The notion of highest weight state depends on the choice of the raising
generators. For rectangular representations $a!(n-a)!$ such choices
correspond to the same highest weight state.%
} for $\vert I\vert=n-a$. The number of highest/lowest weight states
for such representations is $\binom{n}{a}$ and exactly coincides
with the number of operators $\mathcal{R}_{I}$ with $\vert I\vert=n-a$.
Each $\mathcal{R}_{0,I}$ and $\tilde{\mathcal{R}}_{0,I}$ of cardinality
$n-a$ projects on a different highest weight state depending on the
elements in $I$. With an appropriate normalization discussed in section~\ref{sub:On-the-normalization}
we find for $\vert I\vert=n-a$ that 
\begin{equation}
\mathcal{R}_{0,I}(\hat{z})=\vert hws\rangle\langle hws\vert\quad\text{and\quad}\tilde{\mathcal{R}}_{0,I}(\check{z})=\vert hws\rangle\langle hws\vert\,.\label{eq:hwsProjector-1}
\end{equation}
 As a direct consequence of (\ref{eq:hwsProjector-1}) we obtain that
the $\mathcal{R}$-operator at the special points $\hat{z}$ and $\check{z}$
can be written as 
\begin{equation}
\mathcal{R}_{I}(\hat{z})=e^{\bar{\mathbf{a}}_{c}^{\dot{c}}J_{\dot{c}}^{c}}\cdot\vert hws\rangle\langle hws\vert\cdot e^{-\mathbf{a}_{\dot{c}}^{c}J_{c}^{\dot{c}}}\quad\text{and\quad}\mathcal{R}_{I}(\check{z})=e^{\bar{\mathbf{a}}_{c}^{\dot{c}}J_{\dot{c}}^{c}}\circ\vert hws\rangle\langle hws\vert\circ e^{-\mathbf{a}_{\dot{c}}^{c}J_{c}^{\dot{c}}}.\label{eq:RProjector}
\end{equation}
As we will see, these properties carry over to non-compact representations
with highest weight that fulfill a \textit{generalized rectangularity
condition}, see section~\ref{sub:Reduction} and appendix~\ref{sec:Shifted-weights}.
However, not all $\mathcal{R}$-operators of a certain cardinality
$\vert I\vert$ share the projection property. See discussion in appendix~\texttt{\textcolor{green}{\ref{sec:Su(2,2)}.}}
It will become clear in section~\ref{sec:Diagrammatics-and-local}
that as a consequence of (\ref{eq:RProjector}) the Q-operators at
the special points $\hat{z}$ and $\check{z}$ are related by the
shift operator, see (\ref{eq:UQgQ-1}).

\subsection{Reduction\label{sub:Reduction}}

It emerged in the discussion of the fundamental representation that
at special values of the spectral parameter the operators $\mathcal{R}_{0,I}$
and $\tilde{\mathcal{R}}_{0,I}$ become projectors on a certain subspace.
In the following we show when and how this happens. The analysis is
clearly connected to the pattern of the decomposition of the $\mathfrak{gl}(n)$
representation at a site 
\begin{equation}
\Lambda\rightarrow\oplus_{\alpha}\, m_{\alpha}\,(\Lambda_{\alpha}^{I},\Lambda_{\alpha}^{\bar{I}})
\end{equation}
under the restriction $\mathfrak{gl}(n)\downarrow\mathfrak{gl}(I)\oplus\mathfrak{gl}(\bar{I})$.
We are specifically interested in representations $\Lambda$ and a
set $I$ such that $\hat{\ell}_{k}^{\bar{I}}$ are bounded from above
and $\hat{\ell}_{k}^{I}$ are bounded from below%
\footnote{For any finite dimensional representation this is true for any set
$I$.%
}. The bound is saturated for all $k$ by the subspace $(\Lambda_{\alpha_{0}}^{I},\Lambda_{\alpha_{0}}^{\bar{I}})$
of $\Lambda$ annihilated by the action of generators $J_{a}^{\dot{a}}$

\begin{equation}
J_{a}^{\dot{a}}\vert\Lambda_{\alpha_{0}}^{I},\Lambda_{\alpha_{0}}^{\bar{I}}\rangle=0\,,
\end{equation}
where the indices take values according to (\ref{eq:sets}). The subspace
$(\Lambda_{\alpha_{0}}^{I},\Lambda_{\alpha_{0}}^{\bar{I}})$ is nothing
but the $\mathfrak{gl}(I)\oplus\mathfrak{gl}(\bar{I})$ irreducible
representation generated by the action of \textbf{$J_{b}^{a}$} and
$J_{\dot{b}}^{\dot{a}}$ on the $\mathfrak{gl}(n)$ highest weight
state%
\footnote{The raising generators that enter the $\mathfrak{gl}(n)$ highest
weight condition for $\vert\Lambda\rangle$ are chosen to include
$J_{a}^{\dot{a}}$. %
} $\vert\Lambda\rangle$. Moreover, the eigenvalues of any fixed $\hat{\ell}_{k}^{K}$
are integer spaced. The fact that the operators $\mathcal{R}_{0,I}$
and $\tilde{\mathcal{R}}_{0,I}$ become projectors on this subspace
for special values of the spectral parameter is an immediate consequence
of the properties of the operators $\hat{\ell}_{k}^{K}$ together
with the pole structure of the gamma function.

The class of representations considered at the end of the previous
section (here referred to as generalized rectangular representations)
have a number of remarkable features, see appendix \ref{sec:Shifted-weights}.
In particular, there is at least one set $I$ such that the subspace
on which $\mathcal{R}_{0,I}$ and $\tilde{\mathcal{R}}_{0,I}$ project
is one-dimensional. This fact is equivalent to the existence of a
state such that 
\begin{equation}
J_{a}^{\dot{a}}\vert\Lambda_{0}^{I},\Lambda_{0}^{\bar{I}}\rangle=0,\quad\quad J_{b}^{a}\vert\Lambda_{0}^{I},\Lambda_{0}^{\bar{I}}\rangle=\lambda_{I}\,\delta_{b}^{a}\vert\Lambda_{0}^{I},\Lambda_{0}^{\bar{I}}\rangle,\quad\quad J_{\dot{b}}^{\dot{a}}\vert\Lambda_{0}^{I},\Lambda_{0}^{\bar{I}}\rangle=\bar{\lambda}_{I}\,\delta_{\dot{b}}^{\dot{a}}\vert\Lambda_{0}^{I},\Lambda_{0}^{\bar{I}}\rangle\label{eq:sate}
\end{equation}
for a properly chosen set $I$ and some $\bar{\lambda}_{I},\lambda_{I}$,
see appendix \texttt{\textcolor{green}{\ref{sec:Shifted-weights}}}
for details. For convenience the state defined in (\ref{eq:sate})
will be denoted as $\vert hws\rangle$.

\subsection{On the normalization of $\mathcal{R}$-operators\label{sub:On-the-normalization}}

Besides the ratio of $\rho_{I}$ and $\tilde{\rho}_{I}$ which is
fixed by (\ref{eq:R_conjugation-1}) an overall normalization of the
$\mathcal{R}$-operators was not yet chosen. In our previous analysis
we determined the one-dimensional subspace which saturates the bound
of $\hat{\ell}_{k}^{\bar{I}}$ and $\hat{\ell}_{k}^{I}$. The action
of the shifted weights on this subspace is given in appendix~\ref{sec:Shifted-weights}.
As already mentioned, for our purposes it is convenient to choose
a normalization such that (\ref{eq:RProjector}) holds, i.e.

\begin{equation}
\mathcal{R}_{0,I}(z)=\kappa_{I}(z)\,\prod_{k=1}^{\vert\bar{I}\vert}\,\frac{\Gamma(z-\sfrac{\vert\bar{I}\vert}{2}-\hat{\ell}_{k}^{\bar{I}}+1)}{\Gamma(z-\sfrac{\vert\bar{I}\vert}{2}+k-\bar{\lambda}_{I})},\quad\quad\tilde{\mathcal{R}}_{0,I}(z)=\tilde{\kappa}_{I}(z)\,\prod_{k=1}^{\vert I\vert}\,\frac{\Gamma(z+\sfrac{\vert\bar{I}\vert}{2}+k-\lambda_{I})}{\Gamma(z+\sfrac{\vert\bar{I}\vert}{2}-\hat{\ell}_{k}^{I}+1)}\,,
\end{equation}
compare to (\ref{eq:R0}) and (\ref{Rt0-2}). Above, $\kappa_{I}$
and $\tilde{\kappa}_{I}$ are periodic functions of $\hat{\ell}_{k}^{\bar{I}}$,
$\hat{\ell}_{k}^{I}$ of period one, respectively. Furthermore, they
coincide on the highest weight $\kappa_{I}(z)\vert hws\rangle=\tilde{\kappa_{I}}(z)\vert hws\rangle=\vert hws\rangle$
and in analogy to $\rho_{I}$ and $\tilde{\rho}_{I}$ are dependent
by (\ref{eq:R_conjugation-1}), see also appendix~\ref{sec:Example:gl2}
for the example of $\mathfrak{gl}(2)$. As discussed in section~\ref{sub:Yang-Baxter-approach},
from the study of the Yang-Baxter equation it seems to be rather natural
to fix the overall normalization such that

\begin{equation}
\tilde{\mathcal{R}}_{0,\bar{I}}(z-\frac{\vert\bar{I}\vert}{2})=\mathcal{R}_{0,I}^{-1}(z+\frac{\vert I\vert}{2})\,.\label{eq:crossing-R0}
\end{equation}
Interestingly, this relation implies the crossing equation

\begin{equation}
\left(\mathcal{R}_{\bar{I}}(z-\frac{\vert\bar{I}\vert}{2})\right)^{*}=\mathcal{R}_{I}^{-1}(z+\frac{\vert I\vert}{2})\label{eq:crossing}
\end{equation}
with $\left(\bar{\mathbf{a}}_{\dot{a}_{1}}^{a_{1}}\cdots\bar{\mathbf{a}}_{\dot{a}_{m}}^{a_{m}}\,\mathbf{a}_{b_{1}}^{\dot{b}_{1}}\cdots\mathbf{a}_{b_{n}}^{\dot{b}_{n}}\right)^{*}=\bar{\mathbf{a}}_{b_{1}}^{\dot{b}_{1}}\cdots\bar{\mathbf{a}}_{b_{m}}^{\dot{b}_{m}}\,\mathbf{a}_{\dot{a}_{1}}^{a_{1}}\cdots\mathbf{a}_{\dot{a}_{n}}^{a_{n}}$.
However, an explicit study of these relations is left to the future.

\section{Diagrammatics and local charges\label{sec:Diagrammatics-and-local}}

As it is customary we denote R-matrices by two crossing lines. In
the construction of generalized transfer matrices each vertical line
corresponds to the quantum space associated to a spin-chain site.
Likewise, horizontal lines represent the auxiliary space. In the following
$\mathcal{R}$-operators generating Q-operators are depicted as 
\begin{equation}
\mathcal{R}_{I}(z)\;%LaTeX with PSTricks extensions
%%Creator: inkscape 0.48.1
%%Please note this file requires PSTricks extensions
\psset{xunit=.5pt,yunit=.5pt,runit=.5pt}
\begin{pspicture}[shift=-50](125.5,111)
{
\newrgbcolor{curcolor}{0 0 1}
\pscustom[linewidth=1,linecolor=curcolor]
{
\newpath
\moveto(75,35.5)
\lineto(75,0.5)
}
}
{
\newrgbcolor{curcolor}{0 0 1}
\pscustom[linewidth=1,linecolor=curcolor]
{
\newpath
\moveto(75,110.5)
\lineto(75,75.5)
}
}
{
\newrgbcolor{curcolor}{1 0 0}
\pscustom[linewidth=1,linecolor=curcolor]
{
\newpath
\moveto(95,55.5)
\curveto(121.97592,55.82787)(125,70.5)(125,110.5)
}
}
{
\newrgbcolor{curcolor}{1 0 0}
\pscustom[linewidth=1,linecolor=curcolor]
{
\newpath
\moveto(55,55.5)
\curveto(28.02408,55.17213)(25,40.5)(25,0.5)
}
}
{
\newrgbcolor{curcolor}{0 0 0}
\pscustom[linestyle=none,fillstyle=solid,fillcolor=curcolor]
{
\newpath
\moveto(12.56250367,57.48466502)
\curveto(12.87500367,57.48466502)(13.25000367,57.48466502)(13.25000367,57.85966502)
\curveto(13.25000367,58.26591502)(12.87500367,58.26591502)(12.59375367,58.26591502)
\lineto(0.65625367,58.26591502)
\curveto(0.37500367,58.26591502)(0.00000367,58.26591502)(0.00000367,57.85966502)
\curveto(0.00000367,57.48466502)(0.37500367,57.48466502)(0.65625367,57.48466502)
\closepath
\moveto(12.59375367,53.60966502)
\curveto(12.87500367,53.60966502)(13.25000367,53.60966502)(13.25000367,54.01591502)
\curveto(13.25000367,54.39091502)(12.87500367,54.39091502)(12.56250367,54.39091502)
\lineto(0.65625367,54.39091502)
\curveto(0.37500367,54.39091502)(0.00000367,54.39091502)(0.00000367,54.01591502)
\curveto(0.00000367,53.60966502)(0.37500367,53.60966502)(0.65625367,53.60966502)
\closepath
\moveto(12.59375367,53.60966502)
}
}
{
\newrgbcolor{curcolor}{1 1 1}
\pscustom[linestyle=none,fillstyle=solid,fillcolor=curcolor]
{
\newpath
\moveto(95,55.5)
\curveto(95,44.454305)(86.045695,35.5)(75,35.5)
\curveto(63.954305,35.5)(55,44.454305)(55,55.5)
\curveto(55,66.545695)(63.954305,75.5)(75,75.5)
\curveto(86.045695,75.5)(95,66.545695)(95,55.5)
\closepath
}
}
{
\newrgbcolor{curcolor}{0 0 0}
\pscustom[linewidth=1,linecolor=curcolor]
{
\newpath
\moveto(95,55.5)
\curveto(95,44.454305)(86.045695,35.5)(75,35.5)
\curveto(63.954305,35.5)(55,44.454305)(55,55.5)
\curveto(55,66.545695)(63.954305,75.5)(75,75.5)
\curveto(86.045695,75.5)(95,66.545695)(95,55.5)
\closepath
}
}
{
\newrgbcolor{curcolor}{0 0 0}
\pscustom[linestyle=none,fillstyle=solid,fillcolor=curcolor]
{
\newpath
\moveto(70.67383367,61.60367732)
\curveto(74.51758367,61.60367732)(75.51758367,60.66617732)(75.51758367,59.35367732)
\curveto(75.51758367,58.13492732)(74.54883367,55.72867732)(71.26758367,55.60367732)
\curveto(70.29883367,55.57242732)(69.61133367,54.88492732)(69.61133367,54.66617732)
\curveto(69.61133367,54.54117732)(69.67383367,54.54117732)(69.70508367,54.54117732)
\curveto(70.54883367,54.38492732)(70.95508367,54.00992732)(71.98633367,51.63492732)
\curveto(72.89258367,49.54117732)(73.42383367,48.63492732)(74.58008367,48.63492732)
\curveto(77.01758367,48.63492732)(79.29883367,51.07242732)(79.29883367,51.50992732)
\curveto(79.29883367,51.66617732)(79.14258367,51.66617732)(79.08008367,51.66617732)
\curveto(78.83008367,51.66617732)(78.04883367,51.38492732)(77.64258367,50.82242732)
\curveto(77.33008367,50.35367732)(76.89258367,49.72867732)(75.92383367,49.72867732)
\curveto(74.89258367,49.72867732)(74.26758367,51.13492732)(73.61133367,52.69742732)
\curveto(73.17383367,53.69742732)(72.83008367,54.44742732)(72.36133367,54.97867732)
\curveto(75.23633367,56.04117732)(77.20508367,58.13492732)(77.20508367,60.19742732)
\curveto(77.20508367,62.69742732)(73.86133367,62.69742732)(70.83008367,62.69742732)
\curveto(68.86133367,62.69742732)(67.73633367,62.69742732)(66.04883367,61.97867732)
\curveto(63.39258367,60.79117732)(63.01758367,59.13492732)(63.01758367,58.97867732)
\curveto(63.01758367,58.85367732)(63.11133367,58.82242732)(63.23633367,58.82242732)
\curveto(63.54883367,58.82242732)(64.01758367,59.10367732)(64.17383367,59.19742732)
\curveto(64.58008367,59.47867732)(64.64258367,59.60367732)(64.76758367,59.97867732)
\curveto(65.04883367,60.79117732)(65.61133367,61.47867732)(68.11133367,61.60367732)
\curveto(68.01758367,60.38492732)(67.83008367,58.50992732)(67.14258367,55.60367732)
\curveto(66.61133367,53.38492732)(65.89258367,51.19742732)(65.01758367,49.04117732)
\curveto(64.89258367,48.82242732)(64.89258367,48.79117732)(64.89258367,48.75992732)
\curveto(64.89258367,48.63492732)(65.04883367,48.63492732)(65.11133367,48.63492732)
\curveto(65.51758367,48.63492732)(66.33008367,49.10367732)(66.54883367,49.47867732)
\curveto(66.61133367,49.60367732)(69.14258367,55.22867732)(69.73633367,61.60367732)
\closepath
\moveto(70.67383367,61.60367732)
}
}
{
\newrgbcolor{curcolor}{0 0 0}
\pscustom[linestyle=none,fillstyle=solid,fillcolor=curcolor]
{
\newpath
\moveto(85.24732,54.49067073)
\curveto(85.37232,54.99067073)(85.40357,55.11567073)(86.46607,55.11567073)
\curveto(86.80982,55.11567073)(86.96607,55.11567073)(86.96607,55.39692073)
\curveto(86.96607,55.52192073)(86.87232,55.61567073)(86.74732,55.61567073)
\curveto(86.46607,55.61567073)(86.09107,55.55317073)(85.80982,55.55317073)
\lineto(84.87232,55.55317073)
\lineto(83.90357,55.55317073)
\curveto(83.59107,55.55317073)(83.24732,55.61567073)(82.93482,55.61567073)
\curveto(82.84107,55.61567073)(82.65357,55.61567073)(82.65357,55.30317073)
\curveto(82.65357,55.11567073)(82.80982,55.11567073)(83.12232,55.11567073)
\curveto(83.12232,55.11567073)(83.40357,55.11567073)(83.62232,55.08442073)
\curveto(83.90357,55.05317073)(84.02857,55.02192073)(84.02857,54.86567073)
\curveto(84.02857,54.80317073)(83.99732,54.70942073)(83.96607,54.58442073)
\lineto(82.09107,47.17817073)
\curveto(81.99732,46.74067073)(81.93482,46.58442073)(80.90357,46.58442073)
\curveto(80.52857,46.58442073)(80.40357,46.58442073)(80.40357,46.27192073)
\curveto(80.40357,46.27192073)(80.40357,46.08442073)(80.62232,46.08442073)
\curveto(81.02857,46.08442073)(82.09107,46.14692073)(82.49732,46.14692073)
\lineto(83.46607,46.11567073)
\curveto(83.74732,46.11567073)(84.12232,46.08442073)(84.40357,46.08442073)
\curveto(84.49732,46.08442073)(84.71607,46.08442073)(84.71607,46.39692073)
\curveto(84.71607,46.58442073)(84.52857,46.58442073)(84.27857,46.58442073)
\curveto(84.24732,46.58442073)(83.93482,46.58442073)(83.65357,46.61567073)
\curveto(83.34107,46.64692073)(83.34107,46.70942073)(83.34107,46.83442073)
\curveto(83.34107,46.83442073)(83.34107,46.92817073)(83.40357,47.11567073)
\closepath
\moveto(85.24732,54.49067073)
}
}
\end{pspicture}\,,
\end{equation}
compare to (\ref{eq:Qop}). We will now develop a pictorial language
for the $\mathcal{R}$-operators, which incorporates all aforementioned
properties, see section \ref{sec:Alternative-presentation-of} and
\ref{sec:Projection-property-of}. One of its main advantages is that
the opposite product (\ref{eq:prod_op}), which might look unfamiliar
in the equations, is translated to a rather natural composition rule.
It is a key ingredient to reveal the emergence of local charges from
Q-operators.

\subsection{Two multiplication rules}

As discussed in section~\ref{sec:Alternative-presentation-of}, it
is natural to introduce two different multiplication rules. Diagrammatically
the product $\cdot$ can then be understood as

\begin{equation}
\mathcal{O}_{1}\cdot\mathcal{O}_{2}\;%LaTeX with PSTricks extensions
%%Creator: inkscape 0.48.1
%%Please note this file requires PSTricks extensions
\psset{xunit=.5pt,yunit=.5pt,runit=.5pt}
\begin{pspicture}[shift=-45](227.5,111)
{
\newrgbcolor{curcolor}{0 0 1}
\pscustom[linewidth=1,linecolor=curcolor]
{
\newpath
\moveto(72,35.5)
\lineto(72,0.5)
}
}
{
\newrgbcolor{curcolor}{0 0 1}
\pscustom[linewidth=1,linecolor=curcolor]
{
\newpath
\moveto(72,75.5)
\curveto(72,85.5)(77,95.5)(92,95.5)
\curveto(107,95.5)(111.72773,85.49629)(112,75.5)
\curveto(112.46098,58.57523)(112.0806,46.03468)(120.34307,34.92015)
\curveto(132.72797,18.2602)(145.93358,3.0848)(155.39765,2.68558)
\curveto(173.76061,1.91098)(177.29455,12.8264)(177.55724,24.67513)
\curveto(177.79807,35.53778)(177.42811,24.72053)(176.75,110.25)
}
}
{
\newrgbcolor{curcolor}{1 0 0}
\pscustom[linewidth=1,linecolor=curcolor]
{
\newpath
\moveto(92,55.5)
\lineto(102.5,55.5)
\curveto(112.5,55.5)(107.7,59.08579)(112.7,60.5)
\curveto(117.87677,58.73224)(112.83876,55.60523)(122.5,55.5)
\lineto(157,55.5)
}
}
{
\newrgbcolor{curcolor}{1 0 0}
\pscustom[linewidth=1,linecolor=curcolor]
{
\newpath
\moveto(52,55.5)
\curveto(25.02408,55.17213)(22,40.5)(22,0.5)
}
}
{
\newrgbcolor{curcolor}{0 0 0}
\pscustom[linestyle=none,fillstyle=solid,fillcolor=curcolor]
{
\newpath
\moveto(12.56250367,56.98466502)
\curveto(12.87500367,56.98466502)(13.25000367,56.98466502)(13.25000367,57.35966502)
\curveto(13.25000367,57.76591502)(12.87500367,57.76591502)(12.59375367,57.76591502)
\lineto(0.65625367,57.76591502)
\curveto(0.37500367,57.76591502)(0.00000367,57.76591502)(0.00000367,57.35966502)
\curveto(0.00000367,56.98466502)(0.37500367,56.98466502)(0.65625367,56.98466502)
\closepath
\moveto(12.59375367,53.10966502)
\curveto(12.87500367,53.10966502)(13.25000367,53.10966502)(13.25000367,53.51591502)
\curveto(13.25000367,53.89091502)(12.87500367,53.89091502)(12.56250367,53.89091502)
\lineto(0.65625367,53.89091502)
\curveto(0.37500367,53.89091502)(0.00000367,53.89091502)(0.00000367,53.51591502)
\curveto(0.00000367,53.10966502)(0.37500367,53.10966502)(0.65625367,53.10966502)
\closepath
\moveto(12.59375367,53.10966502)
}
}
{
\newrgbcolor{curcolor}{1 1 1}
\pscustom[linestyle=none,fillstyle=solid,fillcolor=curcolor]
{
\newpath
\moveto(92.000004,55.5)
\curveto(92.000004,44.454305)(83.045699,35.5)(72.000004,35.5)
\curveto(60.954309,35.5)(52.000004,44.454305)(52.000004,55.5)
\curveto(52.000004,66.545695)(60.954309,75.5)(72.000004,75.5)
\curveto(83.045699,75.5)(92.000004,66.545695)(92.000004,55.5)
\closepath
}
}
{
\newrgbcolor{curcolor}{0 0 0}
\pscustom[linewidth=1,linecolor=curcolor]
{
\newpath
\moveto(92.000004,55.5)
\curveto(92.000004,44.454305)(83.045699,35.5)(72.000004,35.5)
\curveto(60.954309,35.5)(52.000004,44.454305)(52.000004,55.5)
\curveto(52.000004,66.545695)(60.954309,75.5)(72.000004,75.5)
\curveto(83.045699,75.5)(92.000004,66.545695)(92.000004,55.5)
\closepath
}
}
{
\newrgbcolor{curcolor}{0 0 1}
\pscustom[linewidth=1,linecolor=curcolor]
{
\newpath
\moveto(177,110.5)
\lineto(177,75.5)
}
}
{
\newrgbcolor{curcolor}{1 0 0}
\pscustom[linewidth=1,linecolor=curcolor]
{
\newpath
\moveto(197,55.5)
\curveto(223.97592,55.82787)(227,70.5)(227,110.5)
}
}
{
\newrgbcolor{curcolor}{1 1 1}
\pscustom[linestyle=none,fillstyle=solid,fillcolor=curcolor]
{
\newpath
\moveto(197,55.5)
\curveto(197,44.454305)(188.045695,35.5)(177,35.5)
\curveto(165.954305,35.5)(157,44.454305)(157,55.5)
\curveto(157,66.545695)(165.954305,75.5)(177,75.5)
\curveto(188.045695,75.5)(197,66.545695)(197,55.5)
\closepath
}
}
{
\newrgbcolor{curcolor}{0 0 0}
\pscustom[linewidth=1,linecolor=curcolor]
{
\newpath
\moveto(197,55.5)
\curveto(197,44.454305)(188.045695,35.5)(177,35.5)
\curveto(165.954305,35.5)(157,44.454305)(157,55.5)
\curveto(157,66.545695)(165.954305,75.5)(177,75.5)
\curveto(188.045695,75.5)(197,66.545695)(197,55.5)
\closepath
}
}
{
\newrgbcolor{curcolor}{0 0 0}
\pscustom[linestyle=none,fillstyle=solid,fillcolor=curcolor]
{
\newpath
\moveto(75.99001367,58.90453412)
\curveto(75.99001367,61.43578412)(74.86501367,63.34203412)(72.42751367,63.34203412)
\curveto(70.36501367,63.34203412)(68.61501367,61.65453412)(68.45876367,61.52953412)
\curveto(66.86501367,59.96703412)(66.27126367,58.18578412)(66.27126367,58.12328412)
\curveto(66.27126367,57.99828412)(66.36501367,57.99828412)(66.45876367,57.99828412)
\curveto(66.86501367,57.99828412)(67.17751367,58.18578412)(67.45876367,58.40453412)
\curveto(67.83376367,58.65453412)(67.83376367,58.71703412)(68.05251367,59.18578412)
\curveto(68.24001367,59.59203412)(68.70876367,60.56078412)(69.42751367,61.40453412)
\curveto(69.92751367,61.93578412)(70.33376367,62.24828412)(71.11501367,62.24828412)
\curveto(73.05251367,62.24828412)(74.27126367,60.65453412)(74.27126367,58.02953412)
\curveto(74.27126367,53.96703412)(71.39626367,49.93578412)(67.70876367,49.93578412)
\curveto(64.86501367,49.93578412)(63.33376367,52.18578412)(63.33376367,55.02953412)
\curveto(63.33376367,57.74828412)(64.70876367,60.74828412)(67.61501367,62.46703412)
\curveto(67.83376367,62.59203412)(68.42751367,62.93578412)(68.42751367,63.18578412)
\curveto(68.42751367,63.34203412)(68.27126367,63.34203412)(68.24001367,63.34203412)
\curveto(67.52126367,63.34203412)(61.61501367,60.15453412)(61.61501367,54.18578412)
\curveto(61.61501367,51.40453412)(63.08376367,48.84203412)(66.39626367,48.84203412)
\curveto(70.27126367,48.84203412)(75.99001367,52.74828412)(75.99001367,58.90453412)
\closepath
\moveto(75.99001367,58.90453412)
}
}
{
\newrgbcolor{curcolor}{0 0 0}
\pscustom[linestyle=none,fillstyle=solid,fillcolor=curcolor]
{
\newpath
\moveto(81.00826318,55.16652753)
\curveto(81.00826318,55.54152753)(81.00826318,55.54152753)(80.60201318,55.54152753)
\curveto(79.69576318,54.66652753)(78.44576318,54.66652753)(77.88326318,54.66652753)
\lineto(77.88326318,54.16652753)
\curveto(78.19576318,54.16652753)(79.13326318,54.16652753)(79.88326318,54.54152753)
\lineto(79.88326318,47.44777753)
\curveto(79.88326318,46.97902753)(79.88326318,46.79152753)(78.50826318,46.79152753)
\lineto(77.97701318,46.79152753)
\lineto(77.97701318,46.29152753)
\curveto(78.22701318,46.29152753)(79.94576318,46.35402753)(80.44576318,46.35402753)
\curveto(80.88326318,46.35402753)(82.63326318,46.29152753)(82.94576318,46.29152753)
\lineto(82.94576318,46.79152753)
\lineto(82.41451318,46.79152753)
\curveto(81.00826318,46.79152753)(81.00826318,46.97902753)(81.00826318,47.44777753)
\closepath
\moveto(81.00826318,55.16652753)
}
}
{
\newrgbcolor{curcolor}{0 0 0}
\pscustom[linestyle=none,fillstyle=solid,fillcolor=curcolor]
{
\newpath
\moveto(180.64181367,59.61164092)
\curveto(180.64181367,62.14289092)(179.51681367,64.04914092)(177.07931367,64.04914092)
\curveto(175.01681367,64.04914092)(173.26681367,62.36164092)(173.11056367,62.23664092)
\curveto(171.51681367,60.67414092)(170.92306367,58.89289092)(170.92306367,58.83039092)
\curveto(170.92306367,58.70539092)(171.01681367,58.70539092)(171.11056367,58.70539092)
\curveto(171.51681367,58.70539092)(171.82931367,58.89289092)(172.11056367,59.11164092)
\curveto(172.48556367,59.36164092)(172.48556367,59.42414092)(172.70431367,59.89289092)
\curveto(172.89181367,60.29914092)(173.36056367,61.26789092)(174.07931367,62.11164092)
\curveto(174.57931367,62.64289092)(174.98556367,62.95539092)(175.76681367,62.95539092)
\curveto(177.70431367,62.95539092)(178.92306367,61.36164092)(178.92306367,58.73664092)
\curveto(178.92306367,54.67414092)(176.04806367,50.64289092)(172.36056367,50.64289092)
\curveto(169.51681367,50.64289092)(167.98556367,52.89289092)(167.98556367,55.73664092)
\curveto(167.98556367,58.45539092)(169.36056367,61.45539092)(172.26681367,63.17414092)
\curveto(172.48556367,63.29914092)(173.07931367,63.64289092)(173.07931367,63.89289092)
\curveto(173.07931367,64.04914092)(172.92306367,64.04914092)(172.89181367,64.04914092)
\curveto(172.17306367,64.04914092)(166.26681367,60.86164092)(166.26681367,54.89289092)
\curveto(166.26681367,52.11164092)(167.73556367,49.54914092)(171.04806367,49.54914092)
\curveto(174.92306367,49.54914092)(180.64181367,53.45539092)(180.64181367,59.61164092)
\closepath
\moveto(180.64181367,59.61164092)
}
}
{
\newrgbcolor{curcolor}{0 0 0}
\pscustom[linestyle=none,fillstyle=solid,fillcolor=curcolor]
{
\newpath
\moveto(188.03506318,49.52988433)
\lineto(187.56631318,49.52988433)
\curveto(187.53506318,49.21738433)(187.37881318,48.40488433)(187.19131318,48.27988433)
\curveto(187.09756318,48.18613433)(186.03506318,48.18613433)(185.81631318,48.18613433)
\lineto(183.25381318,48.18613433)
\curveto(184.72256318,49.46738433)(185.22256318,49.87363433)(186.03506318,50.52988433)
\curveto(187.06631318,51.34238433)(188.03506318,52.21738433)(188.03506318,53.52988433)
\curveto(188.03506318,55.21738433)(186.56631318,56.24863433)(184.78506318,56.24863433)
\curveto(183.06631318,56.24863433)(181.87881318,55.02988433)(181.87881318,53.74863433)
\curveto(181.87881318,53.06113433)(182.47256318,52.96738433)(182.62881318,52.96738433)
\curveto(182.94131318,52.96738433)(183.34756318,53.21738433)(183.34756318,53.71738433)
\curveto(183.34756318,53.96738433)(183.25381318,54.46738433)(182.53506318,54.46738433)
\curveto(182.97256318,55.43613433)(183.91006318,55.74863433)(184.56631318,55.74863433)
\curveto(185.97256318,55.74863433)(186.69131318,54.65488433)(186.69131318,53.52988433)
\curveto(186.69131318,52.31113433)(185.81631318,51.37363433)(185.37881318,50.87363433)
\lineto(182.03506318,47.52988433)
\curveto(181.87881318,47.40488433)(181.87881318,47.37363433)(181.87881318,46.99863433)
\lineto(187.62881318,46.99863433)
\closepath
\moveto(188.03506318,49.52988433)
}
}
\end{pspicture}\,.\label{eq:O1O2}
\end{equation}
Here the oscillator (\textcolor{red}{red}) and the $\mathcal{U}(\mathfrak{gl}(n))$
(\textcolor{blue}{blue}) components of $\mathcal{O}_{1}$ are both
multiplied from the left to $\mathcal{O}_{2}$. On the other hand,
the product $\circ$ is denoted by
\begin{equation}
\mathcal{O}_{1}\circ\mathcal{O}_{2}\;%LaTeX with PSTricks extensions
%%Creator: inkscape 0.48.1
%%Please note this file requires PSTricks extensions
\psset{xunit=.5pt,yunit=.5pt,runit=.5pt}
\begin{pspicture}[shift=-45](227.5,111)
{
\newrgbcolor{curcolor}{0 0 1}
\pscustom[linewidth=1,linecolor=curcolor]
{
\newpath
\moveto(177,35.5)
\lineto(177,0.5)
}
}
{
\newrgbcolor{curcolor}{0 0 1}
\pscustom[linewidth=1,linecolor=curcolor]
{
\newpath
\moveto(177,75.5)
\curveto(177,85.5)(172,95.5)(157,95.5)
\curveto(142,95.5)(137.27227,85.49629)(137,75.5)
\curveto(136.53902,58.57523)(136.9194,46.03468)(128.65693,34.92015)
\curveto(116.27203,18.2602)(103.06642,3.0848)(93.60235,2.68558)
\curveto(75.23939,1.91098)(71.70545,12.8264)(71.44276,24.67513)
\curveto(71.20193,35.53778)(71.57189,24.72053)(72.25,110.25)
}
}
{
\newrgbcolor{curcolor}{1 0 0}
\pscustom[linewidth=1,linecolor=curcolor]
{
\newpath
\moveto(92,55.5)
\lineto(125.3,55.5)
\curveto(135.3,55.5)(131.3,59.08579)(136.3,60.5)
\curveto(141.47677,58.73224)(137.33876,55.60523)(147,55.5)
\lineto(157,55.5)
}
}
{
\newrgbcolor{curcolor}{1 0 0}
\pscustom[linewidth=1,linecolor=curcolor]
{
\newpath
\moveto(52,55.5)
\curveto(25.02408,55.17213)(22,40.5)(22,0.5)
}
}
{
\newrgbcolor{curcolor}{0 0 0}
\pscustom[linestyle=none,fillstyle=solid,fillcolor=curcolor]
{
\newpath
\moveto(12.56250367,56.98466502)
\curveto(12.87500367,56.98466502)(13.25000367,56.98466502)(13.25000367,57.35966502)
\curveto(13.25000367,57.76591502)(12.87500367,57.76591502)(12.59375367,57.76591502)
\lineto(0.65625367,57.76591502)
\curveto(0.37500367,57.76591502)(0.00000367,57.76591502)(0.00000367,57.35966502)
\curveto(0.00000367,56.98466502)(0.37500367,56.98466502)(0.65625367,56.98466502)
\closepath
\moveto(12.59375367,53.10966502)
\curveto(12.87500367,53.10966502)(13.25000367,53.10966502)(13.25000367,53.51591502)
\curveto(13.25000367,53.89091502)(12.87500367,53.89091502)(12.56250367,53.89091502)
\lineto(0.65625367,53.89091502)
\curveto(0.37500367,53.89091502)(0.00000367,53.89091502)(0.00000367,53.51591502)
\curveto(0.00000367,53.10966502)(0.37500367,53.10966502)(0.65625367,53.10966502)
\closepath
\moveto(12.59375367,53.10966502)
}
}
{
\newrgbcolor{curcolor}{1 0 0}
\pscustom[linewidth=1,linecolor=curcolor]
{
\newpath
\moveto(197,55.5)
\curveto(223.97592,55.82787)(227,70.5)(227,110.5)
}
}
{
\newrgbcolor{curcolor}{1 1 1}
\pscustom[linestyle=none,fillstyle=solid,fillcolor=curcolor]
{
\newpath
\moveto(52,55.5)
\curveto(52,44.454305)(60.954305,35.5)(72,35.5)
\curveto(83.045695,35.5)(92,44.454305)(92,55.5)
\curveto(92,66.545695)(83.045695,75.5)(72,75.5)
\curveto(60.954305,75.5)(52,66.545695)(52,55.5)
\closepath
}
}
{
\newrgbcolor{curcolor}{0 0 0}
\pscustom[linewidth=1,linecolor=curcolor]
{
\newpath
\moveto(52,55.5)
\curveto(52,44.454305)(60.954305,35.5)(72,35.5)
\curveto(83.045695,35.5)(92,44.454305)(92,55.5)
\curveto(92,66.545695)(83.045695,75.5)(72,75.5)
\curveto(60.954305,75.5)(52,66.545695)(52,55.5)
\closepath
}
}
{
\newrgbcolor{curcolor}{1 1 1}
\pscustom[linestyle=none,fillstyle=solid,fillcolor=curcolor]
{
\newpath
\moveto(157,55.5)
\curveto(157,44.454305)(165.954305,35.5)(177,35.5)
\curveto(188.045695,35.5)(197,44.454305)(197,55.5)
\curveto(197,66.545695)(188.045695,75.5)(177,75.5)
\curveto(165.954305,75.5)(157,66.545695)(157,55.5)
\closepath
}
}
{
\newrgbcolor{curcolor}{0 0 0}
\pscustom[linewidth=1,linecolor=curcolor]
{
\newpath
\moveto(157,55.5)
\curveto(157,44.454305)(165.954305,35.5)(177,35.5)
\curveto(188.045695,35.5)(197,44.454305)(197,55.5)
\curveto(197,66.545695)(188.045695,75.5)(177,75.5)
\curveto(165.954305,75.5)(157,66.545695)(157,55.5)
\closepath
}
}
{
\newrgbcolor{curcolor}{0 0 0}
\pscustom[linestyle=none,fillstyle=solid,fillcolor=curcolor]
{
\newpath
\moveto(181.66977367,59.73296122)
\curveto(181.66977367,62.26421122)(180.54477367,64.17046122)(178.10727367,64.17046122)
\curveto(176.04477367,64.17046122)(174.29477367,62.48296122)(174.13852367,62.35796122)
\curveto(172.54477367,60.79546122)(171.95102367,59.01421122)(171.95102367,58.95171122)
\curveto(171.95102367,58.82671122)(172.04477367,58.82671122)(172.13852367,58.82671122)
\curveto(172.54477367,58.82671122)(172.85727367,59.01421122)(173.13852367,59.23296122)
\curveto(173.51352367,59.48296122)(173.51352367,59.54546122)(173.73227367,60.01421122)
\curveto(173.91977367,60.42046122)(174.38852367,61.38921122)(175.10727367,62.23296122)
\curveto(175.60727367,62.76421122)(176.01352367,63.07671122)(176.79477367,63.07671122)
\curveto(178.73227367,63.07671122)(179.95102367,61.48296122)(179.95102367,58.85796122)
\curveto(179.95102367,54.79546122)(177.07602367,50.76421122)(173.38852367,50.76421122)
\curveto(170.54477367,50.76421122)(169.01352367,53.01421122)(169.01352367,55.85796122)
\curveto(169.01352367,58.57671122)(170.38852367,61.57671122)(173.29477367,63.29546122)
\curveto(173.51352367,63.42046122)(174.10727367,63.76421122)(174.10727367,64.01421122)
\curveto(174.10727367,64.17046122)(173.95102367,64.17046122)(173.91977367,64.17046122)
\curveto(173.20102367,64.17046122)(167.29477367,60.98296122)(167.29477367,55.01421122)
\curveto(167.29477367,52.23296122)(168.76352367,49.67046122)(172.07602367,49.67046122)
\curveto(175.95102367,49.67046122)(181.66977367,53.57671122)(181.66977367,59.73296122)
\closepath
\moveto(181.66977367,59.73296122)
}
}
{
\newrgbcolor{curcolor}{0 0 0}
\pscustom[linestyle=none,fillstyle=solid,fillcolor=curcolor]
{
\newpath
\moveto(186.68802318,55.99495463)
\curveto(186.68802318,56.36995463)(186.68802318,56.36995463)(186.28177318,56.36995463)
\curveto(185.37552318,55.49495463)(184.12552318,55.49495463)(183.56302318,55.49495463)
\lineto(183.56302318,54.99495463)
\curveto(183.87552318,54.99495463)(184.81302318,54.99495463)(185.56302318,55.36995463)
\lineto(185.56302318,48.27620463)
\curveto(185.56302318,47.80745463)(185.56302318,47.61995463)(184.18802318,47.61995463)
\lineto(183.65677318,47.61995463)
\lineto(183.65677318,47.11995463)
\curveto(183.90677318,47.11995463)(185.62552318,47.18245463)(186.12552318,47.18245463)
\curveto(186.56302318,47.18245463)(188.31302318,47.11995463)(188.62552318,47.11995463)
\lineto(188.62552318,47.61995463)
\lineto(188.09427318,47.61995463)
\curveto(186.68802318,47.61995463)(186.68802318,47.80745463)(186.68802318,48.27620463)
\closepath
\moveto(186.68802318,55.99495463)
}
}
{
\newrgbcolor{curcolor}{0 0 0}
\pscustom[linestyle=none,fillstyle=solid,fillcolor=curcolor]
{
\newpath
\moveto(76.16977367,58.73296122)
\curveto(76.16977367,61.26421122)(75.04477367,63.17046122)(72.60727367,63.17046122)
\curveto(70.54477367,63.17046122)(68.79477367,61.48296122)(68.63852367,61.35796122)
\curveto(67.04477367,59.79546122)(66.45102367,58.01421122)(66.45102367,57.95171122)
\curveto(66.45102367,57.82671122)(66.54477367,57.82671122)(66.63852367,57.82671122)
\curveto(67.04477367,57.82671122)(67.35727367,58.01421122)(67.63852367,58.23296122)
\curveto(68.01352367,58.48296122)(68.01352367,58.54546122)(68.23227367,59.01421122)
\curveto(68.41977367,59.42046122)(68.88852367,60.38921122)(69.60727367,61.23296122)
\curveto(70.10727367,61.76421122)(70.51352367,62.07671122)(71.29477367,62.07671122)
\curveto(73.23227367,62.07671122)(74.45102367,60.48296122)(74.45102367,57.85796122)
\curveto(74.45102367,53.79546122)(71.57602367,49.76421122)(67.88852367,49.76421122)
\curveto(65.04477367,49.76421122)(63.51352367,52.01421122)(63.51352367,54.85796122)
\curveto(63.51352367,57.57671122)(64.88852367,60.57671122)(67.79477367,62.29546122)
\curveto(68.01352367,62.42046122)(68.60727367,62.76421122)(68.60727367,63.01421122)
\curveto(68.60727367,63.17046122)(68.45102367,63.17046122)(68.41977367,63.17046122)
\curveto(67.70102367,63.17046122)(61.79477367,59.98296122)(61.79477367,54.01421122)
\curveto(61.79477367,51.23296122)(63.26352367,48.67046122)(66.57602367,48.67046122)
\curveto(70.45102367,48.67046122)(76.16977367,52.57671122)(76.16977367,58.73296122)
\closepath
\moveto(76.16977367,58.73296122)
}
}
{
\newrgbcolor{curcolor}{0 0 0}
\pscustom[linestyle=none,fillstyle=solid,fillcolor=curcolor]
{
\newpath
\moveto(83.56302318,48.65120463)
\lineto(83.09427318,48.65120463)
\curveto(83.06302318,48.33870463)(82.90677318,47.52620463)(82.71927318,47.40120463)
\curveto(82.62552318,47.30745463)(81.56302318,47.30745463)(81.34427318,47.30745463)
\lineto(78.78177318,47.30745463)
\curveto(80.25052318,48.58870463)(80.75052318,48.99495463)(81.56302318,49.65120463)
\curveto(82.59427318,50.46370463)(83.56302318,51.33870463)(83.56302318,52.65120463)
\curveto(83.56302318,54.33870463)(82.09427318,55.36995463)(80.31302318,55.36995463)
\curveto(78.59427318,55.36995463)(77.40677318,54.15120463)(77.40677318,52.86995463)
\curveto(77.40677318,52.18245463)(78.00052318,52.08870463)(78.15677318,52.08870463)
\curveto(78.46927318,52.08870463)(78.87552318,52.33870463)(78.87552318,52.83870463)
\curveto(78.87552318,53.08870463)(78.78177318,53.58870463)(78.06302318,53.58870463)
\curveto(78.50052318,54.55745463)(79.43802318,54.86995463)(80.09427318,54.86995463)
\curveto(81.50052318,54.86995463)(82.21927318,53.77620463)(82.21927318,52.65120463)
\curveto(82.21927318,51.43245463)(81.34427318,50.49495463)(80.90677318,49.99495463)
\lineto(77.56302318,46.65120463)
\curveto(77.40677318,46.52620463)(77.40677318,46.49495463)(77.40677318,46.11995463)
\lineto(83.15677318,46.11995463)
\closepath
\moveto(83.56302318,48.65120463)
}
}
\end{pspicture}\,.\label{eq:O1circO2}
\end{equation}
Here the $\mathcal{U}(\mathfrak{gl}(n))$ part of $\mathcal{O}_{1}$
is also multiplied from the left to $\mathcal{O}_{2}$, but the oscillator
part is multiplied to the right. In summary, once the operators $\mathcal{O}_{i}$
are written as (\ref{eq:opdecomp}) the order of the factors (from
left to right) in (\ref{eq:O1O2}) and (\ref{eq:O1circO2}) is obtained
by following the lines from bottom to top.

\subsection{$\mathcal{R}$-operators}

We will now develop a diagrammatic expression for $\mathcal{R}_{I}$
for both (\ref{eq:RI}) and (\ref{eq:RI_anti}). To be pedagogical
we proceed slowly. It is clear that $\mathcal{R}_{I}$ can be regarded
as a composite object of four parts, namely 
\begin{equation}
e^{\bar{\mathbf{a}}_{c}^{\dot{c}}J_{\dot{c}}^{c}}\;%LaTeX with PSTricks extensions
%%Creator: inkscape 0.48.1
%%Please note this file requires PSTricks extensions
\psset{xunit=.5pt,yunit=.5pt,runit=.5pt}
\begin{pspicture}[shift=-50](125.5,111)
{
\newrgbcolor{curcolor}{0 0 1}
\pscustom[linewidth=1,linecolor=curcolor]
{
\newpath
\moveto(75,35.5)
\lineto(75,0.5)
}
}
{
\newrgbcolor{curcolor}{0 0 1}
\pscustom[linewidth=1,linecolor=curcolor]
{
\newpath
\moveto(75,110.5)
\lineto(75,75.5)
}
}
{
\newrgbcolor{curcolor}{1 0 0}
\pscustom[linewidth=1,linecolor=curcolor]
{
\newpath
\moveto(95,55.5)
\curveto(121.97592,55.82787)(125,70.5)(125,110.5)
}
}
{
\newrgbcolor{curcolor}{1 0 0}
\pscustom[linewidth=1,linecolor=curcolor]
{
\newpath
\moveto(55,55.5)
\curveto(28.02408,55.17213)(25,40.5)(25,0.5)
}
}
{
\newrgbcolor{curcolor}{0 0 0}
\pscustom[linestyle=none,fillstyle=solid,fillcolor=curcolor]
{
\newpath
\moveto(12.56250367,57.48466502)
\curveto(12.87500367,57.48466502)(13.25000367,57.48466502)(13.25000367,57.85966502)
\curveto(13.25000367,58.26591502)(12.87500367,58.26591502)(12.59375367,58.26591502)
\lineto(0.65625367,58.26591502)
\curveto(0.37500367,58.26591502)(0.00000367,58.26591502)(0.00000367,57.85966502)
\curveto(0.00000367,57.48466502)(0.37500367,57.48466502)(0.65625367,57.48466502)
\closepath
\moveto(12.59375367,53.60966502)
\curveto(12.87500367,53.60966502)(13.25000367,53.60966502)(13.25000367,54.01591502)
\curveto(13.25000367,54.39091502)(12.87500367,54.39091502)(12.56250367,54.39091502)
\lineto(0.65625367,54.39091502)
\curveto(0.37500367,54.39091502)(0.00000367,54.39091502)(0.00000367,54.01591502)
\curveto(0.00000367,53.60966502)(0.37500367,53.60966502)(0.65625367,53.60966502)
\closepath
\moveto(12.59375367,53.60966502)
}
}
{
\newrgbcolor{curcolor}{1 1 1}
\pscustom[linestyle=none,fillstyle=solid,fillcolor=curcolor]
{
\newpath
\moveto(95,55.5)
\curveto(95,44.454305)(86.045695,35.5)(75,35.5)
\curveto(63.954305,35.5)(55,44.454305)(55,55.5)
\curveto(55,66.545695)(63.954305,75.5)(75,75.5)
\curveto(86.045695,75.5)(95,66.545695)(95,55.5)
\closepath
}
}
{
\newrgbcolor{curcolor}{0 0 0}
\pscustom[linewidth=1,linecolor=curcolor]
{
\newpath
\moveto(95,55.5)
\curveto(95,44.454305)(86.045695,35.5)(75,35.5)
\curveto(63.954305,35.5)(55,44.454305)(55,55.5)
\curveto(55,66.545695)(63.954305,75.5)(75,75.5)
\curveto(86.045695,75.5)(95,66.545695)(95,55.5)
\closepath
}
}
{
\newrgbcolor{curcolor}{0 0 0}
\pscustom[linestyle=none,fillstyle=solid,fillcolor=curcolor]
{
\newpath
\moveto(69.12862967,59.97931122)
\curveto(72.97237967,59.97931122)(73.97237967,59.04181122)(73.97237967,57.72931122)
\curveto(73.97237967,56.51056122)(73.00362967,54.10431122)(69.72237967,53.97931122)
\curveto(68.75362967,53.94806122)(68.06612967,53.26056122)(68.06612967,53.04181122)
\curveto(68.06612967,52.91681122)(68.12862967,52.91681122)(68.15987967,52.91681122)
\curveto(69.00362967,52.76056122)(69.40987967,52.38556122)(70.44112967,50.01056122)
\curveto(71.34737967,47.91681122)(71.87862967,47.01056122)(73.03487967,47.01056122)
\curveto(75.47237967,47.01056122)(77.75362967,49.44806122)(77.75362967,49.88556122)
\curveto(77.75362967,50.04181122)(77.59737967,50.04181122)(77.53487967,50.04181122)
\curveto(77.28487967,50.04181122)(76.50362967,49.76056122)(76.09737967,49.19806122)
\curveto(75.78487967,48.72931122)(75.34737967,48.10431122)(74.37862967,48.10431122)
\curveto(73.34737967,48.10431122)(72.72237967,49.51056122)(72.06612967,51.07306122)
\curveto(71.62862967,52.07306122)(71.28487967,52.82306122)(70.81612967,53.35431122)
\curveto(73.69112967,54.41681122)(75.65987967,56.51056122)(75.65987967,58.57306122)
\curveto(75.65987967,61.07306122)(72.31612967,61.07306122)(69.28487967,61.07306122)
\curveto(67.31612967,61.07306122)(66.19112967,61.07306122)(64.50362967,60.35431122)
\curveto(61.84737967,59.16681122)(61.47237967,57.51056122)(61.47237967,57.35431122)
\curveto(61.47237967,57.22931122)(61.56612967,57.19806122)(61.69112967,57.19806122)
\curveto(62.00362967,57.19806122)(62.47237967,57.47931122)(62.62862967,57.57306122)
\curveto(63.03487967,57.85431122)(63.09737967,57.97931122)(63.22237967,58.35431122)
\curveto(63.50362967,59.16681122)(64.06612967,59.85431122)(66.56612967,59.97931122)
\curveto(66.47237967,58.76056122)(66.28487967,56.88556122)(65.59737967,53.97931122)
\curveto(65.06612967,51.76056122)(64.34737967,49.57306122)(63.47237967,47.41681122)
\curveto(63.34737967,47.19806122)(63.34737967,47.16681122)(63.34737967,47.13556122)
\curveto(63.34737967,47.01056122)(63.50362967,47.01056122)(63.56612967,47.01056122)
\curveto(63.97237967,47.01056122)(64.78487967,47.47931122)(65.00362967,47.85431122)
\curveto(65.06612967,47.97931122)(67.59737967,53.60431122)(68.19112967,59.97931122)
\closepath
\moveto(69.12862967,59.97931122)
}
}
{
\newrgbcolor{curcolor}{0 0 0}
\pscustom[linestyle=none,fillstyle=solid,fillcolor=curcolor]
{
\newpath
\moveto(81.889616,60.83432221)
\lineto(86.170866,60.83432221)
\curveto(86.358366,60.83432221)(86.670866,60.83432221)(86.670866,61.14682221)
\curveto(86.670866,61.52182221)(86.358366,61.52182221)(86.170866,61.52182221)
\lineto(81.889616,61.52182221)
\lineto(81.889616,65.80307221)
\curveto(81.889616,65.95932221)(81.889616,66.30307221)(81.577116,66.30307221)
\curveto(81.233366,66.30307221)(81.233366,65.99057221)(81.233366,65.80307221)
\lineto(81.233366,61.52182221)
\lineto(76.952116,61.52182221)
\curveto(76.764616,61.52182221)(76.420866,61.52182221)(76.420866,61.17807221)
\curveto(76.420866,60.83432221)(76.733366,60.83432221)(76.952116,60.83432221)
\lineto(81.233366,60.83432221)
\lineto(81.233366,56.55307221)
\curveto(81.233366,56.36557221)(81.233366,56.02182221)(81.545866,56.02182221)
\curveto(81.889616,56.02182221)(81.889616,56.36557221)(81.889616,56.55307221)
\closepath
\moveto(81.889616,60.83432221)
}
}
\end{pspicture}\,,\quad\quad e^{-\mathbf{a}_{\dot{c}}^{c}J_{c}^{\dot{c}}}\;%LaTeX with PSTricks extensions
%%Creator: inkscape 0.48.1
%%Please note this file requires PSTricks extensions
\psset{xunit=.5pt,yunit=.5pt,runit=.5pt}
\begin{pspicture}[shift=-50](125.5,111)
{
\newrgbcolor{curcolor}{0 0 1}
\pscustom[linewidth=1,linecolor=curcolor]
{
\newpath
\moveto(75,35.5)
\lineto(75,0.5)
}
}
{
\newrgbcolor{curcolor}{0 0 1}
\pscustom[linewidth=1,linecolor=curcolor]
{
\newpath
\moveto(75,110.5)
\lineto(75,75.5)
}
}
{
\newrgbcolor{curcolor}{1 0 0}
\pscustom[linewidth=1,linecolor=curcolor]
{
\newpath
\moveto(95,55.5)
\curveto(121.97592,55.82787)(125,70.5)(125,110.5)
}
}
{
\newrgbcolor{curcolor}{1 0 0}
\pscustom[linewidth=1,linecolor=curcolor]
{
\newpath
\moveto(55,55.5)
\curveto(28.02408,55.17213)(25,40.5)(25,0.5)
}
}
{
\newrgbcolor{curcolor}{0 0 0}
\pscustom[linestyle=none,fillstyle=solid,fillcolor=curcolor]
{
\newpath
\moveto(12.56250367,57.48466502)
\curveto(12.87500367,57.48466502)(13.25000367,57.48466502)(13.25000367,57.85966502)
\curveto(13.25000367,58.26591502)(12.87500367,58.26591502)(12.59375367,58.26591502)
\lineto(0.65625367,58.26591502)
\curveto(0.37500367,58.26591502)(0.00000367,58.26591502)(0.00000367,57.85966502)
\curveto(0.00000367,57.48466502)(0.37500367,57.48466502)(0.65625367,57.48466502)
\closepath
\moveto(12.59375367,53.60966502)
\curveto(12.87500367,53.60966502)(13.25000367,53.60966502)(13.25000367,54.01591502)
\curveto(13.25000367,54.39091502)(12.87500367,54.39091502)(12.56250367,54.39091502)
\lineto(0.65625367,54.39091502)
\curveto(0.37500367,54.39091502)(0.00000367,54.39091502)(0.00000367,54.01591502)
\curveto(0.00000367,53.60966502)(0.37500367,53.60966502)(0.65625367,53.60966502)
\closepath
\moveto(12.59375367,53.60966502)
}
}
{
\newrgbcolor{curcolor}{1 1 1}
\pscustom[linestyle=none,fillstyle=solid,fillcolor=curcolor]
{
\newpath
\moveto(95,55.5)
\curveto(95,44.454305)(86.045695,35.5)(75,35.5)
\curveto(63.954305,35.5)(55,44.454305)(55,55.5)
\curveto(55,66.545695)(63.954305,75.5)(75,75.5)
\curveto(86.045695,75.5)(95,66.545695)(95,55.5)
\closepath
}
}
{
\newrgbcolor{curcolor}{0 0 0}
\pscustom[linewidth=1,linecolor=curcolor]
{
\newpath
\moveto(95,55.5)
\curveto(95,44.454305)(86.045695,35.5)(75,35.5)
\curveto(63.954305,35.5)(55,44.454305)(55,55.5)
\curveto(55,66.545695)(63.954305,75.5)(75,75.5)
\curveto(86.045695,75.5)(95,66.545695)(95,55.5)
\closepath
}
}
{
\newrgbcolor{curcolor}{0 0 0}
\pscustom[linestyle=none,fillstyle=solid,fillcolor=curcolor]
{
\newpath
\moveto(69.61317367,61.25012392)
\curveto(73.45692367,61.25012392)(74.45692367,60.31262392)(74.45692367,59.00012392)
\curveto(74.45692367,57.78137392)(73.48817367,55.37512392)(70.20692367,55.25012392)
\curveto(69.23817367,55.21887392)(68.55067367,54.53137392)(68.55067367,54.31262392)
\curveto(68.55067367,54.18762392)(68.61317367,54.18762392)(68.64442367,54.18762392)
\curveto(69.48817367,54.03137392)(69.89442367,53.65637392)(70.92567367,51.28137392)
\curveto(71.83192367,49.18762392)(72.36317367,48.28137392)(73.51942367,48.28137392)
\curveto(75.95692367,48.28137392)(78.23817367,50.71887392)(78.23817367,51.15637392)
\curveto(78.23817367,51.31262392)(78.08192367,51.31262392)(78.01942367,51.31262392)
\curveto(77.76942367,51.31262392)(76.98817367,51.03137392)(76.58192367,50.46887392)
\curveto(76.26942367,50.00012392)(75.83192367,49.37512392)(74.86317367,49.37512392)
\curveto(73.83192367,49.37512392)(73.20692367,50.78137392)(72.55067367,52.34387392)
\curveto(72.11317367,53.34387392)(71.76942367,54.09387392)(71.30067367,54.62512392)
\curveto(74.17567367,55.68762392)(76.14442367,57.78137392)(76.14442367,59.84387392)
\curveto(76.14442367,62.34387392)(72.80067367,62.34387392)(69.76942367,62.34387392)
\curveto(67.80067367,62.34387392)(66.67567367,62.34387392)(64.98817367,61.62512392)
\curveto(62.33192367,60.43762392)(61.95692367,58.78137392)(61.95692367,58.62512392)
\curveto(61.95692367,58.50012392)(62.05067367,58.46887392)(62.17567367,58.46887392)
\curveto(62.48817367,58.46887392)(62.95692367,58.75012392)(63.11317367,58.84387392)
\curveto(63.51942367,59.12512392)(63.58192367,59.25012392)(63.70692367,59.62512392)
\curveto(63.98817367,60.43762392)(64.55067367,61.12512392)(67.05067367,61.25012392)
\curveto(66.95692367,60.03137392)(66.76942367,58.15637392)(66.08192367,55.25012392)
\curveto(65.55067367,53.03137392)(64.83192367,50.84387392)(63.95692367,48.68762392)
\curveto(63.83192367,48.46887392)(63.83192367,48.43762392)(63.83192367,48.40637392)
\curveto(63.83192367,48.28137392)(63.98817367,48.28137392)(64.05067367,48.28137392)
\curveto(64.45692367,48.28137392)(65.26942367,48.75012392)(65.48817367,49.12512392)
\curveto(65.55067367,49.25012392)(68.08192367,54.87512392)(68.67567367,61.25012392)
\closepath
\moveto(69.61317367,61.25012392)
}
}
{
\newrgbcolor{curcolor}{0 0 0}
\pscustom[linestyle=none,fillstyle=solid,fillcolor=curcolor]
{
\newpath
\moveto(88.81166,59.10513491)
\curveto(89.03041,59.10513491)(89.37416,59.10513491)(89.37416,59.41763491)
\curveto(89.37416,59.79263491)(89.03041,59.79263491)(88.81166,59.79263491)
\lineto(80.49916,59.79263491)
\curveto(80.28041,59.79263491)(79.93666,59.79263491)(79.93666,59.44888491)
\curveto(79.93666,59.10513491)(80.24916,59.10513491)(80.49916,59.10513491)
\closepath
\moveto(88.81166,59.10513491)
}
}
\end{pspicture},
\end{equation}

\begin{equation}
\mathcal{R}_{0,I}\;%LaTeX with PSTricks extensions
%%Creator: inkscape 0.48.1
%%Please note this file requires PSTricks extensions
\psset{xunit=.5pt,yunit=.5pt,runit=.5pt}
\begin{pspicture}[shift=-50](125.5,111)
{
\newrgbcolor{curcolor}{0 0 1}
\pscustom[linewidth=1,linecolor=curcolor]
{
\newpath
\moveto(85,35)
\lineto(85,0)
}
}
{
\newrgbcolor{curcolor}{0 0 1}
\pscustom[linewidth=1,linecolor=curcolor]
{
\newpath
\moveto(85,110)
\lineto(85,75)
}
}
{
\newrgbcolor{curcolor}{0 0 0}
\pscustom[linestyle=none,fillstyle=solid,fillcolor=curcolor]
{
\newpath
\moveto(12.56250367,57.48466502)
\curveto(12.87500367,57.48466502)(13.25000367,57.48466502)(13.25000367,57.85966502)
\curveto(13.25000367,58.26591502)(12.87500367,58.26591502)(12.59375367,58.26591502)
\lineto(0.65625367,58.26591502)
\curveto(0.37500367,58.26591502)(0.00000367,58.26591502)(0.00000367,57.85966502)
\curveto(0.00000367,57.48466502)(0.37500367,57.48466502)(0.65625367,57.48466502)
\closepath
\moveto(12.59375367,53.60966502)
\curveto(12.87500367,53.60966502)(13.25000367,53.60966502)(13.25000367,54.01591502)
\curveto(13.25000367,54.39091502)(12.87500367,54.39091502)(12.56250367,54.39091502)
\lineto(0.65625367,54.39091502)
\curveto(0.37500367,54.39091502)(0.00000367,54.39091502)(0.00000367,54.01591502)
\curveto(0.00000367,53.60966502)(0.37500367,53.60966502)(0.65625367,53.60966502)
\closepath
\moveto(12.59375367,53.60966502)
}
}
{
\newrgbcolor{curcolor}{1 1 1}
\pscustom[linestyle=none,fillstyle=solid,fillcolor=curcolor]
{
\newpath
\moveto(105,55)
\curveto(105,43.954305)(96.045695,35)(85,35)
\curveto(73.954305,35)(65,43.954305)(65,55)
\curveto(65,66.045695)(73.954305,75)(85,75)
\curveto(96.045695,75)(105,66.045695)(105,55)
\closepath
}
}
{
\newrgbcolor{curcolor}{0 0 0}
\pscustom[linewidth=1,linecolor=curcolor]
{
\newpath
\moveto(105,55)
\curveto(105,43.954305)(96.045695,35)(85,35)
\curveto(73.954305,35)(65,43.954305)(65,55)
\curveto(65,66.045695)(73.954305,75)(85,75)
\curveto(96.045695,75)(105,66.045695)(105,55)
\closepath
}
}
{
\newrgbcolor{curcolor}{1 0 0}
\pscustom[linewidth=1,linecolor=curcolor]
{
\newpath
\moveto(40,109.5)
\lineto(40,-0.5)
}
}
{
\newrgbcolor{curcolor}{0 0 0}
\pscustom[linestyle=none,fillstyle=solid,fillcolor=curcolor]
{
\newpath
\moveto(80.92688367,61.60367732)
\curveto(84.77063367,61.60367732)(85.77063367,60.66617732)(85.77063367,59.35367732)
\curveto(85.77063367,58.13492732)(84.80188367,55.72867732)(81.52063367,55.60367732)
\curveto(80.55188367,55.57242732)(79.86438367,54.88492732)(79.86438367,54.66617732)
\curveto(79.86438367,54.54117732)(79.92688367,54.54117732)(79.95813367,54.54117732)
\curveto(80.80188367,54.38492732)(81.20813367,54.00992732)(82.23938367,51.63492732)
\curveto(83.14563367,49.54117732)(83.67688367,48.63492732)(84.83313367,48.63492732)
\curveto(87.27063367,48.63492732)(89.55188367,51.07242732)(89.55188367,51.50992732)
\curveto(89.55188367,51.66617732)(89.39563367,51.66617732)(89.33313367,51.66617732)
\curveto(89.08313367,51.66617732)(88.30188367,51.38492732)(87.89563367,50.82242732)
\curveto(87.58313367,50.35367732)(87.14563367,49.72867732)(86.17688367,49.72867732)
\curveto(85.14563367,49.72867732)(84.52063367,51.13492732)(83.86438367,52.69742732)
\curveto(83.42688367,53.69742732)(83.08313367,54.44742732)(82.61438367,54.97867732)
\curveto(85.48938367,56.04117732)(87.45813367,58.13492732)(87.45813367,60.19742732)
\curveto(87.45813367,62.69742732)(84.11438367,62.69742732)(81.08313367,62.69742732)
\curveto(79.11438367,62.69742732)(77.98938367,62.69742732)(76.30188367,61.97867732)
\curveto(73.64563367,60.79117732)(73.27063367,59.13492732)(73.27063367,58.97867732)
\curveto(73.27063367,58.85367732)(73.36438367,58.82242732)(73.48938367,58.82242732)
\curveto(73.80188367,58.82242732)(74.27063367,59.10367732)(74.42688367,59.19742732)
\curveto(74.83313367,59.47867732)(74.89563367,59.60367732)(75.02063367,59.97867732)
\curveto(75.30188367,60.79117732)(75.86438367,61.47867732)(78.36438367,61.60367732)
\curveto(78.27063367,60.38492732)(78.08313367,58.50992732)(77.39563367,55.60367732)
\curveto(76.86438367,53.38492732)(76.14563367,51.19742732)(75.27063367,49.04117732)
\curveto(75.14563367,48.82242732)(75.14563367,48.79117732)(75.14563367,48.75992732)
\curveto(75.14563367,48.63492732)(75.30188367,48.63492732)(75.36438367,48.63492732)
\curveto(75.77063367,48.63492732)(76.58313367,49.10367732)(76.80188367,49.47867732)
\curveto(76.86438367,49.60367732)(79.39563367,55.22867732)(79.98938367,61.60367732)
\closepath
\moveto(80.92688367,61.60367732)
}
}
{
\newrgbcolor{curcolor}{0 0 0}
\pscustom[linestyle=none,fillstyle=solid,fillcolor=curcolor]
{
\newpath
\moveto(96.93787,50.52192073)
\curveto(96.93787,52.05317073)(96.75037,53.17817073)(96.12537,54.14692073)
\curveto(95.68787,54.77192073)(94.81287,55.33442073)(93.71912,55.33442073)
\curveto(90.46912,55.33442073)(90.46912,51.52192073)(90.46912,50.52192073)
\curveto(90.46912,49.52192073)(90.46912,45.80317073)(93.71912,45.80317073)
\curveto(96.93787,45.80317073)(96.93787,49.52192073)(96.93787,50.52192073)
\closepath
\moveto(93.71912,46.20942073)
\curveto(93.06287,46.20942073)(92.21912,46.58442073)(91.93787,47.70942073)
\curveto(91.75037,48.52192073)(91.75037,49.67817073)(91.75037,50.70942073)
\curveto(91.75037,51.74067073)(91.75037,52.80317073)(91.93787,53.55317073)
\curveto(92.25037,54.64692073)(93.12537,54.95942073)(93.71912,54.95942073)
\curveto(94.46912,54.95942073)(95.18787,54.49067073)(95.43787,53.67817073)
\curveto(95.65662,52.92817073)(95.68787,51.92817073)(95.68787,50.70942073)
\curveto(95.68787,49.67817073)(95.68787,48.64692073)(95.50037,47.77192073)
\curveto(95.21912,46.49067073)(94.28162,46.20942073)(93.71912,46.20942073)
\closepath
\moveto(93.71912,46.20942073)
}
}
\end{pspicture}\,,\quad\quad\tilde{\mathcal{R}}_{0,I}\;%LaTeX with PSTricks extensions
%%Creator: inkscape 0.48.1
%%Please note this file requires PSTricks extensions
\psset{xunit=.5pt,yunit=.5pt,runit=.5pt}
\begin{pspicture}[shift=-50](125.5,111)
{
\newrgbcolor{curcolor}{0 0 1}
\pscustom[linewidth=1,linecolor=curcolor]
{
\newpath
\moveto(85,35)
\lineto(85,0)
}
}
{
\newrgbcolor{curcolor}{0 0 1}
\pscustom[linewidth=1,linecolor=curcolor]
{
\newpath
\moveto(85,110)
\lineto(85,75)
}
}
{
\newrgbcolor{curcolor}{0 0 0}
\pscustom[linestyle=none,fillstyle=solid,fillcolor=curcolor]
{
\newpath
\moveto(12.56250367,57.48466502)
\curveto(12.87500367,57.48466502)(13.25000367,57.48466502)(13.25000367,57.85966502)
\curveto(13.25000367,58.26591502)(12.87500367,58.26591502)(12.59375367,58.26591502)
\lineto(0.65625367,58.26591502)
\curveto(0.37500367,58.26591502)(0.00000367,58.26591502)(0.00000367,57.85966502)
\curveto(0.00000367,57.48466502)(0.37500367,57.48466502)(0.65625367,57.48466502)
\closepath
\moveto(12.59375367,53.60966502)
\curveto(12.87500367,53.60966502)(13.25000367,53.60966502)(13.25000367,54.01591502)
\curveto(13.25000367,54.39091502)(12.87500367,54.39091502)(12.56250367,54.39091502)
\lineto(0.65625367,54.39091502)
\curveto(0.37500367,54.39091502)(0.00000367,54.39091502)(0.00000367,54.01591502)
\curveto(0.00000367,53.60966502)(0.37500367,53.60966502)(0.65625367,53.60966502)
\closepath
\moveto(12.59375367,53.60966502)
}
}
{
\newrgbcolor{curcolor}{1 1 1}
\pscustom[linestyle=none,fillstyle=solid,fillcolor=curcolor]
{
\newpath
\moveto(105,55)
\curveto(105,43.954305)(96.045695,35)(85,35)
\curveto(73.954305,35)(65,43.954305)(65,55)
\curveto(65,66.045695)(73.954305,75)(85,75)
\curveto(96.045695,75)(105,66.045695)(105,55)
\closepath
}
}
{
\newrgbcolor{curcolor}{0 0 0}
\pscustom[linewidth=1,linecolor=curcolor]
{
\newpath
\moveto(105,55)
\curveto(105,43.954305)(96.045695,35)(85,35)
\curveto(73.954305,35)(65,43.954305)(65,55)
\curveto(65,66.045695)(73.954305,75)(85,75)
\curveto(96.045695,75)(105,66.045695)(105,55)
\closepath
}
}
{
\newrgbcolor{curcolor}{1 0 0}
\pscustom[linewidth=1,linecolor=curcolor]
{
\newpath
\moveto(40,109.5)
\lineto(40,-0.5)
}
}
{
\newrgbcolor{curcolor}{0 0 0}
\pscustom[linestyle=none,fillstyle=solid,fillcolor=curcolor]
{
\newpath
\moveto(87.32827243,66.78613466)
\lineto(87.01577243,67.06738466)
\curveto(87.01577243,67.06738466)(86.26577243,66.12988466)(85.39077243,66.12988466)
\curveto(84.92202243,66.12988466)(84.42202243,66.41113466)(84.07827243,66.62988466)
\curveto(83.54702243,66.94238466)(83.20327243,67.06738466)(82.85952243,67.06738466)
\curveto(82.10952243,67.06738466)(81.70327243,66.62988466)(80.70327243,65.50488466)
\lineto(81.01577243,65.22363466)
\curveto(81.01577243,65.22363466)(81.76577243,66.16113466)(82.64077243,66.16113466)
\curveto(83.10952243,66.16113466)(83.60952243,65.87988466)(83.95327243,65.69238466)
\curveto(84.48452243,65.34863466)(84.85952243,65.22363466)(85.20327243,65.22363466)
\curveto(85.95327243,65.22363466)(86.32827243,65.66113466)(87.32827243,66.78613466)
\closepath
\moveto(87.32827243,66.78613466)
}
}
{
\newrgbcolor{curcolor}{0 0 0}
\pscustom[linestyle=none,fillstyle=solid,fillcolor=curcolor]
{
\newpath
\moveto(81.98754367,61.25012392)
\curveto(85.83129367,61.25012392)(86.83129367,60.31262392)(86.83129367,59.00012392)
\curveto(86.83129367,57.78137392)(85.86254367,55.37512392)(82.58129367,55.25012392)
\curveto(81.61254367,55.21887392)(80.92504367,54.53137392)(80.92504367,54.31262392)
\curveto(80.92504367,54.18762392)(80.98754367,54.18762392)(81.01879367,54.18762392)
\curveto(81.86254367,54.03137392)(82.26879367,53.65637392)(83.30004367,51.28137392)
\curveto(84.20629367,49.18762392)(84.73754367,48.28137392)(85.89379367,48.28137392)
\curveto(88.33129367,48.28137392)(90.61254367,50.71887392)(90.61254367,51.15637392)
\curveto(90.61254367,51.31262392)(90.45629367,51.31262392)(90.39379367,51.31262392)
\curveto(90.14379367,51.31262392)(89.36254367,51.03137392)(88.95629367,50.46887392)
\curveto(88.64379367,50.00012392)(88.20629367,49.37512392)(87.23754367,49.37512392)
\curveto(86.20629367,49.37512392)(85.58129367,50.78137392)(84.92504367,52.34387392)
\curveto(84.48754367,53.34387392)(84.14379367,54.09387392)(83.67504367,54.62512392)
\curveto(86.55004367,55.68762392)(88.51879367,57.78137392)(88.51879367,59.84387392)
\curveto(88.51879367,62.34387392)(85.17504367,62.34387392)(82.14379367,62.34387392)
\curveto(80.17504367,62.34387392)(79.05004367,62.34387392)(77.36254367,61.62512392)
\curveto(74.70629367,60.43762392)(74.33129367,58.78137392)(74.33129367,58.62512392)
\curveto(74.33129367,58.50012392)(74.42504367,58.46887392)(74.55004367,58.46887392)
\curveto(74.86254367,58.46887392)(75.33129367,58.75012392)(75.48754367,58.84387392)
\curveto(75.89379367,59.12512392)(75.95629367,59.25012392)(76.08129367,59.62512392)
\curveto(76.36254367,60.43762392)(76.92504367,61.12512392)(79.42504367,61.25012392)
\curveto(79.33129367,60.03137392)(79.14379367,58.15637392)(78.45629367,55.25012392)
\curveto(77.92504367,53.03137392)(77.20629367,50.84387392)(76.33129367,48.68762392)
\curveto(76.20629367,48.46887392)(76.20629367,48.43762392)(76.20629367,48.40637392)
\curveto(76.20629367,48.28137392)(76.36254367,48.28137392)(76.42504367,48.28137392)
\curveto(76.83129367,48.28137392)(77.64379367,48.75012392)(77.86254367,49.12512392)
\curveto(77.92504367,49.25012392)(80.45629367,54.87512392)(81.05004367,61.25012392)
\closepath
\moveto(81.98754367,61.25012392)
}
}
{
\newrgbcolor{curcolor}{0 0 0}
\pscustom[linestyle=none,fillstyle=solid,fillcolor=curcolor]
{
\newpath
\moveto(97.99853,50.16836733)
\curveto(97.99853,51.69961733)(97.81103,52.82461733)(97.18603,53.79336733)
\curveto(96.74853,54.41836733)(95.87353,54.98086733)(94.77978,54.98086733)
\curveto(91.52978,54.98086733)(91.52978,51.16836733)(91.52978,50.16836733)
\curveto(91.52978,49.16836733)(91.52978,45.44961733)(94.77978,45.44961733)
\curveto(97.99853,45.44961733)(97.99853,49.16836733)(97.99853,50.16836733)
\closepath
\moveto(94.77978,45.85586733)
\curveto(94.12353,45.85586733)(93.27978,46.23086733)(92.99853,47.35586733)
\curveto(92.81103,48.16836733)(92.81103,49.32461733)(92.81103,50.35586733)
\curveto(92.81103,51.38711733)(92.81103,52.44961733)(92.99853,53.19961733)
\curveto(93.31103,54.29336733)(94.18603,54.60586733)(94.77978,54.60586733)
\curveto(95.52978,54.60586733)(96.24853,54.13711733)(96.49853,53.32461733)
\curveto(96.71728,52.57461733)(96.74853,51.57461733)(96.74853,50.35586733)
\curveto(96.74853,49.32461733)(96.74853,48.29336733)(96.56103,47.41836733)
\curveto(96.27978,46.13711733)(95.34228,45.85586733)(94.77978,45.85586733)
\closepath
\moveto(94.77978,45.85586733)
}
}
\end{pspicture}.\label{eq:diag:RRt0}
\end{equation}
$\mathcal{R}_{I,0}$ and $\tilde{\mathcal{R}}_{I,0}$ act trivially
in the auxiliary space, this is depicted by the straight line in (\ref{eq:diag:RRt0}).
The label $I$ is suppressed in the pictures. Let us now construct
the two expressions of $\mathcal{R}_{I}$ given in (\ref{eq:RI})
and (\ref{eq:RI_anti}) . Using the ingredients above and the multiplication
rules (\ref{eq:O1O2}) and (\ref{eq:O1circO2}) one finds

\begin{equation}
\mathcal{R}_{I}\;\input{NewDiagrams/Rdec},\quad\quad\mathcal{R}_{I}\;\input{NewDiagrams/Rtdec}\,.\label{eq:rdec}
\end{equation}
When reading the diagrams from bottom to top it becomes clear that
the expression on the left hand side is normal ordered, while the
expression on the right hand side is anti-normal ordered in the oscillator
space.

\subsection{Projection properties{\normalsize \label{sub:Projection-property}}}

In section~\ref{sec:Projection-property-of} we argued that at the
special points $\hat{z}$ and $\check{z}$ some $\mathcal{R}$-operators
of certain cardinality $\vert I\vert$ decompose into an outer product,
see (\ref{eq:RProjector}). This fact is a consequence of the degeneration
(to rank $1$) of $\mathcal{R}_{0,I}$ and $\tilde{\mathcal{R}}_{0,I}$
for generalized rectangular representations. In the diagrammatics
introduced, the building blocks of (\ref{eq:RProjector}) are denoted
by 
\begin{equation}
e^{\bar{\mathbf{a}}_{c}^{\dot{c}}J_{\dot{c}}^{c}}\vert hws\rangle\;%LaTeX with PSTricks extensions
%%Creator: inkscape 0.48.1
%%Please note this file requires PSTricks extensions
\psset{xunit=.5pt,yunit=.5pt,runit=.5pt}
\begin{pspicture}[shift=-50](123,110.5)
{
\newrgbcolor{curcolor}{1 0 0}
\pscustom[linewidth=1,linecolor=curcolor]
{
\newpath
\moveto(92.14645,55)
\curveto(119.12237,55.32787)(122.14645,70)(122.14645,110)
}
}
{
\newrgbcolor{curcolor}{0 0 1}
\pscustom[linewidth=1,linecolor=curcolor]
{
\newpath
\moveto(72,35.5)
\lineto(72,0.5)
}
}
{
\newrgbcolor{curcolor}{1 0 0}
\pscustom[linewidth=1,linecolor=curcolor]
{
\newpath
\moveto(52,55.5)
\curveto(25.02408,55.17213)(22,40.5)(22,0.5)
}
}
{
\newrgbcolor{curcolor}{0 0 0}
\pscustom[linestyle=none,fillstyle=solid,fillcolor=curcolor]
{
\newpath
\moveto(12.56250367,56.98466502)
\curveto(12.87500367,56.98466502)(13.25000367,56.98466502)(13.25000367,57.35966502)
\curveto(13.25000367,57.76591502)(12.87500367,57.76591502)(12.59375367,57.76591502)
\lineto(0.65625367,57.76591502)
\curveto(0.37500367,57.76591502)(0.00000367,57.76591502)(0.00000367,57.35966502)
\curveto(0.00000367,56.98466502)(0.37500367,56.98466502)(0.65625367,56.98466502)
\closepath
\moveto(12.59375367,53.10966502)
\curveto(12.87500367,53.10966502)(13.25000367,53.10966502)(13.25000367,53.51591502)
\curveto(13.25000367,53.89091502)(12.87500367,53.89091502)(12.56250367,53.89091502)
\lineto(0.65625367,53.89091502)
\curveto(0.37500367,53.89091502)(0.00000367,53.89091502)(0.00000367,53.51591502)
\curveto(0.00000367,53.10966502)(0.37500367,53.10966502)(0.65625367,53.10966502)
\closepath
\moveto(12.59375367,53.10966502)
}
}
{
\newrgbcolor{curcolor}{1 1 1}
\pscustom[linestyle=none,fillstyle=solid,fillcolor=curcolor]
{
\newpath
\moveto(92.000004,55.5)
\curveto(92.000004,44.454305)(83.045699,35.5)(72.000004,35.5)
\curveto(60.954309,35.5)(52.000004,44.454305)(52.000004,55.5)
\curveto(52.000004,66.545695)(60.954309,75.5)(72.000004,75.5)
\curveto(83.045699,75.5)(92.000004,66.545695)(92.000004,55.5)
\closepath
}
}
{
\newrgbcolor{curcolor}{0 0 0}
\pscustom[linewidth=1,linecolor=curcolor]
{
\newpath
\moveto(92.000004,55.5)
\curveto(92.000004,44.454305)(83.045699,35.5)(72.000004,35.5)
\curveto(60.954309,35.5)(52.000004,44.454305)(52.000004,55.5)
\curveto(52.000004,66.545695)(60.954309,75.5)(72.000004,75.5)
\curveto(83.045699,75.5)(92.000004,66.545695)(92.000004,55.5)
\closepath
}
}
{
\newrgbcolor{curcolor}{0 0 0}
\pscustom[linestyle=none,fillstyle=solid,fillcolor=curcolor]
{
\newpath
\moveto(72.57602367,54.95171122)
\lineto(78.13852367,54.95171122)
\curveto(78.41977367,54.95171122)(78.79477367,54.95171122)(78.79477367,55.35796122)
\curveto(78.79477367,55.73296122)(78.41977367,55.73296122)(78.13852367,55.73296122)
\lineto(72.57602367,55.73296122)
\lineto(72.57602367,61.32671122)
\curveto(72.57602367,61.60796122)(72.57602367,61.98296122)(72.16977367,61.98296122)
\curveto(71.76352367,61.98296122)(71.76352367,61.60796122)(71.76352367,61.32671122)
\lineto(71.76352367,55.73296122)
\lineto(66.20102367,55.73296122)
\curveto(65.91977367,55.73296122)(65.54477367,55.73296122)(65.54477367,55.35796122)
\curveto(65.54477367,54.95171122)(65.91977367,54.95171122)(66.20102367,54.95171122)
\lineto(71.76352367,54.95171122)
\lineto(71.76352367,49.35796122)
\curveto(71.76352367,49.07671122)(71.76352367,48.70171122)(72.16977367,48.70171122)
\curveto(72.57602367,48.70171122)(72.57602367,49.07671122)(72.57602367,49.35796122)
\closepath
\moveto(72.57602367,54.95171122)
}
}
\end{pspicture},\quad\quad\langle hws\vert e^{-\mathbf{a}_{\dot{c}}^{c}J_{c}^{\dot{c}}}\;%LaTeX with PSTricks extensions
%%Creator: inkscape 0.48.1
%%Please note this file requires PSTricks extensions
\psset{xunit=.5pt,yunit=.5pt,runit=.5pt}
\begin{pspicture}[shift=-50](122.64644623,111.19999695)
{
\newrgbcolor{curcolor}{1 0 0}
\pscustom[linewidth=1,linecolor=curcolor]
{
\newpath
\moveto(92.14645,54.99999695)
\curveto(119.12237,55.32786695)(122.14645,69.99999695)(122.14645,109.99999695)
}
}
{
\newrgbcolor{curcolor}{0 0 1}
\pscustom[linewidth=1,linecolor=curcolor]
{
\newpath
\moveto(71.8,110.69999695)
\lineto(71.8,75.69999695)
}
}
{
\newrgbcolor{curcolor}{1 0 0}
\pscustom[linewidth=1,linecolor=curcolor]
{
\newpath
\moveto(52,55.49999695)
\curveto(25.02408,55.17212695)(22,40.49999695)(22,0.49999695)
}
}
{
\newrgbcolor{curcolor}{0 0 0}
\pscustom[linestyle=none,fillstyle=solid,fillcolor=curcolor]
{
\newpath
\moveto(12.56250367,56.98466197)
\curveto(12.87500367,56.98466197)(13.25000367,56.98466197)(13.25000367,57.35966197)
\curveto(13.25000367,57.76591197)(12.87500367,57.76591197)(12.59375367,57.76591197)
\lineto(0.65625367,57.76591197)
\curveto(0.37500367,57.76591197)(0.00000367,57.76591197)(0.00000367,57.35966197)
\curveto(0.00000367,56.98466197)(0.37500367,56.98466197)(0.65625367,56.98466197)
\closepath
\moveto(12.59375367,53.10966197)
\curveto(12.87500367,53.10966197)(13.25000367,53.10966197)(13.25000367,53.51591197)
\curveto(13.25000367,53.89091197)(12.87500367,53.89091197)(12.56250367,53.89091197)
\lineto(0.65625367,53.89091197)
\curveto(0.37500367,53.89091197)(0.00000367,53.89091197)(0.00000367,53.51591197)
\curveto(0.00000367,53.10966197)(0.37500367,53.10966197)(0.65625367,53.10966197)
\closepath
\moveto(12.59375367,53.10966197)
}
}
{
\newrgbcolor{curcolor}{1 1 1}
\pscustom[linestyle=none,fillstyle=solid,fillcolor=curcolor]
{
\newpath
\moveto(92.000004,55.49999695)
\curveto(92.000004,44.45430195)(83.045699,35.49999695)(72.000004,35.49999695)
\curveto(60.954309,35.49999695)(52.000004,44.45430195)(52.000004,55.49999695)
\curveto(52.000004,66.54569194)(60.954309,75.49999695)(72.000004,75.49999695)
\curveto(83.045699,75.49999695)(92.000004,66.54569194)(92.000004,55.49999695)
\closepath
}
}
{
\newrgbcolor{curcolor}{0 0 0}
\pscustom[linewidth=1,linecolor=curcolor]
{
\newpath
\moveto(92.000004,55.49999695)
\curveto(92.000004,44.45430195)(83.045699,35.49999695)(72.000004,35.49999695)
\curveto(60.954309,35.49999695)(52.000004,44.45430195)(52.000004,55.49999695)
\curveto(52.000004,66.54569194)(60.954309,75.49999695)(72.000004,75.49999695)
\curveto(83.045699,75.49999695)(92.000004,66.54569194)(92.000004,55.49999695)
\closepath
}
}
{
\newrgbcolor{curcolor}{0 0 0}
\pscustom[linestyle=none,fillstyle=solid,fillcolor=curcolor]
{
\newpath
\moveto(77.50410367,54.95538787)
\curveto(77.84785367,54.95538787)(78.22285367,54.95538787)(78.22285367,55.36163787)
\curveto(78.22285367,55.73663787)(77.84785367,55.73663787)(77.50410367,55.73663787)
\lineto(66.72285367,55.73663787)
\curveto(66.37910367,55.73663787)(66.03535367,55.73663787)(66.03535367,55.36163787)
\curveto(66.03535367,54.95538787)(66.37910367,54.95538787)(66.72285367,54.95538787)
\closepath
\moveto(77.50410367,54.95538787)
}
}
\end{pspicture}.\label{eq:ApAm}
\end{equation}
As the above expressions are {}``vector'' and {}``covector'' in
the quantum space, a quantum space operator acts on them by left and
right multiplication, respectively. This is indicated by one missing
ingoing/outgoing vertical line. Using the notions of the two defined
products we see that according to (\ref{eq:RProjector}) at the special
points (\ref{eq:rdec}) simplifies to 

\begin{equation}
\mathcal{R}_{I}(\hat{z})\;%LaTeX with PSTricks extensions
%%Creator: inkscape 0.48.1
%%Please note this file requires PSTricks extensions
\psset{xunit=.5pt,yunit=.5pt,runit=.5pt}
\begin{pspicture}[shift=-50](193,111)
{
\newrgbcolor{curcolor}{0 0 1}
\pscustom[linewidth=1,linecolor=curcolor]
{
\newpath
\moveto(71.99999633,35.5)
\lineto(71.99999633,0.5)
}
}
{
\newrgbcolor{curcolor}{1 0 0}
\pscustom[linewidth=1,linecolor=curcolor]
{
\newpath
\moveto(91.99999633,55.5)
\curveto(114.41354633,55.63359)(138.29021633,55.508)(156.99999633,55.5)
}
}
{
\newrgbcolor{curcolor}{1 0 0}
\pscustom[linewidth=1,linecolor=curcolor]
{
\newpath
\moveto(51.99999633,55.5)
\curveto(25.02407633,55.17213)(21.99999633,40.5)(21.99999633,0.5)
}
}
{
\newrgbcolor{curcolor}{0 0 0}
\pscustom[linestyle=none,fillstyle=solid,fillcolor=curcolor]
{
\newpath
\moveto(12.5625,56.98466502)
\curveto(12.875,56.98466502)(13.25,56.98466502)(13.25,57.35966502)
\curveto(13.25,57.76591502)(12.875,57.76591502)(12.59375,57.76591502)
\lineto(0.65625,57.76591502)
\curveto(0.375,57.76591502)(0,57.76591502)(0,57.35966502)
\curveto(0,56.98466502)(0.375,56.98466502)(0.65625,56.98466502)
\closepath
\moveto(12.59375,53.10966502)
\curveto(12.875,53.10966502)(13.25,53.10966502)(13.25,53.51591502)
\curveto(13.25,53.89091502)(12.875,53.89091502)(12.5625,53.89091502)
\lineto(0.65625,53.89091502)
\curveto(0.375,53.89091502)(0,53.89091502)(0,53.51591502)
\curveto(0,53.10966502)(0.375,53.10966502)(0.65625,53.10966502)
\closepath
\moveto(12.59375,53.10966502)
}
}
{
\newrgbcolor{curcolor}{1 1 1}
\pscustom[linestyle=none,fillstyle=solid,fillcolor=curcolor]
{
\newpath
\moveto(92.00000033,55.5)
\curveto(92.00000033,44.454305)(83.04569532,35.5)(72.00000033,35.5)
\curveto(60.95430533,35.5)(52.00000033,44.454305)(52.00000033,55.5)
\curveto(52.00000033,66.545695)(60.95430533,75.5)(72.00000033,75.5)
\curveto(83.04569532,75.5)(92.00000033,66.545695)(92.00000033,55.5)
\closepath
}
}
{
\newrgbcolor{curcolor}{0 0 0}
\pscustom[linewidth=1,linecolor=curcolor]
{
\newpath
\moveto(92.00000033,55.5)
\curveto(92.00000033,44.454305)(83.04569532,35.5)(72.00000033,35.5)
\curveto(60.95430533,35.5)(52.00000033,44.454305)(52.00000033,55.5)
\curveto(52.00000033,66.545695)(60.95430533,75.5)(72.00000033,75.5)
\curveto(83.04569532,75.5)(92.00000033,66.545695)(92.00000033,55.5)
\closepath
}
}
{
\newrgbcolor{curcolor}{0 0 0}
\pscustom[linestyle=none,fillstyle=solid,fillcolor=curcolor]
{
\newpath
\moveto(72.57602,54.95171122)
\lineto(78.13852,54.95171122)
\curveto(78.41977,54.95171122)(78.79477,54.95171122)(78.79477,55.35796122)
\curveto(78.79477,55.73296122)(78.41977,55.73296122)(78.13852,55.73296122)
\lineto(72.57602,55.73296122)
\lineto(72.57602,61.32671122)
\curveto(72.57602,61.60796122)(72.57602,61.98296122)(72.16977,61.98296122)
\curveto(71.76352,61.98296122)(71.76352,61.60796122)(71.76352,61.32671122)
\lineto(71.76352,55.73296122)
\lineto(66.20102,55.73296122)
\curveto(65.91977,55.73296122)(65.54477,55.73296122)(65.54477,55.35796122)
\curveto(65.54477,54.95171122)(65.91977,54.95171122)(66.20102,54.95171122)
\lineto(71.76352,54.95171122)
\lineto(71.76352,49.35796122)
\curveto(71.76352,49.07671122)(71.76352,48.70171122)(72.16977,48.70171122)
\curveto(72.57602,48.70171122)(72.57602,49.07671122)(72.57602,49.35796122)
\closepath
\moveto(72.57602,54.95171122)
}
}
{
\newrgbcolor{curcolor}{0 0 1}
\pscustom[linewidth=1,linecolor=curcolor]
{
\newpath
\moveto(142.49999633,110.5)
\lineto(142.49999633,75.5)
}
}
{
\newrgbcolor{curcolor}{1 0 0}
\pscustom[linewidth=1,linecolor=curcolor]
{
\newpath
\moveto(162.49999633,55.5)
\curveto(189.47591633,55.82787)(192.49999633,70.5)(192.49999633,110.5)
}
}
{
\newrgbcolor{curcolor}{1 1 1}
\pscustom[linestyle=none,fillstyle=solid,fillcolor=curcolor]
{
\newpath
\moveto(162.49999633,55.5)
\curveto(162.49999633,44.454305)(153.54569132,35.5)(142.49999633,35.5)
\curveto(131.45430133,35.5)(122.49999633,44.454305)(122.49999633,55.5)
\curveto(122.49999633,66.545695)(131.45430133,75.5)(142.49999633,75.5)
\curveto(153.54569132,75.5)(162.49999633,66.545695)(162.49999633,55.5)
\closepath
}
}
{
\newrgbcolor{curcolor}{0 0 0}
\pscustom[linewidth=1,linecolor=curcolor]
{
\newpath
\moveto(162.49999633,55.5)
\curveto(162.49999633,44.454305)(153.54569132,35.5)(142.49999633,35.5)
\curveto(131.45430133,35.5)(122.49999633,44.454305)(122.49999633,55.5)
\curveto(122.49999633,66.545695)(131.45430133,75.5)(142.49999633,75.5)
\curveto(153.54569132,75.5)(162.49999633,66.545695)(162.49999633,55.5)
\closepath
}
}
{
\newrgbcolor{curcolor}{0 0 0}
\pscustom[linestyle=none,fillstyle=solid,fillcolor=curcolor]
{
\newpath
\moveto(149.04275584,54.70171122)
\curveto(149.38650584,54.70171122)(149.76150584,54.70171122)(149.76150584,55.10796122)
\curveto(149.76150584,55.48296122)(149.38650584,55.48296122)(149.04275584,55.48296122)
\lineto(138.26150584,55.48296122)
\curveto(137.91775584,55.48296122)(137.57400584,55.48296122)(137.57400584,55.10796122)
\curveto(137.57400584,54.70171122)(137.91775584,54.70171122)(138.26150584,54.70171122)
\closepath
\moveto(149.04275584,54.70171122)
}
}
\end{pspicture},\quad\quad\mathcal{R}_{I}(\check{z})\;%LaTeX with PSTricks extensions
%%Creator: inkscape 0.48.1
%%Please note this file requires PSTricks extensions
\psset{xunit=.5pt,yunit=.5pt,runit=.5pt}
\begin{pspicture}[shift=-50](193,111)
{
\newrgbcolor{curcolor}{0 0 1}
\pscustom[linewidth=1,linecolor=curcolor]
{
\newpath
\moveto(71.99999633,110.5)
\lineto(71.99999633,75.5)
}
}
{
\newrgbcolor{curcolor}{1 0 0}
\pscustom[linewidth=1,linecolor=curcolor]
{
\newpath
\moveto(91.99999633,55.5)
\curveto(114.41354633,55.63359)(138.29021633,55.508)(156.99999633,55.5)
}
}
{
\newrgbcolor{curcolor}{1 0 0}
\pscustom[linewidth=1,linecolor=curcolor]
{
\newpath
\moveto(51.99999633,55.5)
\curveto(25.02407633,55.17213)(21.99999633,40.5)(21.99999633,0.5)
}
}
{
\newrgbcolor{curcolor}{0 0 0}
\pscustom[linestyle=none,fillstyle=solid,fillcolor=curcolor]
{
\newpath
\moveto(12.5625,56.98466502)
\curveto(12.875,56.98466502)(13.25,56.98466502)(13.25,57.35966502)
\curveto(13.25,57.76591502)(12.875,57.76591502)(12.59375,57.76591502)
\lineto(0.65625,57.76591502)
\curveto(0.375,57.76591502)(0,57.76591502)(0,57.35966502)
\curveto(0,56.98466502)(0.375,56.98466502)(0.65625,56.98466502)
\closepath
\moveto(12.59375,53.10966502)
\curveto(12.875,53.10966502)(13.25,53.10966502)(13.25,53.51591502)
\curveto(13.25,53.89091502)(12.875,53.89091502)(12.5625,53.89091502)
\lineto(0.65625,53.89091502)
\curveto(0.375,53.89091502)(0,53.89091502)(0,53.51591502)
\curveto(0,53.10966502)(0.375,53.10966502)(0.65625,53.10966502)
\closepath
\moveto(12.59375,53.10966502)
}
}
{
\newrgbcolor{curcolor}{1 1 1}
\pscustom[linestyle=none,fillstyle=solid,fillcolor=curcolor]
{
\newpath
\moveto(92.00000033,55.5)
\curveto(92.00000033,44.454305)(83.04569532,35.5)(72.00000033,35.5)
\curveto(60.95430533,35.5)(52.00000033,44.454305)(52.00000033,55.5)
\curveto(52.00000033,66.545695)(60.95430533,75.5)(72.00000033,75.5)
\curveto(83.04569532,75.5)(92.00000033,66.545695)(92.00000033,55.5)
\closepath
}
}
{
\newrgbcolor{curcolor}{0 0 0}
\pscustom[linewidth=1,linecolor=curcolor]
{
\newpath
\moveto(92.00000033,55.5)
\curveto(92.00000033,44.454305)(83.04569532,35.5)(72.00000033,35.5)
\curveto(60.95430533,35.5)(52.00000033,44.454305)(52.00000033,55.5)
\curveto(52.00000033,66.545695)(60.95430533,75.5)(72.00000033,75.5)
\curveto(83.04569532,75.5)(92.00000033,66.545695)(92.00000033,55.5)
\closepath
}
}
{
\newrgbcolor{curcolor}{0 0 1}
\pscustom[linewidth=1,linecolor=curcolor]
{
\newpath
\moveto(142.49999633,35.5)
\lineto(142.49999633,0.5)
}
}
{
\newrgbcolor{curcolor}{1 0 0}
\pscustom[linewidth=1,linecolor=curcolor]
{
\newpath
\moveto(162.49999633,55.5)
\curveto(189.47591633,55.82787)(192.49999633,70.5)(192.49999633,110.5)
}
}
{
\newrgbcolor{curcolor}{1 1 1}
\pscustom[linestyle=none,fillstyle=solid,fillcolor=curcolor]
{
\newpath
\moveto(162.49999633,55.5)
\curveto(162.49999633,44.454305)(153.54569132,35.5)(142.49999633,35.5)
\curveto(131.45430133,35.5)(122.49999633,44.454305)(122.49999633,55.5)
\curveto(122.49999633,66.545695)(131.45430133,75.5)(142.49999633,75.5)
\curveto(153.54569132,75.5)(162.49999633,66.545695)(162.49999633,55.5)
\closepath
}
}
{
\newrgbcolor{curcolor}{0 0 0}
\pscustom[linewidth=1,linecolor=curcolor]
{
\newpath
\moveto(162.49999633,55.5)
\curveto(162.49999633,44.454305)(153.54569132,35.5)(142.49999633,35.5)
\curveto(131.45430133,35.5)(122.49999633,44.454305)(122.49999633,55.5)
\curveto(122.49999633,66.545695)(131.45430133,75.5)(142.49999633,75.5)
\curveto(153.54569132,75.5)(162.49999633,66.545695)(162.49999633,55.5)
\closepath
}
}
{
\newrgbcolor{curcolor}{0 0 0}
\pscustom[linestyle=none,fillstyle=solid,fillcolor=curcolor]
{
\newpath
\moveto(77.79275584,55.20171122)
\curveto(78.13650584,55.20171122)(78.51150584,55.20171122)(78.51150584,55.60796122)
\curveto(78.51150584,55.98296122)(78.13650584,55.98296122)(77.79275584,55.98296122)
\lineto(67.01150584,55.98296122)
\curveto(66.66775584,55.98296122)(66.32400584,55.98296122)(66.32400584,55.60796122)
\curveto(66.32400584,55.20171122)(66.66775584,55.20171122)(67.01150584,55.20171122)
\closepath
\moveto(77.79275584,55.20171122)
}
}
{
\newrgbcolor{curcolor}{0 0 0}
\pscustom[linestyle=none,fillstyle=solid,fillcolor=curcolor]
{
\newpath
\moveto(143.07602,54.45171122)
\lineto(148.63852,54.45171122)
\curveto(148.91977,54.45171122)(149.29477,54.45171122)(149.29477,54.85796122)
\curveto(149.29477,55.23296122)(148.91977,55.23296122)(148.63852,55.23296122)
\lineto(143.07602,55.23296122)
\lineto(143.07602,60.82671122)
\curveto(143.07602,61.10796122)(143.07602,61.48296122)(142.66977,61.48296122)
\curveto(142.26352,61.48296122)(142.26352,61.10796122)(142.26352,60.82671122)
\lineto(142.26352,55.23296122)
\lineto(136.70102,55.23296122)
\curveto(136.41977,55.23296122)(136.04477,55.23296122)(136.04477,54.85796122)
\curveto(136.04477,54.45171122)(136.41977,54.45171122)(136.70102,54.45171122)
\lineto(142.26352,54.45171122)
\lineto(142.26352,48.85796122)
\curveto(142.26352,48.57671122)(142.26352,48.20171122)(142.66977,48.20171122)
\curveto(143.07602,48.20171122)(143.07602,48.57671122)(143.07602,48.85796122)
\closepath
\moveto(143.07602,54.45171122)
}
}
\end{pspicture}\,.\label{eq:rdecproj}
\end{equation}

\subsection{Baxter Q-operators\label{sub:Baxter-Q-operators}}

Baxter Q-operators can be built from the monodromy of the $\mathcal{R}$-operators
as reviewed in section~\ref{sec:Review}. Here, we will concentrate
on the Q-operators constructed out of the $\mathcal{R}$-operators
that satisfy condition (\ref{eq:RProjector}), see section~\ref{sec:Projection-property-of}
for more details. Hereafter, the index $I$ of the chosen Q-operator
will be omitted. The diagrammatics for these $\mathcal{R}$-operators
was developed above. From (\ref{eq:rdecproj}) it is clear that the
corresponding Q-operators at the special points are given by 
\begin{equation}
{\bf Q}(\hat{z})\;\input{NewDiagrams/Qpresize}\label{eq:qp}
\end{equation}
and
\begin{equation}
{\bf Q}(\check{z})\;\input{NewDiagrams/Qmresize}.\label{eq:qm}
\end{equation}
Here $\mathcal{D}$ denotes the regulator in (\ref{eq:regulator}).
For convenience we recall that in (\ref{eq:qp}) and (\ref{eq:qm})
there is one ingoing and one outgoing vertical line for each spin-chain
site. As indicated in the picture, the auxiliary space is closed by
the trace, see (\ref{eq:Qop}).

\subsection{Shift mechanism\label{sub:Shift-mechanism-and}}

The homogeneous spin-chain has the property of being translationally
invariant; the shift operator defined as 
\begin{equation}
\mathbf{U}\, X_{n}\,\mathbf{U}^{-1}=X_{n+1}\,,\quad\quad\mathbf{U}\, X_{L}\,\mathbf{U}^{-1}=f_{1}(\phi)\, X_{1}\, f_{1}^{-1}(\phi)\,,\label{eq:shiftX}
\end{equation}
commutes with the Hamiltonian and all generalized transfer matrices.
The operator $f_{1}(\phi)$ arises from the twisted boundary conditions
and is explicitly given below.

The shift operator can be written as (see e.g. \cite{Sklyanin2008a,Faddeev2007})
\begin{equation}
\mathbf{U=}f_{1}(\phi)\,\mathbf{P}_{1,2}\mathbf{P}_{2,3}\cdots\mathbf{P}_{L-1,L}\,,
\end{equation}
 where $\mathbf{P}_{i,i+1}$ acts as a permutation on site $i$ and
$i+1$ in the quantum space. The main result of this subsection is
to show the relation 
\begin{equation}
\mathbf{Q}(\check{z})=\mathbf{U}\,\mathbf{Q}(\hat{z})\,.\label{eq:UQgQ-1}
\end{equation}
The label $I$ has been omitted following the logic as in section~\ref{sub:Projection-property}
and \ref{sub:Baxter-Q-operators}. This equation is immediately proven
once it is rewritten in the diagrammatic language developed previously:
\begin{equation}
\input{NewDiagrams/QeqUQ}\,.\label{eq:uqpic}
\end{equation}
The only non-trivial step in the proof is to move the last term $%LaTeX with PSTricks extensions
%%Creator: inkscape 0.48.1
%%Please note this file requires PSTricks extensions
\psset{xunit=.2pt,yunit=.2pt,runit=.2pt}
\begin{pspicture}(41,41)
{
\newrgbcolor{curcolor}{1 1 1}
\pscustom[linestyle=none,fillstyle=solid,fillcolor=curcolor]
{
\newpath
\moveto(40.5,20.5)
\curveto(40.5,9.454305)(31.545695,0.5)(20.5,0.5)
\curveto(9.454305,0.5)(0.5,9.454305)(0.5,20.5)
\curveto(0.5,31.545695)(9.454305,40.5)(20.5,40.5)
\curveto(31.545695,40.5)(40.5,31.545695)(40.5,20.5)
\closepath
}
}
{
\newrgbcolor{curcolor}{0 0 0}
\pscustom[linewidth=1,linecolor=curcolor]
{
\newpath
\moveto(40.5,20.5)
\curveto(40.5,9.454305)(31.545695,0.5)(20.5,0.5)
\curveto(9.454305,0.5)(0.5,9.454305)(0.5,20.5)
\curveto(0.5,31.545695)(9.454305,40.5)(20.5,40.5)
\curveto(31.545695,40.5)(40.5,31.545695)(40.5,20.5)
\closepath
}
}
{
\newrgbcolor{curcolor}{0 0 0}
\pscustom[linestyle=none,fillstyle=solid,fillcolor=curcolor]
{
\newpath
\moveto(20.86829663,19.34981438)
\lineto(31.3141774,19.34981438)
\curveto(31.84233991,19.34981438)(32.54655659,19.34981438)(32.54655659,20.11271578)
\curveto(32.54655659,20.81693246)(31.84233991,20.81693246)(31.3141774,20.81693246)
\lineto(20.86829663,20.81693246)
\lineto(20.86829663,31.32149796)
\curveto(20.86829663,31.84966047)(20.86829663,32.55387715)(20.10539522,32.55387715)
\curveto(19.34249382,32.55387715)(19.34249382,31.84966047)(19.34249382,31.32149796)
\lineto(19.34249382,20.81693246)
\lineto(8.89661304,20.81693246)
\curveto(8.36845053,20.81693246)(7.66423385,20.81693246)(7.66423385,20.11271578)
\curveto(7.66423385,19.34981438)(8.36845053,19.34981438)(8.89661304,19.34981438)
\lineto(19.34249382,19.34981438)
\lineto(19.34249382,8.84524888)
\curveto(19.34249382,8.31708637)(19.34249382,7.61286969)(20.10539522,7.61286969)
\curveto(20.86829663,7.61286969)(20.86829663,8.31708637)(20.86829663,8.84524888)
\closepath
\moveto(20.86829663,19.34981438)
}
}
\end{pspicture}$
in the left hand side of (\ref{eq:uqpic}) through the regulator $%LaTeX with PSTricks extensions
%%Creator: inkscape 0.48.1
%%Please note this file requires PSTricks extensions
\psset{xunit=.2pt,yunit=.2pt,runit=.2pt}
\begin{pspicture}(41,41)
{
\newrgbcolor{curcolor}{1 1 1}
\pscustom[linestyle=none,fillstyle=solid,fillcolor=curcolor]
{
\newpath
\moveto(40.5,20.5)
\curveto(40.5,9.454305)(31.545695,0.5)(20.5,0.5)
\curveto(9.454305,0.5)(0.5,9.454305)(0.5,20.5)
\curveto(0.5,31.545695)(9.454305,40.5)(20.5,40.5)
\curveto(31.545695,40.5)(40.5,31.545695)(40.5,20.5)
\closepath
}
}
{
\newrgbcolor{curcolor}{0 0 0}
\pscustom[linewidth=1,linecolor=curcolor]
{
\newpath
\moveto(40.5,20.5)
\curveto(40.5,9.454305)(31.545695,0.5)(20.5,0.5)
\curveto(9.454305,0.5)(0.5,9.454305)(0.5,20.5)
\curveto(0.5,31.545695)(9.454305,40.5)(20.5,40.5)
\curveto(31.545695,40.5)(40.5,31.545695)(40.5,20.5)
\closepath
}
}
{
\newrgbcolor{curcolor}{0 0 0}
\pscustom[linestyle=none,fillstyle=solid,fillcolor=curcolor]
{
\newpath
\moveto(14.64935033,8.91556116)
\curveto(22.65924018,8.91556116)(33.94834668,15.04393326)(33.94834668,24.39776436)
\curveto(33.94834668,27.46195041)(32.49689013,29.34346816)(30.83040298,30.41862116)
\curveto(27.87373223,32.35389656)(24.70203088,32.35389656)(21.42281423,32.35389656)
\curveto(18.51990113,32.35389656)(16.47711043,32.35389656)(13.57419733,31.11747061)
\curveto(9.05855473,29.07467991)(8.35970528,26.22552446)(8.35970528,25.95673621)
\curveto(8.35970528,25.74170561)(8.52097823,25.68794796)(8.73600883,25.68794796)
\curveto(9.27358533,25.68794796)(10.07995008,26.17176681)(10.34873833,26.33303976)
\curveto(11.04758778,26.81685861)(11.15510308,27.03188921)(11.37013368,27.67698101)
\curveto(11.85395253,29.07467991)(12.82159023,30.25734821)(17.12220223,30.47237881)
\curveto(16.58462573,23.26885371)(14.81062328,16.71042041)(12.44528668,10.74332126)
\curveto(11.15510308,10.31326006)(10.34873833,9.50689531)(10.34873833,9.13059176)
\curveto(10.34873833,8.96931881)(10.34873833,8.91556116)(11.10134543,8.91556116)
\closepath
\moveto(14.81062328,10.79707891)
\curveto(18.73493173,20.36594061)(19.43378118,26.33303976)(19.91760003,30.47237881)
\curveto(22.22917898,30.47237881)(30.99167593,30.47237881)(30.99167593,22.89255016)
\curveto(30.99167593,16.17284391)(24.97081913,10.79707891)(16.69214103,10.79707891)
\closepath
\moveto(14.81062328,10.79707891)
}
}
\end{pspicture}$.
This is done using the relation

\begin{equation}
\input{NewDiagrams/eqn}\,.
\end{equation}
A direct computation shows that 
\begin{equation}
f(\phi)=\exp\left\{ i\sum_{c\in I}\phi_{c}(J_{c}^{c}-\lambda_{I})+i\sum_{\dot{c}\in\bar{I}}\phi_{\dot{c}}(J_{\dot{c}}^{\dot{c}}-\bar{\lambda}_{I})\right\} \,.
\end{equation}
This proves relation (\ref{eq:UQgQ-1}) for the large class of generalized
rectangular representations. Using (\ref{eq:UQgQ-1}) and the form
of the Q-operator eigenvalues\footnote{Criteria concerning the diagonalizability of  Q-operators can be obtained following the arguments in \cite{Reshetikhin1989}. One finds sufficient conditions on $\mathcal{R}$-operators such that the corresponding transfer matrix is a normal operator.
This analysis requires a case-by-case study being a property connected to the choice of quantum space. One can show that the Q-operators considered in this section are diagonalizable.
} in terms of Bethe roots $\left\{ z_{i}\right\} _{i=1}^{M}$
\begin{equation}
Q(z)=e^{iz\phi_{I}}\prod_{i=1}^{M}(z-z_{i})\,.\label{eq:qbethe}
\end{equation}
 The eigenvalues of the shift operators are written as
\begin{equation}
U(\{z_{i}\})=e^{i(\check{z}-\hat{z})\phi_{I}}\prod_{i=1}^{M}\frac{\check{z}-z_{i}}{\hat{z}-z_{i}}\,.\label{eq:shiftfunction-1}
\end{equation}
The identification of the special points $\hat{z}$ and $\check{z}$
is particularly important as it reveals how higher local charges may
be extracted from Q-operators. In the next subsection this is elucidated
for the case of the nearest-neighbor Hamiltonian.

\subsection{The nearest-neighbor Hamiltonian and its action\label{sub:The-Hamiltonian-density}}

We identified two special points $\hat{z}$ and $\check{z}$ at which
the Q-operators are related by the shift operator, see (\ref{eq:UQgQ-1}).
This enables us to give a direct operatorial derivation of (\ref{eq:intofm}).
The main result of this section is that 
\begin{equation}
\mathbf{H}=\frac{\mathbf{Q}'(\check{z})}{\mathbf{Q}(\check{z})}-\frac{\mathbf{Q}'(\hat{z})}{\mathbf{Q}(\hat{z})}\,,\label{eq:H=00003DQ/Q-Q/Q}
\end{equation}
is a nearest-neighbor Hamiltonian, i.e. 
\begin{equation}
\mathbf{H}=\sum_{i=1}^{L}\mathcal{H}_{i,i+1}\,.\label{eq:density}
\end{equation}
It is of prime importance as it yields the total energy of the spin-chain
and determines the time-evolution of the system. An important step
in the derivation of the Hamiltonian (\ref{eq:density}) is to rewrite
(\ref{eq:H=00003DQ/Q-Q/Q}) as 
\begin{equation}
\mathbf{H}\,\mathbf{Q}(\check{z})=\mathbf{Q}'(\check{z})-\mathbf{U}\,\mathbf{Q}'(\hat{z})\,.\label{eq:HQ-1}
\end{equation}
Here we used (\ref{eq:UQgQ-1}). The derivation of (\ref{eq:H=00003DQ/Q-Q/Q})
is a direct consequence of the truly remarkable identity 
\begin{equation}
\input{NewDiagrams/Haction}\,\,.\label{eq:haction}
\end{equation}
The significance of this equation is twofold. Firstly, it contains
the non-trivial statement that the right-hand side of (\ref{eq:haction})
can be written as the left-hand side for {}``some'' $\mathcal{H}$
that, as encoded in the picture, acts non-trivially only in the quantum
space. This is proven in appendix~\ref{sec:The-action-of} using
the so called Sutherland equation, originally introduced to provide
a criterion for a local Hamiltonian to commute with a given tranfer
matrix \cite{Sutherland1970}, and the special properties of the $\mathcal{R}$-operator.
Secondly, (\ref{eq:haction}) defines $\mathcal{H}$ uniquely in terms
of the $\mathcal{R}$-operators for Q-operators. The fact that this
way of defining $\mathcal{H}$ can be particularly convenient for
practical purposes is supported by the non-trivial example of $\mathfrak{sl}(2)$
spin~$-\frac{1}{2}$ in appendix \ref{sec:The-spin-half}. 

Using (\ref{eq:haction}), (\ref{eq:HQ-1}) can be shown quickly.
The derivative of the Q-operators follows immediately from the definition
in (\ref{eq:Qop}) for any set $I$ 
\begin{equation}
\mathbf{Q}_{I}^{'}(z)=i\,\phi_{I}\,\mathbf{Q}_{I}(z)+e^{iz\,\phi_{I}}\,\sum_{k=1}^{L}\,\widehat{\text{Tr}}_{{\mathcal{H}}^{(I,\bar{I})}}\big\{\mathcal{D}_{I}\,\mathcal{R}_{I}(z)\otimes\ldots\otimes\underbrace{\mathcal{R}'_{I}(z)}_{\text{k-th side}}\otimes\ldots\otimes\mathcal{R}_{I}(z)\big\}\,.\label{eq:Qopprime-1}
\end{equation}
Taking a closer look at (\ref{eq:HQ-1}), one finds that the right
hand side can be rearranged as a sum of local contributions corresponding
to the Hamiltonian density $\mathcal{H}$. This is done by pairing
terms according to (\ref{eq:haction}). Furthermore, from (\ref{eq:HQ-1})
it follows that 
\begin{equation}
\mathcal{H}_{L,L+1}=f_{1}(\phi)\mathcal{H}_{L,1}f_{1}^{-1}(\phi)\,.\label{eq:htwist}
\end{equation}
Thus, we have shown (\ref{eq:HQ-1}). It is worth to mention that
an analogous and equivalent relation as (\ref{eq:haction}) holds
for the action of $\mathcal{H}$ from the right, see also (\ref{eq:HactionmodR}).
On the level of eigenvalues, upon using (\ref{eq:qbethe}), (\ref{eq:H=00003DQ/Q-Q/Q})
gives the famous energy formula 
\begin{equation}
E(\{z_{i}\})=\sum_{i=1}^{M}\left(\frac{1}{\check{z}-z_{i}}-\frac{1}{\hat{z}-z_{i}}\right)\,.\label{eq:energyformula}
\end{equation}
This coincides with (\ref{eq:intofm}) for $k=1$. We would like to
stress again that the auxiliary and quantum space of the $\mathcal{R}$-operators
are of different nature. The mechanism by which the Hamiltonian density
can be extracted from the $\mathcal{R}$-operators is encoded in (\ref{eq:haction}).
The explicit expression for $\mathcal{H}$ for generalized rectangular
representations in the quantum space is provided in the appendix~\ref{sub:Hamiltonian-Density}.
If we further restrict to certain representations one obtains rather
convenient expressions for the Hamiltonian density. This is done for
the fundamental representation and the $\mathfrak{sl}(2)$ spin~$-\frac{1}{2}$
case.

\section{Concluding remarks\label{sub:conclusion}}

In this paper we studied how the local charges directly enter the
hierarchy of Q-operators. Our studies provide a transparent derivation
of the operatorial version of the eigenvalue formula for the first
two local charges in (\ref{eq:intofm}) by employing the recent construction
of Q-operators \cite{Frassek2011}. The method states clearly which
Q-operators%
\footnote{The zeros of the eigenvalues of those distinguished Q-operators are
usually referred to as momentum carrying Bethe roots.%
} in the hierarchy can be used to extract local charges, given a rectangular
representation and their non-compact generalizations in the quantum
space $\Lambda$. It further implies the validity of the dispersion
relation for these representations. 

We found that the Q-operators constructed in \cite{Frassek2011} provide
an intuitive way to extract local charges. The mechanism relies on
special features of the novel $\mathcal{R}$-operators used as generating
objects for the Q-operators. Each $\mathcal{R}$-operator admits two
alternative presentations corresponding to a normal or anti-normal
ordered form in the auxiliary oscillator part. Some properties of
the $\mathcal{R}$-operators that are manifest in one presentation
are hidden in the other and vice versa. Following this paradigm, we
identified \textit{two} special points at which the $\mathcal{R}$-operators
degenerate under the two products defined in section~(\ref{sub:Yang-Baxter-approach}).
This fact can be traced back to the reduction of the oscillator independent
part $\mathcal{R}_{0,I}$ and $\tilde{\mathcal{R}}_{0,I}$ to rank
$1$, respectively. They become projectors on a highest weight state.
The diagrammatics developed incorporate this reduction. On the level
of Q-operators this leads to the shift operator which relates the
Q-operator at these two points, see (\ref{eq:UQgQ-1}). Furthermore,
it yields to the identification of the Hamiltonian density for the
nearest neighbor Hamiltonian. Here, we would like to stress that in
this method, the quantum and auxiliary spaces are in general of different
nature. An explicit formula of the Hamiltonian density and its action
in terms of the quantum space generators at the corresponding two
sites was obtained. In appendix~\ref{sec:The-spin-half} we showed
that this formulation is particularly convenient for the non-compact
$\mathfrak{sl}(2)$ spin-chain of representation $s=-1/2$. Interestingly,
spin-chains of this type emerge in the study of certain four dimensional
gauge theories, see e.g. \cite{Korchemsky2010}.

\begin{figure}
\begin{centering}
$\input{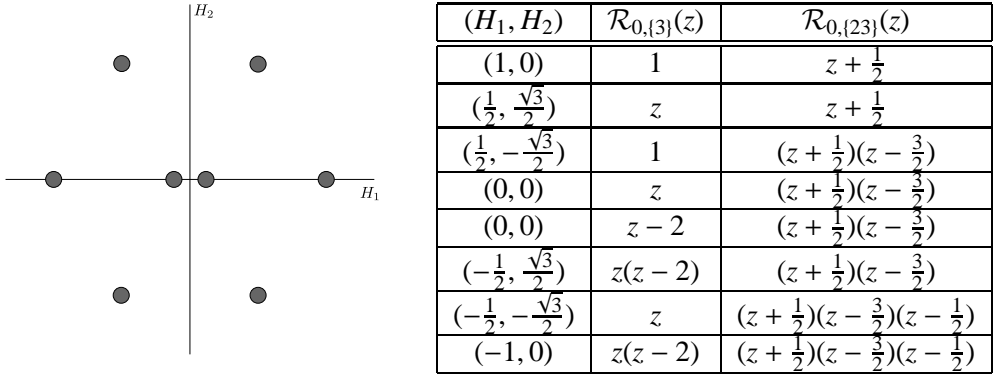}\quad\quad$%
\begin{tabular}{|c|c|c|}
\hline 
$(H_{1},H_{2})$ & $\mathcal{R}_{0,\{3\}}(z)$ & $\mathcal{R}_{0,\{23\}}(z)$\tabularnewline
\hline 
\hline 
$(1,0)$ & $1$ & $z+\frac{1}{2}$\tabularnewline
\hline 
$(\frac{1}{2},\frac{\sqrt{3}}{2})$ & $z$ & $z+\frac{1}{2}$\tabularnewline
\hline 
$(\frac{1}{2},-\frac{\sqrt{3}}{2})$ & $1$ & $(z+\frac{1}{2})(z-\frac{3}{2})$\tabularnewline
\hline 
$(0,0)$ & $z$ & $(z+\frac{1}{2})(z-\frac{3}{2})$\tabularnewline
\hline 
$(0,0)$ & $z-2$ & $(z+\frac{1}{2})(z-\frac{3}{2})$\tabularnewline
\hline 
$(-\frac{1}{2},\frac{\sqrt{3}}{2})$ & $z(z-2)$ & $(z+\frac{1}{2})(z-\frac{3}{2})$\tabularnewline
\hline 
$(-\frac{1}{2},-\frac{\sqrt{3}}{2})$ & $z$ & $(z+\frac{1}{2})(z-\frac{3}{2})(z-\frac{1}{2})$\tabularnewline
\hline 
$(-1,0)$ & $z(z-2)$ & $(z+\frac{1}{2})(z-\frac{3}{2})(z-\frac{1}{2})$\tabularnewline
\hline 
\end{tabular}
\par\end{centering}

\caption{\label{fig:Polynomial-structure-of}Polynomial structure of the $\mathcal{R}$-operators
using the normalization in \cite{Frassek2011}.}
\end{figure}

The generalization to non-rectangular representations remains open.
It is known that the nearest-neighbor Hamiltonian is non-hermitian
and the transfer matrices are non-normal operators for these representations
\cite{Reshetikhin1989}. In the example of the adjoint representation
of $\mathfrak{gl}(3)$ it is easy to show that the oscillator independent
part of the $\mathcal{R}$-operator does not reduce to rank $1$ at
any point of the spectral parameter $z$. We collected their polynomial
structure in figure~\ref{fig:Polynomial-structure-of}. It would
be interesting to study this problem more carefully to define local
charges using the developed method as a guiding principle. This might
also shed some light on representations without highest- nor lowest-weight
state. Besides the $\mathfrak{gl}(n)$ homogeneous spin chains there
exists a large zoo of quantum integrable models. Clearly, it would
be very interesting to study the presented mechanism in those. In
particular, it is certainly interesting to apply the method to the case of $\mathfrak{gl}(n\vert m)$.

The $\mathcal{R}_{0}$-operators at special values of the spectral
parameter discussed in section~\ref{sec:Projection-property-of}
share striking similarities with so called extremal-projectors, see
\cite{Tolstoy2010}. This connection deserves further investigation.
Besides, there is more territory to be explored in this direction.
In particular, it would be interesting to recover the $\mathcal{R}$-operators
used in this paper from the formula of Khoroshkin and Tolstoy for
the universal R-matrix for the Yangian double \cite{Khoroshkin2008},
see also \cite{Boos2010,Ridout2011} for recent applications in the
case of quantum affine algebras. This exercise, conceptually interesting
on its own, will allow to \textquotedbl{}dress\textquotedbl{} the
$\mathcal{R}$-operators used in this paper with their preferred normalization
factors. The relation between such factors and crossing symmetry has
briefly been discussed in section~\ref{sub:On-the-normalization}.
Remarkably, it appears that proper normalizations are necessary in
order to exploit the symmetry between the two presentations of $\mathcal{R}$-operators
used in this work.

The program of developing a systematic approach to the theory of quantum
integrable models based on the Q-operator method received increasing
attention in the last years, see e.g.~\cite{Boos2010,Tsuboi2012a}
and references therein. Despite constant progress a number of questions
remain open. We believe that the properties of the $\mathcal{R}$-operators
emphasized in this work play an important role in this program. Furthermore,
the calculation of correlation functions in quantum integrable models
remains an outstanding problem. It is believed that the Q-operator
method plays a prominent role in this investigation \cite{BoosCommun.Math.Phys.272:263-2812007}.
Also, it is worth mentioning potential applications of such integrability
techniques in the study of structure constants of $\mathcal{N}=4$
super Yang-Mills theory (SYM) along the lines presented in \cite{Escobedo2010}
and references therein.

Recently, an intriguing connection has been observed between tree
level scattering amplitudes in $\mathcal{N}=4$ SYM and certain contributions
to the $\mathfrak{psu}(2,2\vert4)$ integrable Hamiltonian corresponding
to the dilatation operator of the theory \cite{Zwiebel2011}. It would
be interesting to bring together the new point of view on integrable
Hamiltonians (\ref{eq:haction}) presented in this paper with this
remarkable connection. This might shed some light on the role the
degenerate representations of the Yangian algebra, crucial in the
Q-operator construction, play in the surprisingly rich structure\textcolor{black}{{}
\cite{Arkani-Hamed2010} }behind scattering amplitudes of $\mathcal{N}=4$
SYM.

\subsection*{Acknowledgments}

We would like to thank Matthias Staudacher for proposing this interesting
problem. Furthermore, we thank Yuri Aisaka, Nils Kanning, Tomasz \L{}ukowski
and Matthias Staudacher for useful discussions. We also like to thank
Tomasz \L{}ukowski for providing \texttt{Mathematica} code for the
construction of $\mathcal{R}$-matrices for compact representations.
Graphics were done using \texttt{Inkscape}. R.F. thanks the IAS Jerusalem
for hospitality during the course of this work. C.M. is partially
supported by a DFG grant in the framework of the SFB 676 \textquotedblleft{}\textit{Particles,
Strings and the Early Universe}\textquotedbl{}. 

\appendix

\section{Shifted weights\label{sec:Shifted-weights}}

The quantities $\hat{\ell}_{k}^{K}$ are important building blocks
for the $\mathcal{R}$-operators used in this paper. Spelling out
their characteristics is an essential step in the study of the properties
of Q-operators. The labels of $\hat{\ell}_{k}^{K}$ correspond to
a subset $K$ of $\{1,2,\dots,n\}$ and an index $k=1,2,\dots,|K|$.
For $\mathfrak{gl}(n)$ there are $n\cdot2^{n-1}$ such $\hat{\ell}_{k}^{K}$.
The set $K$ identifies a natural embedding of $\mathfrak{gl}(K)$
in $\mathfrak{gl}(n)$. The Casimirs of $\mathfrak{gl}(K)$ defined
as 
\begin{equation}
C_{i}^{(K)}\,=\, J_{a_{i}}^{a_{1}}\, J_{a_{1}}^{a_{2}}\,\dots J_{a_{i-1}}^{a_{i}}\quad\quad\text{with }a_{j}\in K\,,\label{eq:Casimir}
\end{equation}
 are symmetric polynomials of $\hat{\ell}_{k}^{K}$ via the following
formula %
\footnote{We refer to the previous paper \cite{Frassek2011}.%
} 
\begin{equation}
C_{i}^{(K)}\,=\,\sum_{k\in K}\,\prod_{j\neq k}\,\left(1+\frac{1}{\hat{\ell}_{k}^{K}-\hat{\ell}_{j}^{K}}\right)\,(\hat{\ell}_{k}^{K})^{i}\,.\label{eq:casimirofl}
\end{equation}
In general not all $\hat{\ell}_{k}^{K}$ do commute among themselves.
For a chosen path in the Hasse diagram, i.e. a sequence of sets $\mathcal{P}\equiv\emptyset\subset\{a\}\subset\{a,b\}\subset\dots\subset\{1,2,\dots,n\}$
ordered by inclusion, all the $\frac{n(n+1)}{2}$ corresponding $\hat{\ell}_{k}^{K}$
commute among themselves. In particular, for a given irreducible representation
of $\mathfrak{gl}(n)$ there exists a basis such that all $\hat{\ell}_{k}^{K}$
corresponding to the chosen path $\mathcal{P}$ act diagonally\footnote{An interesting class of infinite dimensional representations of $\mathfrak{gl}(n)$ for which $\hat{\ell}_{k}^{K}$ act as multiplication operators was introduced in \cite{Gerasimov2007} in connection with the method of separation of variables.}. This
basis coincides with the famous Gelfand-Tsetlin basis (see e.g. \cite{MolevM.HazewinkelEd.2006pp.109-170}
for a nice review and collection of references). The algebra $\mathfrak{gl}(n)$
admits a large zoo of representations. In the following we consider
some specific examples in more details. In the previous paper \cite{Frassek2011}
we defined the generalized rectangularity condition as $J_{C}^{A}J_{B}^{C}=\alpha J_{B}^{A}+\beta\delta_{B}^{A}$.
In this case, using (\ref{eq:Casimir}) and (\ref{eq:casimirofl})
for the full set $K=\{1,\ldots,n\}$, one can show that

\begin{equation}
\ell_{i}=\begin{cases}
(\bar{\lambda}_{I}-i+1)\, & i\leq a\\
(\lambda_{I}-i+1)\, & i>a
\end{cases}\,,\label{eq:ellls}
\end{equation}
where $a$ is an integer with $0\leq a\leq n$, compare to (\ref{eq:tableaux}),
and $\lambda_{I},\bar{\lambda}_{I}$ are in general complex numbers
related to $\alpha$ and $\beta$ via $\alpha=\lambda_{I}+\bar{\lambda}_{I}+\vert I\vert$,
$\beta=-\lambda_{I}(\bar{\lambda}_{I}+\vert I\vert)$ . The label
$I$ is introduced for consistency with section~\ref{sub:Reduction},
where $\vert I\vert=n-a$. Moreover, (\ref{eq:ellls}) should be understood
up to permutation of $\ell_{i}$. Generalized rectangular representations
have a number of remarkable features. Among others the tensor product
of such representations is multiplicity free, the weight diagram is
multiplicity free and the corresponding $\mathcal{L}$-operator (\ref{eq:Lax})
satisfies $\mathcal{L}(z)\mathcal{L}(\alpha-z)=z(\alpha-z)+\beta$.
If we further restrict to generalized rectangular representations
with highest weight we have

\begin{equation}
\hat{\ell}_{i}^{\bar{I}}\vert\Lambda_{0}^{I},\Lambda_{0}^{\bar{I}}\rangle=\ell_{i}\vert\Lambda_{0}^{I},\Lambda_{0}^{\bar{I}}\rangle=(\bar{\lambda}_{I}-i+1)\vert\Lambda_{0}^{I},\Lambda_{0}^{\bar{I}}\rangle,
\end{equation}
\begin{equation}
\hat{\ell}_{i}^{I}\vert\Lambda_{0}^{I},\Lambda_{0}^{\bar{I}}\rangle=(\ell_{i+\vert\bar{I}\vert}+\vert\bar{I}\vert)\vert\Lambda_{0}^{I},\Lambda_{0}^{\bar{I}}\rangle=(\lambda_{I}-i+1)\vert\Lambda_{0}^{I},\Lambda_{0}^{\bar{I}}\rangle\,.
\end{equation}
Where the state $\vert\Lambda_{0}^{I},\Lambda_{0}^{\bar{I}}\rangle$
was defined in (\ref{eq:sate}). It is worth to stress that for generalized
rectangular representations, for any set $K$, the shifted weight
operators $\hat{\ell}_{k}^{K},\hat{\ell}_{k}^{\bar{K}}$ contain the
same information. This is no longer the case for more general representations.

\section{Hamiltonian density}

\subsection{The action of the Hamiltonian density\label{sec:The-action-of}}

In this appendix we explain the origin of (\ref{eq:haction}). The
starting point is the equation 
\begin{equation}
[\mathcal{H}_{1,2},\mathcal{R}_{I,1}(z)\mathcal{R}_{I,2}(z)]=\mathcal{R}_{I,1}(z)\mathcal{R}'_{I,2}(z)-\mathcal{R}'_{I,1}(z)\mathcal{R}_{I,2}(z)\,.\label{eq:sutherland}
\end{equation}
It is a special case of Sutherland's equation \cite{Sutherland1970,Sklyanin2008a}
and ensures the commutativity of $\mathbf{H}$ with $\mathbf{Q}_{I}$
and the complete hierarchy of commuting operators. This equation plays a crucial role in the construction of higher charges using the boost operator approach \cite{tetelman}. See also \cite{Bargheer} where it was applied in a systematic study of integrable long-range spin-chains. Equation (\ref{eq:sutherland})
follows from the Yang-Baxter equation 
\begin{equation}
\mathbf{R}_{1,2}(z_{1}-z_{2})\mathcal{R}_{I,1}(z_{1})\mathcal{R}_{I,2}(z_{2})=\mathcal{R}_{I,2}(z_{2})\mathcal{R}_{I,1}(z_{1})\mathbf{R}_{1,2}(z_{1}-z_{2})\label{eq:lamlamint}
\end{equation}
where $\mathbf{R}_{1,2}$ denotes the fundamental R-matrix entering
(\ref{eq:regularity}) with equal the representation $\Lambda$ of
$\mathfrak{gl}(n)$ in $1$ and $2$. $\mathcal{R}_{I,1}$ and $\mathcal{R}_{I,2}$
are the $\mathcal{R}$-operators defined via (\ref{eq:YBE}). Expanding
(\ref{eq:lamlamint}) around $z_{1}=z_{2}=z$ and using the regularity
condition (\ref{eq:regularity}) together with $\mathcal{H}_{1,2}=\mathbf{P}_{1,2}\mathbf{R}'_{1,2}(0)$
one obtains (\ref{eq:sutherland}). Equation (\ref{eq:sutherland})
contains the free parameter $z$. To prove (\ref{eq:haction}) we
will focus on the values $\hat{z}$ and $\check{z}$. Instead of writing
formulas we rely on the developed diagrammatics. Without loss of generality
we may write 
\begin{equation}
\input{NewDiagrams/Hactionright}\,\,\label{eq:hactionmodL}
\end{equation}
and 
\begin{equation}
\input{NewDiagrams/Hactionleft}\,\,.\label{eq:HactionmodR}
\end{equation}
Inserting these equations in (\ref{eq:sutherland}) specified to the
values $\hat{z}$ and $\check{z}$ one obtains 
\begin{equation}
\input{NewDiagrams/sklyanin1}\,,\quad\quad\input{NewDiagrams/sklyanin2}\,,\label{eq:sklyanin}
\end{equation}
respectively. Observing that the operators
\begin{equation}
%LaTeX with PSTricks extensions
%%Creator: inkscape 0.48.1
%%Please note this file requires PSTricks extensions
\psset{xunit=.5pt,yunit=.5pt,runit=.5pt}
\begin{pspicture}[shift=-30](71.0246048,71.0246048)
{
\newrgbcolor{curcolor}{0 0 1}
\pscustom[linestyle=none,fillstyle=solid,fillcolor=curcolor]
{
\newpath
\moveto(35.5123,0.5123048)
\curveto(35.39332,26.9318048)(35.5053,48.4585448)(35.5123,70.5123048)
}
}
{
\newrgbcolor{curcolor}{0 0 1}
\pscustom[linewidth=1.02460253,linecolor=curcolor]
{
\newpath
\moveto(35.5123,0.5123048)
\curveto(35.39332,26.9318048)(35.5053,48.4585448)(35.5123,70.5123048)
}
}
{
\newrgbcolor{curcolor}{1 1 1}
\pscustom[linestyle=none,fillstyle=solid,fillcolor=curcolor]
{
\newpath
\moveto(30.51231409,45.51229285)
\lineto(40.51231409,45.51229285)
\lineto(40.51231409,70.51229285)
\lineto(30.51231409,70.51229285)
\closepath
}
}
{
\newrgbcolor{curcolor}{1 1 1}
\pscustom[linewidth=1,linecolor=curcolor]
{
\newpath
\moveto(30.51231409,45.51229285)
\lineto(40.51231409,45.51229285)
\lineto(40.51231409,70.51229285)
\lineto(30.51231409,70.51229285)
\closepath
}
}
{
\newrgbcolor{curcolor}{1 0 0}
\pscustom[linewidth=1.02460253,linecolor=curcolor]
{
\newpath
\moveto(0.5123,35.5123048)
\curveto(26.9318,35.6312848)(48.45854,35.5193048)(70.5123,35.5123048)
}
}
{
\newrgbcolor{curcolor}{1 1 1}
\pscustom[linestyle=none,fillstyle=solid,fillcolor=curcolor]
{
\newpath
\moveto(50.14644305,35.87816064)
\curveto(50.14644305,27.79594486)(43.59451262,21.24401444)(35.51229685,21.24401444)
\curveto(27.43008108,21.24401444)(20.87815065,27.79594486)(20.87815065,35.87816064)
\curveto(20.87815065,43.96037641)(27.43008108,50.51230684)(35.51229685,50.51230684)
\curveto(43.59451262,50.51230684)(50.14644305,43.96037641)(50.14644305,35.87816064)
\closepath
}
}
{
\newrgbcolor{curcolor}{0 0 0}
\pscustom[linewidth=0.73170731,linecolor=curcolor]
{
\newpath
\moveto(50.14644305,35.87816064)
\curveto(50.14644305,27.79594486)(43.59451262,21.24401444)(35.51229685,21.24401444)
\curveto(27.43008108,21.24401444)(20.87815065,27.79594486)(20.87815065,35.87816064)
\curveto(20.87815065,43.96037641)(27.43008108,50.51230684)(35.51229685,50.51230684)
\curveto(43.59451262,50.51230684)(50.14644305,43.96037641)(50.14644305,35.87816064)
\closepath
}
}
{
\newrgbcolor{curcolor}{0 0 0}
\pscustom[linestyle=none,fillstyle=solid,fillcolor=curcolor]
{
\newpath
\moveto(35.474567,34.88939819)
\lineto(40.19744434,34.88939819)
\curveto(40.43624151,34.88939819)(40.75463774,34.88939819)(40.75463774,35.23432744)
\curveto(40.75463774,35.55272366)(40.43624151,35.55272366)(40.19744434,35.55272366)
\lineto(35.474567,35.55272366)
\lineto(35.474567,40.30213402)
\curveto(35.474567,40.54093119)(35.474567,40.85932741)(35.12963776,40.85932741)
\curveto(34.78470852,40.85932741)(34.78470852,40.54093119)(34.78470852,40.30213402)
\lineto(34.78470852,35.55272366)
\lineto(30.06183118,35.55272366)
\curveto(29.82303401,35.55272366)(29.50463779,35.55272366)(29.50463779,35.23432744)
\curveto(29.50463779,34.88939819)(29.82303401,34.88939819)(30.06183118,34.88939819)
\lineto(34.78470852,34.88939819)
\lineto(34.78470852,30.13998784)
\curveto(34.78470852,29.90119067)(34.78470852,29.58279444)(35.12963776,29.58279444)
\curveto(35.474567,29.58279444)(35.474567,29.90119067)(35.474567,30.13998784)
\closepath
\moveto(35.474567,34.88939819)
}
}
\end{pspicture}\,:\mathcal{F}\rightarrow\mathcal{F}\otimes V_{\Lambda}\,,\quad\quad%LaTeX with PSTricks extensions
%%Creator: inkscape 0.48.1
%%Please note this file requires PSTricks extensions
\psset{xunit=.5pt,yunit=.5pt,runit=.5pt}
\begin{pspicture}[shift=-30](71.0246048,71.0246048)
{
\newrgbcolor{curcolor}{0 0 1}
\pscustom[linestyle=none,fillstyle=solid,fillcolor=curcolor]
{
\newpath
\moveto(35.5123,70.5123048)
\curveto(35.39332,44.0928048)(35.5053,22.5660648)(35.5123,0.5123048)
}
}
{
\newrgbcolor{curcolor}{0 0 1}
\pscustom[linewidth=1.02460253,linecolor=curcolor]
{
\newpath
\moveto(35.5123,70.5123048)
\curveto(35.39332,44.0928048)(35.5053,22.5660648)(35.5123,0.5123048)
}
}
{
\newrgbcolor{curcolor}{1 1 1}
\pscustom[linestyle=none,fillstyle=solid,fillcolor=curcolor]
{
\newpath
\moveto(30.5123,25.51230204)
\lineto(40.5123,25.51230204)
\lineto(40.5123,0.51230204)
\lineto(30.5123,0.51230204)
\closepath
}
}
{
\newrgbcolor{curcolor}{1 1 1}
\pscustom[linewidth=1,linecolor=curcolor]
{
\newpath
\moveto(30.5123,25.51230204)
\lineto(40.5123,25.51230204)
\lineto(40.5123,0.51230204)
\lineto(30.5123,0.51230204)
\closepath
}
}
{
\newrgbcolor{curcolor}{1 0 0}
\pscustom[linewidth=1.02460253,linecolor=curcolor]
{
\newpath
\moveto(0.5123,35.5123048)
\curveto(26.9318,35.3933248)(48.45854,35.5053048)(70.5123,35.5123048)
}
}
{
\newrgbcolor{curcolor}{1 1 1}
\pscustom[linestyle=none,fillstyle=solid,fillcolor=curcolor]
{
\newpath
\moveto(50.14644305,35.14644896)
\curveto(50.14644305,43.22866473)(43.59451262,49.78059516)(35.51229685,49.78059516)
\curveto(27.43008108,49.78059516)(20.87815065,43.22866473)(20.87815065,35.14644896)
\curveto(20.87815065,27.06423318)(27.43008108,20.51230276)(35.51229685,20.51230276)
\curveto(43.59451262,20.51230276)(50.14644305,27.06423318)(50.14644305,35.14644896)
\closepath
}
}
{
\newrgbcolor{curcolor}{0 0 0}
\pscustom[linewidth=0.73170731,linecolor=curcolor]
{
\newpath
\moveto(50.14644305,35.14644896)
\curveto(50.14644305,43.22866473)(43.59451262,49.78059516)(35.51229685,49.78059516)
\curveto(27.43008108,49.78059516)(20.87815065,43.22866473)(20.87815065,35.14644896)
\curveto(20.87815065,27.06423318)(27.43008108,20.51230276)(35.51229685,20.51230276)
\curveto(43.59451262,20.51230276)(50.14644305,27.06423318)(50.14644305,35.14644896)
\closepath
}
}
{
\newrgbcolor{curcolor}{0 0 0}
\pscustom[linestyle=none,fillstyle=solid,fillcolor=curcolor]
{
\newpath
\moveto(39.3409106,35.67968206)
\curveto(39.59243498,35.67968206)(39.86682522,35.67968206)(39.86682522,35.38242597)
\curveto(39.86682522,35.10803574)(39.59243498,35.10803574)(39.3409106,35.10803574)
\lineto(31.45219127,35.10803574)
\curveto(31.20066689,35.10803574)(30.9491425,35.10803574)(30.9491425,35.38242597)
\curveto(30.9491425,35.67968206)(31.20066689,35.67968206)(31.45219127,35.67968206)
\closepath
\moveto(39.3409106,35.67968206)
}
}
\end{pspicture}\,:\mathcal{F}\rightarrow\mathcal{F}\otimes V_{\Lambda}^{*}\,,
\end{equation}
have no kernel. According to (\ref{eq:ApAm}), $\mathcal{F}$ is associated
to the oscillator horizontal line while $V_{\Lambda}$ and $V_{\Lambda}^{*}$
correspond to the quantum space at a site and its dual. Then, (\ref{eq:sklyanin})
immediately implies
\begin{equation}
\input{NewDiagrams/italian}\,\,,\quad\quad\input{NewDiagrams/italianY}\,\,,
\end{equation}
where $c$ is an arbitrary constant and can be reabsorbed into the
definition of $\mathbf{H}$ in (\ref{eq:hactionmodL}), (\ref{eq:HactionmodR}).
This concludes the proof of (\ref{eq:haction}).

\subsection{A plug-in formula for the Hamiltonian density\label{sub:Hamiltonian-Density}}

For practical purposes we give a plug-in formula for the Hamiltonian
density in this appendix. By multiplying (\ref{eq:haction}) in the
auxiliary space with $%LaTeX with PSTricks extensions
%%Creator: inkscape 0.48.1
%%Please note this file requires PSTricks extensions
\psset{xunit=.3pt,yunit=.3pt,runit=.3pt}
\begin{pspicture}[shift=-30](71.0246048,71.0246048)
{
\newrgbcolor{curcolor}{0 0 1}
\pscustom[linestyle=none,fillstyle=solid,fillcolor=curcolor]
{
\newpath
\moveto(35.5123,70.5123048)
\curveto(35.39332,44.0928048)(35.5053,22.5660648)(35.5123,0.5123048)
}
}
{
\newrgbcolor{curcolor}{0 0 1}
\pscustom[linewidth=1.02460253,linecolor=curcolor]
{
\newpath
\moveto(35.5123,70.5123048)
\curveto(35.39332,44.0928048)(35.5053,22.5660648)(35.5123,0.5123048)
}
}
{
\newrgbcolor{curcolor}{1 1 1}
\pscustom[linestyle=none,fillstyle=solid,fillcolor=curcolor]
{
\newpath
\moveto(30.5123,25.51230204)
\lineto(40.5123,25.51230204)
\lineto(40.5123,0.51230204)
\lineto(30.5123,0.51230204)
\closepath
}
}
{
\newrgbcolor{curcolor}{1 1 1}
\pscustom[linewidth=1,linecolor=curcolor]
{
\newpath
\moveto(30.5123,25.51230204)
\lineto(40.5123,25.51230204)
\lineto(40.5123,0.51230204)
\lineto(30.5123,0.51230204)
\closepath
}
}
{
\newrgbcolor{curcolor}{1 0 0}
\pscustom[linewidth=1.02460253,linecolor=curcolor]
{
\newpath
\moveto(0.5123,35.5123048)
\curveto(26.9318,35.3933248)(48.45854,35.5053048)(70.5123,35.5123048)
}
}
{
\newrgbcolor{curcolor}{1 1 1}
\pscustom[linestyle=none,fillstyle=solid,fillcolor=curcolor]
{
\newpath
\moveto(50.14644305,35.14644896)
\curveto(50.14644305,43.22866473)(43.59451262,49.78059516)(35.51229685,49.78059516)
\curveto(27.43008108,49.78059516)(20.87815065,43.22866473)(20.87815065,35.14644896)
\curveto(20.87815065,27.06423318)(27.43008108,20.51230276)(35.51229685,20.51230276)
\curveto(43.59451262,20.51230276)(50.14644305,27.06423318)(50.14644305,35.14644896)
\closepath
}
}
{
\newrgbcolor{curcolor}{0 0 0}
\pscustom[linewidth=0.73170731,linecolor=curcolor]
{
\newpath
\moveto(50.14644305,35.14644896)
\curveto(50.14644305,43.22866473)(43.59451262,49.78059516)(35.51229685,49.78059516)
\curveto(27.43008108,49.78059516)(20.87815065,43.22866473)(20.87815065,35.14644896)
\curveto(20.87815065,27.06423318)(27.43008108,20.51230276)(35.51229685,20.51230276)
\curveto(43.59451262,20.51230276)(50.14644305,27.06423318)(50.14644305,35.14644896)
\closepath
}
}
{
\newrgbcolor{curcolor}{0 0 0}
\pscustom[linestyle=none,fillstyle=solid,fillcolor=curcolor]
{
\newpath
\moveto(39.3409106,35.67968206)
\curveto(39.59243498,35.67968206)(39.86682522,35.67968206)(39.86682522,35.38242597)
\curveto(39.86682522,35.10803574)(39.59243498,35.10803574)(39.3409106,35.10803574)
\lineto(31.45219127,35.10803574)
\curveto(31.20066689,35.10803574)(30.9491425,35.10803574)(30.9491425,35.38242597)
\curveto(30.9491425,35.67968206)(31.20066689,35.67968206)(31.45219127,35.67968206)
\closepath
\moveto(39.3409106,35.67968206)
}
}
\end{pspicture}$ from the right
one finds that 

\begin{equation}
\mathcal{H}_{i,i+1}\mathcal{R}_{I,i}(\check{z})\,\mathcal{R}_{I,i+1}(\check{z})=\mathcal{R}_{I,i}(\check{z})\,\mathcal{R}'_{I,i+1}(\check{z})-\mathbf{P}_{i,i+1}e^{\bar{\mathbf{a}}_{c}^{\dot{c}}J(i)_{\dot{c}}^{c}}\circ\vert hws\rangle_{i}\mathcal{R}'_{I,i+1}(\hat{z})\langle hws\vert_{i}\circ e^{-\mathbf{a}_{\dot{c}}^{c}J(i)_{c}^{\dot{c}}}.\label{eq:HLL}
\end{equation}
Interestingly, $\mathcal{R}_{I,i}(\check{z})\,\mathcal{R}_{I,i+1}(\check{z})$
can be inverted under $\cdot$ to obtain $\mathcal{H}_{i,i+1}$. As
$\mathcal{H}_{i,i+1}$ does not depend on the auxiliary space all
oscillators can be removed in a consistent way. In this way one can
write

\begin{multline}
\mathcal{H}_{i,i+1}\mathcal{R}_{0,i}(\check{z})e^{-J(i)_{a}^{\dot{a}}J(i+1)_{\dot{a}}^{a}}\mathcal{R}_{0,i+1}(\check{z})=\mathcal{R}_{0,i}(\check{z})e^{-J(i)_{a}^{\dot{a}}J(i+1)_{\dot{a}}^{a}}\mathcal{R}'_{0,i+1}(\check{z})\\
-\mathbf{P}_{i,i+1}\sum_{\{k_{c\dot{c}}\},\{m_{c\dot{c}}\}=0}^{\infty}\prod_{c\in I,\dot{c}\in\bar{I}}\frac{1}{k_{c\dot{c}}!m_{c\dot{c}}!}\,(J_{\dot{c}}^{c}(i)+J_{\dot{c}}^{c}(i+1))^{k_{\dot{c}c}+m_{c\dot{c}}}\,\mathcal{R}_{0,I}(\hat{z})\,\mathcal{\tilde{R}}'_{0,i+1}(\hat{z})\,\prod_{c\in I,\dot{c}\in\bar{I}}(J_{c}^{\dot{c}}(i))^{k_{c\dot{c}}}(J_{c}^{\dot{c}}(i+1))^{m_{c\dot{c}}}.
\end{multline}
In analogy to (\ref{eq:HLL}) this yields the Hamiltonian density.

\section{Examples}

\subsection{The $\mathfrak{gl}(2)$ case: Reordering and projection in full detail\label{sec:Example:gl2}}

In this section we exploit the properties mentioned in the previous
sections for the example of $\mathfrak{gl}(2)$ with $\vert I\vert=\{\dot{c}\}$,
$I=\{c\}$ and $c,\dot{c}=1,2$. In this case $\mathcal{R}_{0,\{c\}}$
and $\tilde{\mathcal{R}}_{0,\{c\}}$ are given by

\begin{equation}
\mathcal{R}_{0,\{c\}}(z)=\kappa_{\{c\}}(z)\,\frac{\Gamma(z+\frac{1}{2}-\ell_{1}^{\{\dot{c}\}})}{\Gamma(z+\frac{1}{2}-\bar{\lambda}_{c})},\quad\quad\tilde{\mathcal{R}}_{0,\{c\}}(z)=\tilde{\kappa}_{\{c\}}(z)\,\frac{\Gamma(z+\frac{3}{2}-\lambda_{c})}{\Gamma(z+\frac{3}{2}-\ell_{1}^{\{c\}})}\,;\label{eq:gl2_rtilde}
\end{equation}
see (\ref{eq:RI}) and (\ref{eq:RI_anti}), respectively. We will
now determine the explicit relation between $\rho$ and $\tilde{\rho}$
as discussed in section~\ref{sub:Direct-approach}. For the $\mathfrak{gl}(2)$
case $\mathcal{R}_{I}$ contains only one pair of oscillators. From
this follows that the conjugation in (\ref{eq:R_conjugation-1}) has
to be performed only once 
\begin{equation}
\tilde{\mathcal{R}}_{0,\{c\}}(z)=\sum_{n=0}^{\infty}\frac{1}{n!}(J_{\dot{c}}^{c})^{n}\mathcal{R}_{0,\{c\}}(z)(J_{c}^{\dot{c}})^{n}\,.\label{eq:gl2_conjugation}
\end{equation}
We can sum up this expression using the relation 
\begin{equation}
\left(J_{\dot{a}}^{a}\right)^{k}\left(J_{a}^{\dot{a}}\right)^{k}=(-1)^{k}\frac{\Gamma(J_{\dot{a}}^{\dot{a}}-\ell_{1}+k)\Gamma(J_{\dot{a}}^{\dot{a}}-\ell_{2}+k)}{\Gamma(J_{\dot{a}}^{\dot{a}}-\ell_{1})\Gamma(J_{\dot{a}}^{\dot{a}}-\ell_{2})},\label{eq:zenter-1}
\end{equation}
where $\ell_{i}=\ell_{i}^{\{a,\dot{a}\}}=\ell_{i}^{\{1,2\}}$. Applying
the reflection formula for Gamma functions 
\begin{equation}
\Gamma(1-z)\Gamma(z)=\frac{\pi}{\sin\pi z}\label{eq:Gamma_reflection}
\end{equation}
one finds that
\begin{equation}
\tilde{\mathcal{R}}_{0,\{c\}}(z)=-\kappa_{\{c\}}(z)\,\frac{\sin\pi(z+\frac{1}{2}-\ell_{1})\sin\pi(z+\frac{1}{2}-\ell_{2})}{\sin\pi(z+\frac{1}{2}-\ell_{1}^{\{\dot{c}\}})\sin\pi(z+\frac{1}{2}-\ell_{1}^{\{c\}})}\,\frac{\Gamma(z+\frac{3}{2}-\lambda_{c})}{\Gamma(z+\frac{3}{2}-\ell_{1}^{\{c\}})}\,,\label{eq:rtgl2exp}
\end{equation}
using $C_{1}=J_{c}^{c}+J_{\dot{c}}^{\dot{c}}=1+\ell_{1}+\ell_{2}$
and that up to permutation of $\ell_{1}$ and $\ell_{2}$ it holds
that $\ell_{1}=\bar{\lambda}_{c}$, $\ell_{2}=\lambda_{c}-1$, see
(\ref{eq:ellls}). This is exactly what we expected from the analysis
of the Yang-Baxter equation, compare (\ref{eq:gl2_rtilde}). Furthermore,
it fixes the relative normalization 
\begin{equation}
\tilde{\kappa}_{\{c\}}(z)=-\,\frac{\sin\pi(z+\frac{1}{2}-\ell_{1})\sin\pi(z+\frac{1}{2}-\ell_{2})}{\sin\pi(z+\frac{1}{2}-\ell_{1}^{\{\dot{c}\}})\sin\pi(z+\frac{1}{2}-\ell_{1}^{\{c\}})}\kappa_{\{c\}}(z)\,.\label{eq:rho_relation}
\end{equation}
Let us now look for the projection point as discussed section~\ref{sec:Projection-property-of}.
For $\hat{z}=\lambda_{\dot{c}}-\frac{1}{2}$ one obtains%
\footnote{Here we take $\kappa_{\{c\}}(z)\vert hws\rangle=\vert hws\rangle$. %
} 
\begin{equation}
\mathcal{R}_{0,\{c\}}(\hat{z})=\vert hws\rangle\langle hws\vert\,.
\end{equation}
On the other hand at $\check{z}=\lambda_{c}-\frac{3}{2}$ we find
\begin{equation}
\tilde{\mathcal{R}}_{0,\{c\}}(\check{z})=\vert hws\rangle\langle hws\vert.
\end{equation}
The total trigonometric prefactor reduces to $1$. Note that this
is the case for arbitrary $z$ on any state if spectrum of $\ell^{\{\dot{c}\}}$
and $\ell^{\{c\}}$ is integer spaced.

\subsection{The Hamiltonian action for the non-compact spin $s=-\frac{1}{2}$
chain\label{sec:The-spin-half} }

In this appendix we study how the Hamiltonian density for the non-compact
spin $-\frac{1}{2}$ spin-chain emerges in the presented formalism.
This spin-chain received special interest in the context of the $AdS_{5}/CFT_{4}$
correspondence \cite{BeisertNucl.Phys.B676:3-422004}.
The $\mathcal{R}$-operators for $\mathfrak{gl}(2)$ were discussed
in appendix~\ref{sec:Example:gl2} in great detail. Restricting to
$\mathfrak{sl}(2)$ one finds that one of the two $\mathcal{R}$-operators
can be written as

\begin{equation}
\mathcal{R}_{+}(z)\,=e^{\bar{\mathbf{a}}\, J_{+}}\,\,\mathcal{R}_{0,+}(z)\,\, e^{\mathbf{a}\, J_{-}}\quad\quad\text{with}\quad\quad\mathcal{R}_{0,+}(z)=\,\frac{\Gamma(z+J_{0})}{\Gamma(z+\sfrac{1}{2})}\,,\label{eq:RI-half}
\end{equation}
compare \ref{sec:Example:gl2}. The usual $\mathfrak{sl}(2)$ commutation
relations are 

\[
[J_{0},J_{\pm}]=\pm J_{\pm}\quad\quad[J_{+},J_{-}]=-2J_{0}\,.
\]
Furthermore, we define the action on module via the common relations

\begin{equation}
J_{+}\vert m\rangle=(m+1)\vert m+1\rangle\quad\quad J_{-}\vert m\rangle=m\vert m-1\rangle\quad\quad J_{0}\vert m\rangle=(m+\sfrac{1}{2})\vert m\rangle\,.
\end{equation}
It follows that 

\begin{equation}
\mathcal{R}_{0,+}(z)=\sum_{m=0}^{\infty}\frac{\Gamma(z+\sfrac{1}{2}+m)}{\Gamma(z+\sfrac{1}{2})}\,\vert m\rangle\langle m\vert\,,\quad\quad\mathcal{R}'_{0,+}(\hat{z})=\sum_{m=1}^{\infty}\Gamma(m)\,\vert m\rangle\langle m\vert\,,\label{eq:R0-half_mel}
\end{equation}
\begin{equation}
\mathcal{\tilde{R}}_{0,+}(z)=\sum_{m=0}^{\infty}(-1)^{m}\frac{\Gamma(-z+\sfrac{1}{2}+m)}{\Gamma(-z+\sfrac{1}{2})}\,\vert m\rangle\langle m\vert\,,\quad\quad\mathcal{\tilde{R}}'_{0,+}(\check{z})=\sum_{m=1}^{\infty}(-1)^{m+1}\Gamma(m)\,\vert m\rangle\langle m\vert\,,\label{eq:Rt0-half_mel-1}
\end{equation}
with $\hat{z}=-\sfrac{1}{2}$ and $\check{z}=\sfrac{1}{2}$. The relevant
terms in (\ref{eq:haction}) are then given by

\begin{equation}
%LaTeX with PSTricks extensions
%%Creator: inkscape 0.48.1
%%Please note this file requires PSTricks extensions
\psset{xunit=.5pt,yunit=.5pt,runit=.5pt}
\begin{pspicture}[shift=-80](170.890625,170.890625)
{
\newrgbcolor{curcolor}{0 0 1}
\pscustom[linewidth=0.89062893,linecolor=curcolor]
{
\newpath
\moveto(85.44531447,170.44531053)
\lineto(85.00000447,99.99999053)
}
}
{
\newrgbcolor{curcolor}{1 0 0}
\pscustom[linewidth=0.89062893,linecolor=curcolor]
{
\newpath
\moveto(112.55443447,86.33595053)
\curveto(132.51659447,86.45493053)(153.78184447,86.34295053)(170.44531447,86.33595053)
}
}
{
\newrgbcolor{curcolor}{1 0 0}
\pscustom[linewidth=0.89062893,linecolor=curcolor]
{
\newpath
\moveto(40.44531447,86.33595053)
\curveto(60.40746447,86.45493053)(81.67272447,86.34295053)(98.33619447,86.33595053)
}
}
{
\newrgbcolor{curcolor}{1 0 0}
\pscustom[linewidth=0.89062893,linecolor=curcolor]
{
\newpath
\moveto(0.44531447,86.33595053)
\curveto(20.40746447,86.45493053)(36.67272447,86.34295053)(53.33619447,86.33595053)
}
}
{
\newrgbcolor{curcolor}{1 1 1}
\pscustom[linestyle=none,fillstyle=solid,fillcolor=curcolor]
{
\newpath
\moveto(49.71361145,86.33595329)
\curveto(49.71361145,78.25373747)(43.16168099,71.70180701)(35.07946517,71.70180701)
\curveto(26.99724936,71.70180701)(20.4453189,78.25373747)(20.4453189,86.33595329)
\curveto(20.4453189,94.4181691)(26.99724936,100.97009956)(35.07946517,100.97009956)
\curveto(43.16168099,100.97009956)(49.71361145,94.4181691)(49.71361145,86.33595329)
\closepath
}
}
{
\newrgbcolor{curcolor}{0 0 0}
\pscustom[linewidth=0.73170731,linecolor=curcolor]
{
\newpath
\moveto(49.71361145,86.33595329)
\curveto(49.71361145,78.25373747)(43.16168099,71.70180701)(35.07946517,71.70180701)
\curveto(26.99724936,71.70180701)(20.4453189,78.25373747)(20.4453189,86.33595329)
\curveto(20.4453189,94.4181691)(26.99724936,100.97009956)(35.07946517,100.97009956)
\curveto(43.16168099,100.97009956)(49.71361145,94.4181691)(49.71361145,86.33595329)
\closepath
}
}
{
\newrgbcolor{curcolor}{0 0 0}
\pscustom[linestyle=none,fillstyle=solid,fillcolor=curcolor]
{
\newpath
\moveto(35.26063126,85.93476537)
\lineto(39.33075315,85.93476537)
\curveto(39.53654583,85.93476537)(39.81093606,85.93476537)(39.81093606,86.23202147)
\curveto(39.81093606,86.50641171)(39.53654583,86.50641171)(39.33075315,86.50641171)
\lineto(35.26063126,86.50641171)
\lineto(35.26063126,90.59939945)
\curveto(35.26063126,90.80519213)(35.26063126,91.07958237)(34.96337516,91.07958237)
\curveto(34.66611907,91.07958237)(34.66611907,90.80519213)(34.66611907,90.59939945)
\lineto(34.66611907,86.50641171)
\lineto(30.59599718,86.50641171)
\curveto(30.3902045,86.50641171)(30.11581427,86.50641171)(30.11581427,86.23202147)
\curveto(30.11581427,85.93476537)(30.3902045,85.93476537)(30.59599718,85.93476537)
\lineto(34.66611907,85.93476537)
\lineto(34.66611907,81.84177763)
\curveto(34.66611907,81.63598495)(34.66611907,81.36159471)(34.96337516,81.36159471)
\curveto(35.26063126,81.36159471)(35.26063126,81.63598495)(35.26063126,81.84177763)
\closepath
\moveto(35.26063126,85.93476537)
}
}
{
\newrgbcolor{curcolor}{1 0 0}
\pscustom[linewidth=0.89062893,linecolor=curcolor]
{
\newpath
\moveto(95.44531447,86.33595053)
\curveto(115.40746447,86.45493053)(136.67271447,86.34295053)(153.33619447,86.33595053)
}
}
{
\newrgbcolor{curcolor}{1 1 1}
\pscustom[linestyle=none,fillstyle=solid,fillcolor=curcolor]
{
\newpath
\moveto(99.71361138,86.33595329)
\curveto(99.71361138,78.25373747)(93.16168092,71.70180701)(85.07946511,71.70180701)
\curveto(76.99724929,71.70180701)(70.44531883,78.25373747)(70.44531883,86.33595329)
\curveto(70.44531883,94.4181691)(76.99724929,100.97009956)(85.07946511,100.97009956)
\curveto(93.16168092,100.97009956)(99.71361138,94.4181691)(99.71361138,86.33595329)
\closepath
}
}
{
\newrgbcolor{curcolor}{0 0 0}
\pscustom[linewidth=0.73170731,linecolor=curcolor]
{
\newpath
\moveto(99.71361138,86.33595329)
\curveto(99.71361138,78.25373747)(93.16168092,71.70180701)(85.07946511,71.70180701)
\curveto(76.99724929,71.70180701)(70.44531883,78.25373747)(70.44531883,86.33595329)
\curveto(70.44531883,94.4181691)(76.99724929,100.97009956)(85.07946511,100.97009956)
\curveto(93.16168092,100.97009956)(99.71361138,94.4181691)(99.71361138,86.33595329)
\closepath
}
}
{
\newrgbcolor{curcolor}{0 0 0}
\pscustom[linestyle=none,fillstyle=solid,fillcolor=curcolor]
{
\newpath
\moveto(89.35863951,86.09549997)
\curveto(89.6101639,86.09549997)(89.88455414,86.09549997)(89.88455414,86.39275606)
\curveto(89.88455414,86.6671463)(89.6101639,86.6671463)(89.35863951,86.6671463)
\lineto(81.46992012,86.6671463)
\curveto(81.21839574,86.6671463)(80.96687135,86.6671463)(80.96687135,86.39275606)
\curveto(80.96687135,86.09549997)(81.21839574,86.09549997)(81.46992012,86.09549997)
\closepath
\moveto(89.35863951,86.09549997)
}
}
{
\newrgbcolor{curcolor}{1 1 1}
\pscustom[linestyle=none,fillstyle=solid,fillcolor=curcolor]
{
\newpath
\moveto(150.07946431,85.97009929)
\curveto(150.07946431,77.88788347)(143.52753385,71.33595301)(135.44531804,71.33595301)
\curveto(127.36310222,71.33595301)(120.81117176,77.88788347)(120.81117176,85.97009929)
\curveto(120.81117176,94.0523151)(127.36310222,100.60424556)(135.44531804,100.60424556)
\curveto(143.52753385,100.60424556)(150.07946431,94.0523151)(150.07946431,85.97009929)
\closepath
}
}
{
\newrgbcolor{curcolor}{0 0 0}
\pscustom[linewidth=0.73170731,linecolor=curcolor]
{
\newpath
\moveto(150.07946431,85.97009929)
\curveto(150.07946431,77.88788347)(143.52753385,71.33595301)(135.44531804,71.33595301)
\curveto(127.36310222,71.33595301)(120.81117176,77.88788347)(120.81117176,85.97009929)
\curveto(120.81117176,94.0523151)(127.36310222,100.60424556)(135.44531804,100.60424556)
\curveto(143.52753385,100.60424556)(150.07946431,94.0523151)(150.07946431,85.97009929)
\closepath
}
}
{
\newrgbcolor{curcolor}{0 0 0}
\pscustom[linestyle=none,fillstyle=solid,fillcolor=curcolor]
{
\newpath
\moveto(135.62648576,85.56891137)
\lineto(139.69660765,85.56891137)
\curveto(139.90240033,85.56891137)(140.17679057,85.56891137)(140.17679057,85.86616747)
\curveto(140.17679057,86.14055771)(139.90240033,86.14055771)(139.69660765,86.14055771)
\lineto(135.62648576,86.14055771)
\lineto(135.62648576,90.23354545)
\curveto(135.62648576,90.43933813)(135.62648576,90.71372837)(135.32922967,90.71372837)
\curveto(135.03197358,90.71372837)(135.03197358,90.43933813)(135.03197358,90.23354545)
\lineto(135.03197358,86.14055771)
\lineto(130.96185169,86.14055771)
\curveto(130.75605901,86.14055771)(130.48166877,86.14055771)(130.48166877,85.86616747)
\curveto(130.48166877,85.56891137)(130.75605901,85.56891137)(130.96185169,85.56891137)
\lineto(135.03197358,85.56891137)
\lineto(135.03197358,81.47592363)
\curveto(135.03197358,81.27013095)(135.03197358,80.99574071)(135.32922967,80.99574071)
\curveto(135.62648576,80.99574071)(135.62648576,81.27013095)(135.62648576,81.47592363)
\closepath
\moveto(135.62648576,85.56891137)
}
}
{
\newrgbcolor{curcolor}{0 0 1}
\pscustom[linewidth=0.89062893,linecolor=curcolor]
{
\newpath
\moveto(35.44531447,71.34563053)
\lineto(35.00000447,0.90031053)
}
}
{
\newrgbcolor{curcolor}{0 0 1}
\pscustom[linewidth=0.89062893,linecolor=curcolor]
{
\newpath
\moveto(135.44531447,71.34563053)
\lineto(135.00000447,0.90031053)
}
}
\end{pspicture}=\sum_{m_{1},m_{2}=0}^{\infty}\bar{\mathbf{a}}^{m_{1}}\left(\vert m_{1},m_{2}\rangle\right)\langle0\vert e^{\mathbf{a}J_{2}^{-}}\bar{\mathbf{a}}^{m_{2}}\,,
\end{equation}
\begin{equation}
\input{NewDiagrams/Rpp}=-\sum_{m_{1},m_{2}=0}^{\infty}\bar{\mathbf{a}}^{m_{1}}\left(h(m_{1})\,\vert m_{1},m_{2}\rangle-\sum_{\ell=1}^{m_{2}}\frac{1}{\ell}\vert m_{1}+\ell,m_{2}-\ell\rangle\right)\langle0\vert e^{\mathbf{a}J_{2}^{-}}\bar{\mathbf{a}}^{m_{2}}\,,
\end{equation}

\begin{equation}
\input{NewDiagrams/pR}=\sum_{m_{1},m_{2}=0}^{\infty}\bar{\mathbf{a}}^{m_{1}}\left(h(m_{2})\,\vert m_{1},m_{2}\rangle-\sum_{\ell=1}^{m_{1}}\frac{1}{\ell}\vert m_{1}-\ell,m_{2}+\ell\rangle\right)\langle0\vert e^{\mathbf{a}J_{2}^{-}}\bar{\mathbf{a}}^{m_{2}}\,.
\end{equation}
From this we find that
\begin{equation}
\mathcal{H}\vert m_{1}m_{2}\rangle=\left(h(m_{1})+h(m_{2})\right)\vert m_{1},m_{2}\rangle-\sum_{\ell=1}^{m_{1}}\frac{1}{\ell}\vert m_{1}-\ell,m_{2}+\ell\rangle-\sum_{\ell=1}^{m_{2}}\frac{1}{\ell}\vert m_{1}+\ell,m_{2}-\ell\rangle\,.
\end{equation}
Note that the constant discussed in appendix~\ref{sec:The-action-of}
is fixed to be $c=0$.

\subsection{Projection properties of $\mathcal{R}$-operators for $\mathfrak{su}(2,2)$\label{sec:Su(2,2)}}

\begin{figure}
\begin{centering}
$\input{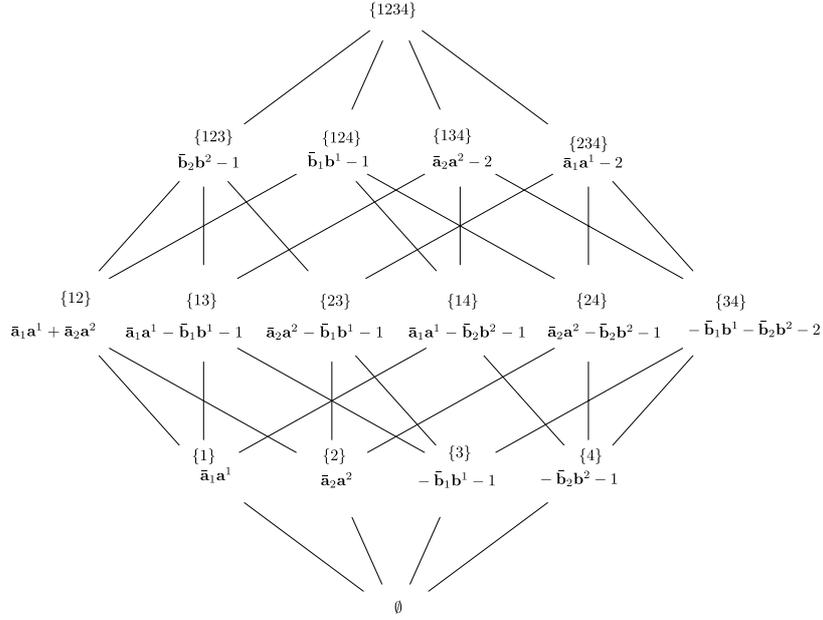}$
\par\end{centering}

\caption{\label{fig:Hasse-diagramm-with}Hasse diagramm for $\mathfrak{su}(2,2)$
including the shifted weights of the subalgebra $\hat{\ell}_{1}^{I}-\gamma$
according to (\ref{eq:twelve}). }
\end{figure}
It is instructive to present the structure of $\hat{\ell}_{k}^{K}$
in full detail for the interesting class of representations usually
referred to as oscillator representations. These representations represent
a subfamily of the generalized rectangular representations. The $\mathfrak{gl}(n)$
generators take the simple form 
\begin{equation}
J_{B}^{A}\,=\,\bar{b}^{A}\, b_{B}+\gamma\,\delta_{B}^{A}\,,\qquad[b_{B},\bar{b}^{A}]\,=\,\delta_{B}^{A}\,,
\end{equation}
where $\gamma$ commutes with all the generators. Using purely algebraic
manipulations, one can show that for any fixed set $K\subseteq\{1,\ldots,n\}$
the corresponding set of $\hat{\ell}_{k}^{K}$ is given by 
\begin{equation}
\{\gamma+\widehat{N}^{K},\gamma-1,\gamma-2,\dots,\gamma-|K|+1\}\,,\qquad\widehat{N}^{K}\,\equiv\,\sum_{c\in K}\,\bar{b}^{c}\, b_{c}\,.\label{eq:twelve}
\end{equation}
Notice that for each set $K$ only one $\hat{\ell}_{k}^{K}$ is a
non-trivial operator. The spectrum of $\hat{\ell}_{k}^{K}$ thus follows
from the spectrum of $\widehat{N}^{K}$. The spectrum of $\widehat{N}^{K}$
in turn depends on the choice of the vacuum for the oscillator algebra.
After renaming the oscillators according to 
\begin{equation}
\bar{b}^{A}\,=\,(\bar{a}^{\alpha},b^{\dot{\alpha}})\,,\qquad b_{A}\,=\,(a_{\alpha},-\bar{b}_{\dot{\alpha}})\,,
\end{equation}
 where $\alpha=1,\ldots,p$ and $\dot{\alpha}=p+1,\ldots,n$, the
vacuum is defined by 
\begin{equation}
a_{\alpha}\,\vert0\rangle=\,0\,=b^{\dot{\alpha}}\,\vert0\rangle\,.
\end{equation}
The representations obtained in this way are not irreducible, the
operator $\widehat{N}^{\{1,\dots,n\}}$ is central and its eigenvalues
label an infinite family of unitary irreducible representations of
$\mathfrak{su}(p,n-p)$. The case of $\mathfrak{su}(2,2)$ for $\widehat{N}^{\{1,\dots,4\}}=0$
is given in figure~\ref{fig:Hasse-diagramm-with}. We conclude that
only the Q-operator corresponding to the set $I=\{3,4\}$ fulfills
the criteria given in section~\ref{sub:Reduction}. The algebra $\mathfrak{su}(2,2)$
is the conformal algebra in four dimensions and the representation
chosen in the example corresponds to the so called massless scalar
field.

\subsection{The Hamiltonian for the fundamental representation }

For the fundamental representation one has $a=1$, see (\ref{eq:tableaux}).
Therefore the $\mathcal{R}$-operators of cardinality $\vert I\vert=n-1$
carry the information about the Hamiltonian. In this case the special
points are located at $\hat{z}=+\frac{1}{2}$ and $\check{z}=-\frac{1}{2}$,
compare (\ref{eq:degenerate_fund}). The derivative \textbf{$\mathbf{L}_{I}'$
}does not depend on the spectral parameter $z$ and does not contain
oscillators. It follows that equation (\ref{eq:HLL}) simplifies to

\begin{equation}
\mathcal{H}_{i,i+1}\mathbf{L}_{I,i}(\check{z})\,\mathbf{L}_{I,i+1}(\check{z})=(\mathbf{1}-\mathbf{P})_{i,i+1}\mathbf{L}_{I,i}(\check{z})\mathbf{L}'_{I,i+1}\,.\label{eq:HLL_restricted}
\end{equation}
Furthermore, $\mathbf{L}'_{I}$ can be written as 
\begin{equation}
\mathbf{L}'_{I}=\mathbf{L}_{I}(\hat{z})-\mathbf{L}_{I}(\check{z})\,.
\end{equation}
The well known expression for the Hamiltonian density 
\begin{equation}
\mathcal{H}_{i,i+1}=(\mathbf{P}-\mathbf{1})_{i,i+1}
\end{equation}
follows noting that
\begin{equation}\label{eq:ident}
(\mathbf{1}-\mathbf{P})_{i,i+1}\mathbf{L}_{I,i}(\check{z})\mathbf{L}_{I,i+1}(\hat{z})=0\,.
\end{equation}
Interestingly, as a consequence of (\ref{eq:rdecproj}), identity (\ref{eq:ident})
holds true for any generalized rectangular representation.

\section{Reordering formula\label{sub:Reorder}}

The reordering of the oscillators in the auxiliary space we are interested
in is of the form
\begin{equation}
e^{\bar{\mathbf{a}}A}\cdot B\cdot e^{\mathbf{a}C}=e^{\bar{\mathbf{a}}A}\circ\tilde{B}\circ e^{\mathbf{a}C}\,.\label{eq:reorder2}
\end{equation}
Using $e^{-\bar{\mathbf{a}}A}\mathbf{a}{}^{n}e^{\bar{\mathbf{a}}A}=(\mathbf{a}+A)^{n}$
we find that 
\begin{equation}
\tilde{B}=\sum_{n=0}^{\infty}\frac{(-1)^{n}}{n!}\, A^{n}\, B\, C^{n}\,,\quad\quad B=\sum_{n=0}^{\infty}\frac{1}{n!}\, A^{n}\,\tilde{B}\, C^{n}\,.
\end{equation}
Here we did not specifying any commutation relations among $A,B,C$.

\bibliographystyle{utcaps}
\bibliography{Qshift_v3_grp}

\end{document}